\documentclass[twocolumn]{aastex63}

\usepackage{enumerate}

\newcommand{\sigs}{$\sigma'_{\star,\rm{int}}$}
\newcommand{\sigg}{$\sigma'_{g,\rm{int}}$}
\makeatletter
\g@addto@macro{\UrlBreaks}{\UrlOrds}
\makeatother
\usepackage{CJKutf8}

\usepackage{import}
\usepackage{amsmath}
\usepackage{amssymb}
\usepackage{natbib}
\defcitealias{muzzin13}{M13} 

\newcommand\s{$\sim$}
\newcommand{\e}{\'{e}}

\usepackage{xcolor}
\usepackage[normalem]{ulem}
\usepackage{blindtext}
\shorttitle{LEGA-C DR3}
\shortauthors{van der Wel et al.}

\begin{document}

\title{
The Large Early Galaxy Astrophysics Census (LEGA-C) Data Release 3: 3000 High-Quality Spectra of $K_s$-selected galaxies at $z>0.6$
}
\correspondingauthor{Arjen van der Wel}
\email{arjen.vanderwel@ugent.be}

\author{Arjen van der Wel}
\affil{Sterrenkundig Observatorium, Universiteit Gent, Krijgslaan 281 S9, 9000 Gent, Belgium}
\affil{Max-Planck Institut f\"{u}r Astronomie K\"{o}nigstuhl, D-69117, Heidelberg, Germany}

\author{Rachel Bezanson}
\affil{University of Pittsburgh, Department of Physics and Astronomy, 100 Allen Hall, 3941 O'Hara St, Pittsburgh PA 15260, USA}

\author{Francesco D'Eugenio}
\affil{Sterrenkundig Observatorium, Universiteit Gent, Krijgslaan 281 S9, 9000 Gent, Belgium}

\author{Caroline Straatman}
\affil{Sterrenkundig Observatorium, Universiteit Gent, Krijgslaan 281 S9, 9000 Gent, Belgium}

\author{Marijn Franx}
\affil{Leiden Observatory, Leiden University, P.O.Box 9513, NL-2300 AA Leiden, The Netherlands}

\author{Josha van Houdt}
\affil{Max-Planck Institut f\"{u}r Astronomie K\"{o}nigstuhl, D-69117, Heidelberg, Germany}

\author{Michael V.~Maseda}
\affil{Leiden Observatory, Leiden University, P.O.Box 9513, NL-2300 AA Leiden, The Netherlands}

\author{Anna Gallazzi}
\affil{INAF-Osservatorio Astrofisico di Arcetri, Largo Enrico Fermi 5, I-50125 Firenze, Italy}

\author{Po-Feng Wu \begin{CJK*}{UTF8}{bkai}(吳柏鋒)\end{CJK*}}
\affil{National Astronomical Observatory of Japan, Osawa 2-21-1, Mitaka, Tokyo 181-8588, Japan}

\author{Camilla Pacifici}
\affil{Space Telescope Science Institute, 3700 San Martin Drive, Baltimore, MD 21218, USA}

\author{Ivana Barisic}
\affil{Max-Planck Institut f\"{u}r Astronomie K\"{o}nigstuhl, D-69117, Heidelberg, Germany}

\author{Gabriel B. Brammer}
\affil{Cosmic Dawn Center (DAWN)}
\affil{Niels Bohr Institute, University of Copenhagen, Jagtvej 128, København N, DK-2200, Denmark}

\author{Juan Carlos Munoz-Mateos}
\affil{European Southern Observatory, Karl-Schwarzchild Straße 2, D-85748 Garching bei Munchen, Germany}

\author{Sarah Vervalcke}
\affil{Sterrenkundig Observatorium, Universiteit Gent, Krijgslaan 281 S9, 9000 Gent, Belgium}

\author{Stefano Zibetti}
\affil{INAF-Osservatorio Astrofisico di Arcetri, Largo Enrico Fermi 5, I-50125 Firenze, Italy}

\author{David Sobral}
\affil{Department of Physics, Lancaster University, Lancaster LA1 4YB, UK}

\author{Anna de Graaff}
\affil{Leiden Observatory, Leiden University, P.O.Box 9513, NL-2300 AA Leiden, The Netherlands}

\author{Joao Calhau}
\affil{Instituto de Astrofísica de Canarias, E-38200 La Laguna, Tenerife, Spain}
\affil{Departamento de Astrofísica, Universidad de La Laguna, E-38206 La Laguna, Spain}

\author{Yasha Kaushal}
\affil{University of Pittsburgh, Department of Physics and Astronomy, 100 Allen Hall, 3941 O'Hara St, Pittsburgh PA 15260, USA}

\author{Adam Muzzin}
\affil{Department of Physics and Astronomy, York University, 4700 Keele St., Toronto, Ontario, M3J 1P3, Canada}

\author{Eric F.~Bell}
\affil{Department of Astronomy, University of Michigan, 1085 South University Ave., Ann Arbor, MI 48109, USA}

\author{Pieter G.~van Dokkum}
\affil{Astronomy Department, Yale University, 52 Hillhouse Ave, New Haven, CT 06511, USA}

\begin{abstract}
We present the third and final data release of the Large Early Galaxy Astrophysics Census (LEGA-C), an ESO/VLT public spectroscopic survey targeting $0.6 < z < 1.0$,  Ks-selected galaxies. The data release contains 3528 spectra with measured stellar velocity dispersions and stellar population properties, a 25-fold increase in sample size compared to previous work. This $K_s$-selected sample probes the galaxy population down to $\sim0.3 L^*$, for all colors and morphological types. Along with the spectra we publish a value-added catalog with stellar and ionized gas velocity dispersions, stellar absorption line indices, emission line fluxes and equivalent widths, complemented with structural parameters measured from HST/ACS imaging.  With its combination of high precision and large sample size, LEGA-C provides a new benchmark for galaxy evolution studies.

\end{abstract}

\keywords{galaxies: high-redshift -- galaxies: kinematics and dynamics -- galaxies: structure}

\section{Introduction}
\label{section:intro}

The  successful tenure of ESO's VLT/VIMOS spectrograph \citep{le-fevre03} was completed by the implementation of the two most recent Public Spectroscopic Surveys: LEGA-C \citep{van-der-wel16, straatman18} and VANDELS \citep{mclure18, pentericci18}. Both surveys, carried out from 2015 to 2018, are characterized by a focus on depth and data quality, representing a clear departure from VIMOS' previous use as the `redshift machine' for which it was designed. The detector upgrade and instrument refurbishment in 2010 made this possible. 

The practical goal of the LEGA-C survey is to collect high signal-to-noise, high resolution spectra for thousands of galaxies in the redshift range $0.6\lesssim z\lesssim1$. This allows, for the first time for the general galaxy population, to probe stellar populations (ages and metallicities) and stellar kinematics (velocity dispersions) of thousands of galaxies at a look-back time of $\sim7$~Gyr. 

The primary science goals span all key open questions in the field of galaxy formation and evolution. First, what are the star-formation histories of galaxies? The spectra constrain the bulk formation age of the stellar population, the metal enrichment history and how bursty the star formation history is. Crucially, at large look-back time it is easier to resolve the main formation phase; in the present-day Universe most stars formed many Gyr ago, and old stellar populations are difficult to dissect due to their slow spectral evolution. Second, how does star formation quench in massive galaxies? Does this occur rapidly due to strong outflows, possibly associated with a preceding increase in star formation, or relatively slowly due to a lack of cooling to supplement fuel? The role of environment, Active Galactic Nuclei and mergers can be examined thanks to the large sample size. Third, how do galaxies evolve after the cessation of star formation? They can evolve merely passively, or continue growth and structural evolution through merging. The kinematic properties and chemical composition constrain this process.

These topics are interconnected and also connect to quantifying the fundamental properties of the galaxy population as a whole. Given the dynamical mass measurements and spectroscopy-based estimates with unprecedented quality of the stellar mass, we will be able to assess the contribution of dark matter (and/or a bottom-heavy stellar initial mass function) across galaxy types, and how this changes with cosmic time. The abundances of heavy elements can be constrained and compared with those of the (ionized) interstellar medium to indirectly constrain the importance of in- and outflows of galaxies.

Initial results based on the 1st and 2nd data releases \citep[][hereafter, DR2]{straatman18} illustrate the broad range of applications of LEGA-C data. \citet{bezanson18} showed that $z \sim 1 $ quiescent galaxies show more rotation in their stellar body than equally massive counterparts at $z \sim 0$, providing strong evidence for a reduction in net angular momentum through merging after galaxies become quiescent, with modest environmental dependence \citep{cole20}. At the same time, the question whether galaxies change their structure upon `quenching' -- the transition from an actively star-forming state to a quiescent state -- remains open.  \citet{wu20} and \citet{deugenio20} show that post-starburst galaxies rapidly formed a large number of stars in the central regions of galaxies, producing compact galaxies. But, at the same time, larger galaxies are generally younger \citep{wu18b}, reflecting a complexity in galaxy evolution that can only be revealed by high-quality spectroscopic data.  The specific implication here is that there are multiple evolutionary pathways to reach a quiescent state. This is just one example of how constraints on stellar populations \citep{wu18a} increase our physical insight into the evolutionary process, and our first attempts to quantify this information demonstrated that the `downsizing' trend -- that galaxies with higher stellar masses are older -- was already evident at $z\sim 1$ \citep{chauke18}. The quality of the LEGA-C spectroscopy is also such that, for the first time, we could detect and quantify bursts of star formation at $z \sim 1$ that occurred after an initial period of quiescence at $z\sim 2$ \citep{chauke19}, as well as a novel approach to quantifying dust attenuation in star-forming galaxies; \citet{barisic20} shows steep attenuation curves, especially for face-on galaxies, and frequently detect the 2175\AA bump in individual galaxies for the first time.

 The ultimate goal to jointly analyse the evolution of stellar populations and the kinematic state of the stellar body is now within reach, which starts with examining the evolution and detailed properties of scaling relations such as the Fundamental Plane \citep{de-graaff20, de-graaff21}. Detailed comparisons with the latest cosmological simulations of galaxy formation will then improve the fidelity of the physical models used in those simulations. Finally, the large sample size also allows to probe rare classes of objects such as radio-loud Actice Galactic Nuclei and explore their kinematic and stellar population properties of the first time \citep{barisic17}.

With the full LEGA-C sample, where we increase the sample of spectra with measured redshifts and stellar velocity dispersions from 1447 to 3528 compared with DR2, we enable the community to pursue this line of research. In this paper we will first present the updated sample and the improvements in the data reduction pipeline compared to DR2, as well as an analysis of the \textit{Hubble Space Telescope/Advanced Camera for Surveys (HST/ACS)} data that are available for nearly all objects in our sample (Section \ref{section:data}). We then describe the basic measurements that we are publishing with this paper: stellar and ionized gas velocity dispersions, spectral indices, and emission line fluxes, as well as inferred quantities such as dynamical mass estimates (Section \ref{section:measurements}). Our published measurements are independent of cosmological parameters, with the exception of dynamical mass, which has one element (galaxy size) that requires a conversion from observed to physical units. There we adopt a flat $\Lambda$CDM cosmology with $\Omega_m = 0.3$ and $H_0 = 70$~km~s$^{-1}$.

\newpage
\section{Data}
\label{section:data}

\begin{table*}
\begin{center}
  \caption{ Sample selection and success rate}
 \begin{tabular}{|l|c|c|c|c|c|c|}
\hline
Sample & Redshift & $K$-band limit & \#~in UltraVISTA$^1$ & \#~in LEGA-C & with $z$ & with \sigs \\
\hline
Parent         &     Any     & $< 23.4$ (90\% compl.)           &  157,667 &    4,081    & 3,937 & 3,528 \\
Primary        &     $0.6<z<1.0$ & $< 20.7-7.5\log((1+z)/1.8)$  &        8,655 &    3,029    & 3,018 & 2,932 \\
Filler I       &     $z>1$       & $< 20.36$                     &       3,511 &      260    &   184 &   163 \\
Filler II      &     $0.6<z<1.0$ & $> 20.7-7.5\log((1+z)/1.8)$  &       32,261 &      590    &   553 &   380 \\
Filler III     &     Other       & Other                      &      113,240 &      202    &   182 &    53 \\
\hline
\end{tabular}
\tablecomments{Sample selection criteria for the primary sample and the 3 categories of fillers. Numbers of spectra with successfully measured redshifts and stellar velocity dispersions (Section \ref{section:measurements} are given in the last two columns. $^1$: in \citet{muzzin13a} catalog and within LEGA-C footprint (Fig.~\ref{fig:radec}).}
\label{tab:sample}
\end{center}
\end{table*}

\subsection{The LEGA-C Survey: Observations}

\begin{figure}[t]
\epsscale{1.2}
\plotone{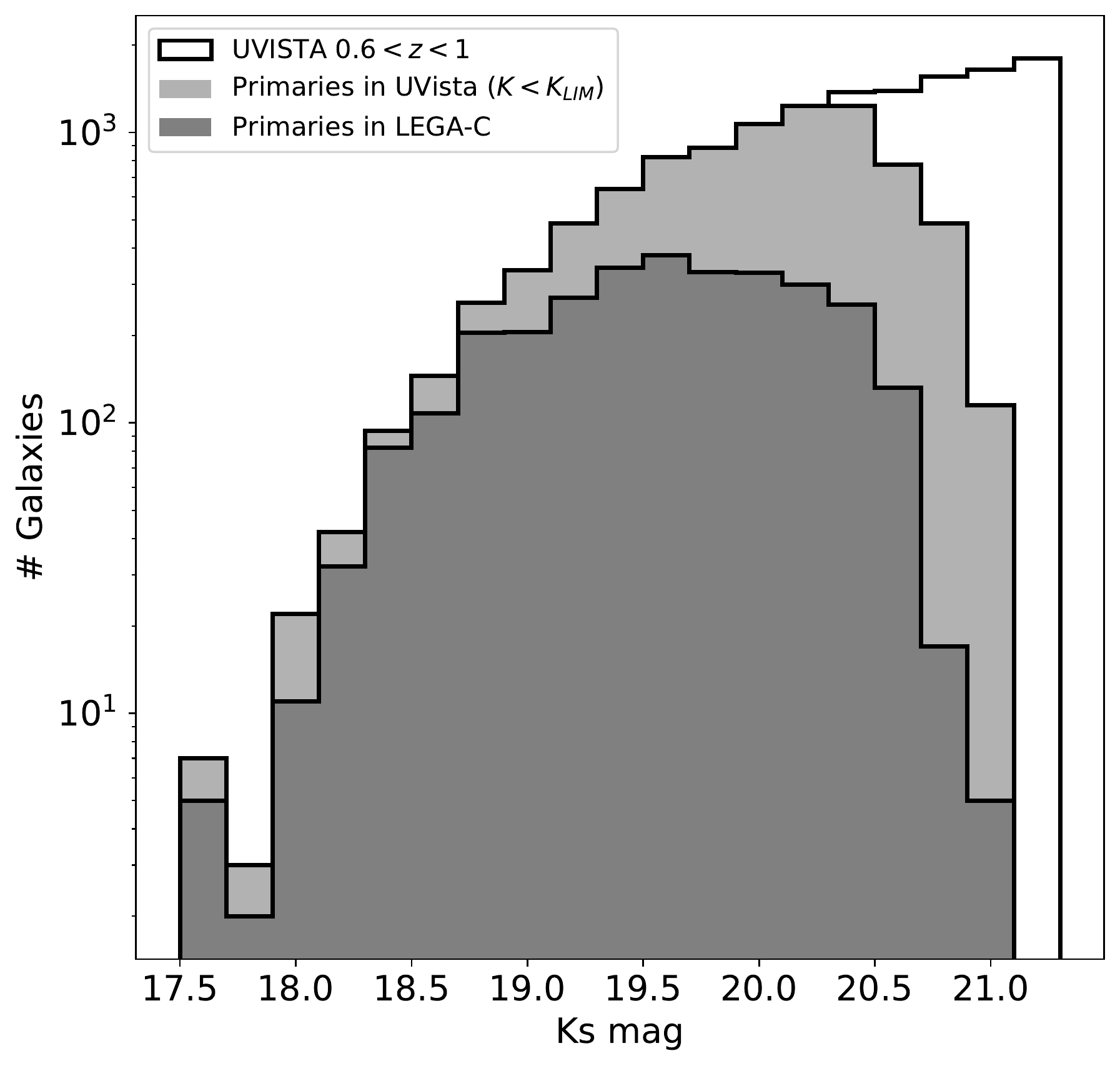} 
    \caption{$K_s$ band distribution of the Ultra-VISTA catalog for galaxies with (spectroscopic or photometric) redshifts in the range $0.6<z<1$ compared to the $K_s$-band selected parent sample of candidates for the LEGA-C survey. The gradual decrease with $K$ magnitude for the parent sample is the result of the redshift-dependent $K_s$ limit. The darkest histogram gives the $K_s$-band distribution for targets with LEGA-C observations.  Masks were designed by including objects sorted by $K_s$ flux (brightest first). As a result, down to $K_s\sim 19.5$ LEGA-C observed 50\% or more of the population, while for $K_s=20-20.5$ it is around 30\%. \label{fig:K_hist}}
\end{figure}

\begin{figure}[t]
\epsscale{1.2}
\plotone{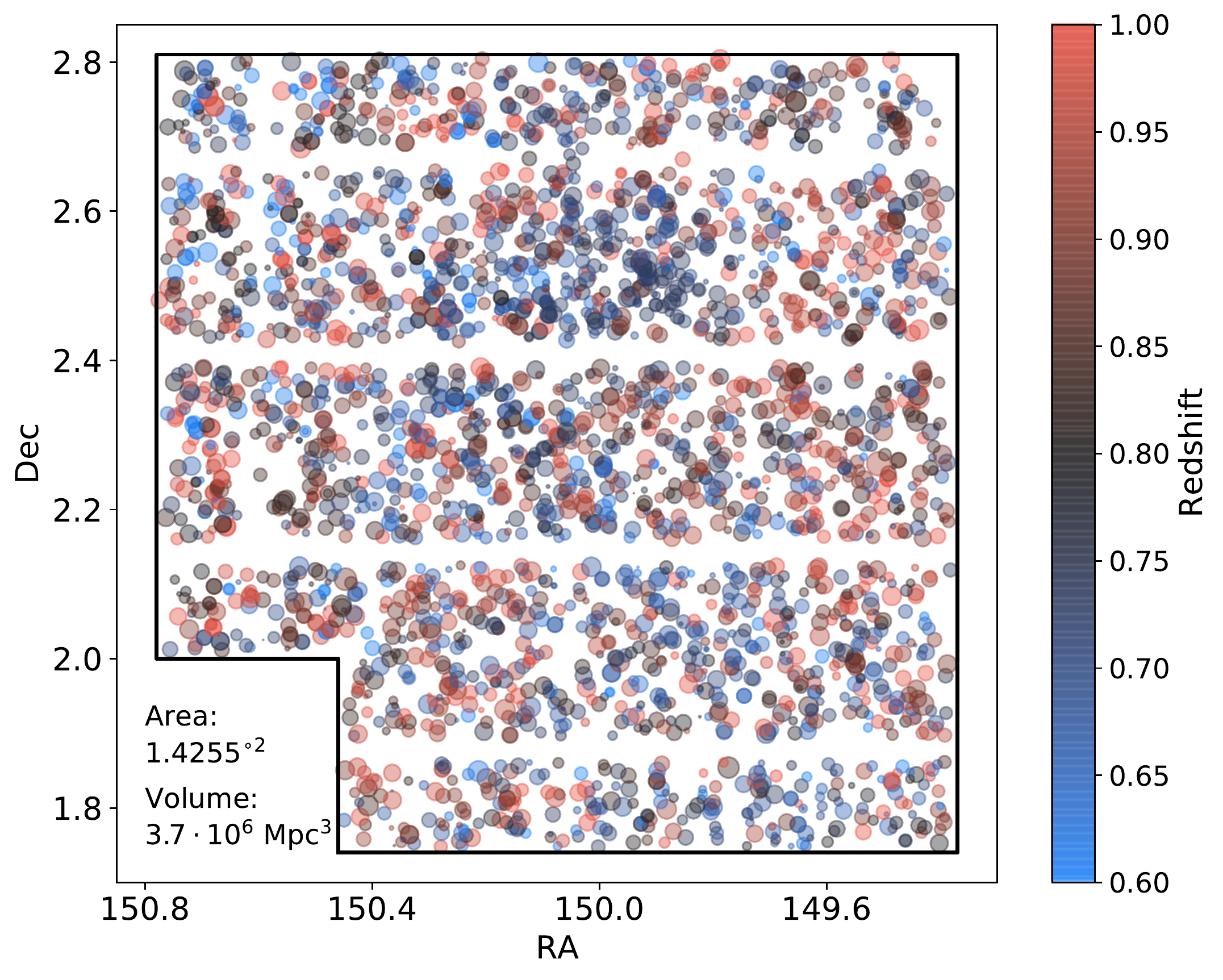} 
    \caption{Survey footprint and source distribution in the COSMOS field. Symbol size reflects the stellar velocity dispersion (Section \ref{section:kinematics}) in order to emphasize the more massive galaxies.
    \label{fig:radec}}
\end{figure}

\begin{figure}[t]
\epsscale{1.2}
\plotone{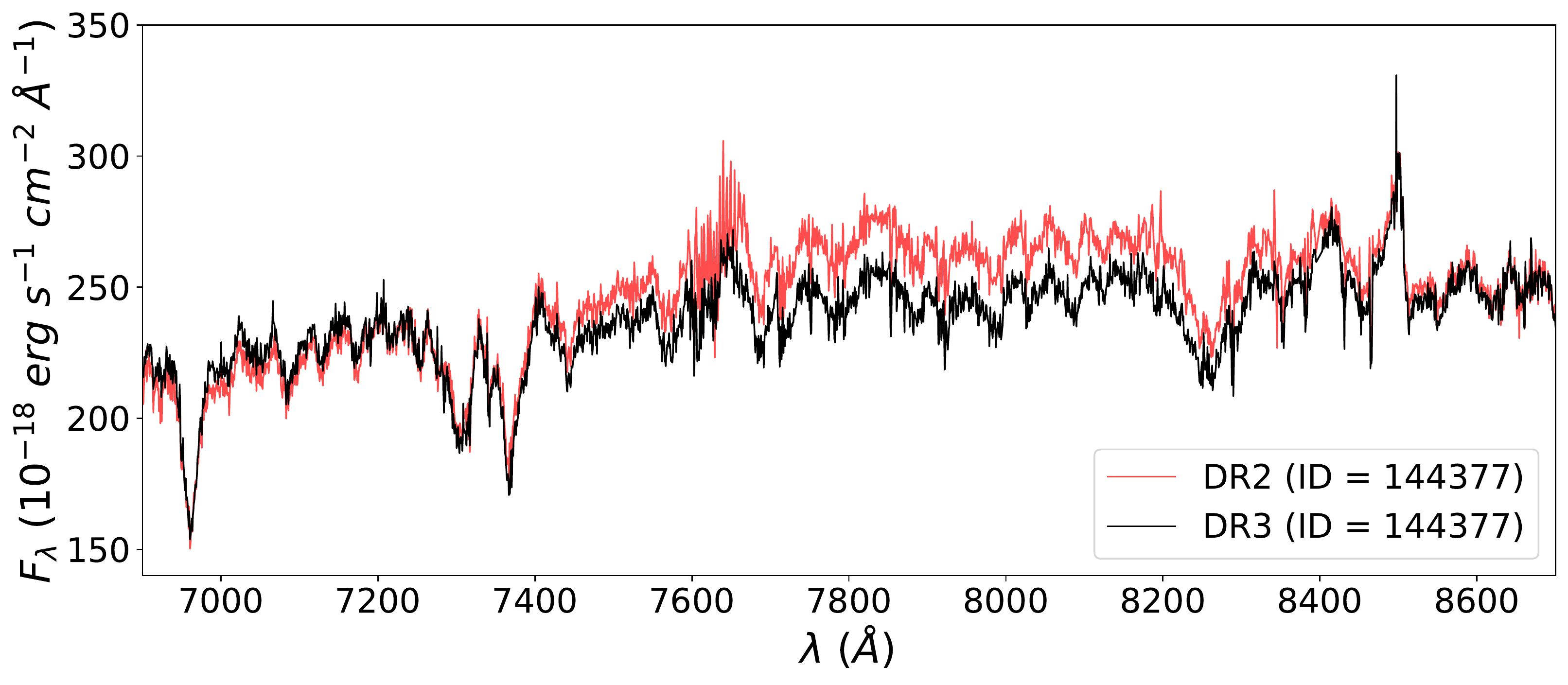} 
    \caption{Example comparison of DR2 (red) and DR3 (black) versions of the same spectrum, illustrating the improvements and changes to the data processing techniques. The overall normalization is affected (on the level of a few percent) by the zero point changes in the photometry used for the flux calibration (Appendix \ref{AppendixA}). The shape of the continuum is affected by the changes in the photometric SED fit (\ref{section:fluxcal}). The improved telluric absorption correction (\ref{sec:telluric}) is most clearly seen in the wavelength range $7600-7700$~\AA, where the systematic residual in the form of high-frequency `beating' is reduced in the DR3 version of the spectrum.\label{fig:dr23_spec}}
\end{figure}

The survey goals and design were extensively discussed by \citet{van-der-wel16} and \citet{straatman18}. The key characteristic of the survey is that it is $K_s$-band selected. As we will see, the long integration times of $\sim$20 hours lead to a very high success rate in measurement redshifts, stellar kinematics, and stellar population properties, even for the faintest, reddest targets. From the UltraVISTA catalog \citep{muzzin13} we first create a parent sample of galaxies with (spectroscopic or photometric) redshifts in the range $0.6<z<1$ and with $K_s$ magnitudes brighter than a redshift-dependent limit $K_{s,\rm{LIM}} = 20.7-7.5\log((1+z)/1.8)$ (Fig.~\ref{fig:K_hist}). When designing the slit masks we included targets ordered by $K_s$ magnitude (brightest first).  As the mask fills up, slit collisions prevent the inclusion of fainter galaxies, leading to higher (and precisely quantifiable) completeness for brighter targets. Any remaining mask space is used to include fillers: $K_s$-bright objects at $z>1$, fainter targets in the LEGA-C redshift range, and a mixed bag objects at lower redshifts and faint, higher-redshift objects. We refer to Table \ref{tab:sample} and, for comparison, to the equivalent Table 1 in the survey paper \citep{van-der-wel16} for an overview of the LEGA-C sample. The success rate of the redshift and stellar velocity dispersions (discussed in Section \ref{section:measurements}) is also given.

The final survey footprint is created by 30 slightly overlapping pointings, together covering 1.4255 square degrees (Figure \ref{fig:radec}), giving a total survey co-moving volume of $3.664\times10^6$~Mpc$^{3}$ between $z=0.6$ and $z=1.0$, equivalent to the local volume out to $\sim$100 Mpc ($z\sim 0.023$). 

Two additional pointings, contained within the main footprint, were chosen for specific purposes. One of these has the same selection approach as the 30 main pointings but the slits are oriented in the East-West direction where all other masks and slits in the North-South direction. The goal here is to obtain kinematic information along two perpendicular axes.  The other extra mask has a different selection function and is optimized to include red, low-mass galaxies that are otherwise rarely included in the survey. This increases the dynamic range in scaling relations.

The 32 masks produced 4081 galaxy spectra, with 3741 unique objects (340 are observed more than once). We have 3029 spectra of primary targets (compare their $K$ magnitude distribution with the $K$ magnitude distribution of the parent sample in Figure \ref{fig:K_hist}), and 1052 of fillers. The success rate (in terms of making key measurements) is very high; the fraction of primary spectra with successful redshift and stellar velocity dispersion measurements is close to 100\% as described in detail below (Section \ref{section:measurements}). The completeness of the LEGA-C sample in relation to both the parent sample and the full galaxy population is discussed in detail in Appendix \ref{AppendixA1}.

Observations were carried out between December 2014 and March 2018.  In total, 211 (mostly partial) nights were spent observing in Visitor Mode, for a total on-source integration time of 695 hours out of the allocated 1107 hours. The resulting efficiency of 63\% includes not only the usual overheads, but also weather and technical losses. The anticipated efficiency in the survey management plan was also precisely 63\% (including weather losses);  from a scheduling and execution perspective LEGA-C can then be considered a success. The motivation for executing the survey in Visitor Mode is obvious: even without weather and technical losses Service Mode observations would have at most 50\% efficiency given standard calibration requirements and scheduling constraints. The increased efficiency is mainly due to an efficient calibration plan and flexibility in designing the observations, circumventing the need for hourly re-acquisition and alignment.  On a general note, given that VIMOS was not originally designed for long exposures and tracking objects for many hours, the performance of instrument and telescope has been very good.

\subsection{Custom LEGA-C pipeline}\label{sec:pipeline}
As described in the DR2 paper we designed a custom pipeline built on top of ESO-provided data handling infrastructure. Here, for DR3, we make three improvements in the data analysis related to 1) sky subtraction/object extraction, 2) telluric absorption correction, and 3) flux calibration.

\subsubsection{Sky Subtraction / Object Extraction}
For DR2 we created sky$+$object models in the spatial direction where the object was represented by a Gaussian. We knew that a Gaussian imperfectly describes seeing-convolved galaxy light profiles, and we had already addressed this by refitting the object in the co-added, 2-dimensional spectra with a Moffat profile. For most galaxies this worked fine, but for the many bright galaxies that illuminate more than half of the slit this approach was insufficient. For DR3 we make use of the fact that most galaxies have been imaged by HST, for which we derived S\'ersic light profiles (see Section \ref{sec:hst}). Based on the observed light profile in the LEGA-C spectra, collapsed in the wavelength direction, we construct a galaxy$+$seeing profile by convolving the S\'ersic profile with a Moffat profile, and propagating the resulting profile through a rectangular slit with width 1 arcsec. The two Moffat parameters are optimized in this fit, producing a spectral extraction kernel that encompasses both the galaxy and the seeing. As before (see DR2) this kernel is used at each wavelength bin to fit for the sky and the object flux.

After co-addition of the sky-subtracted 2-dimensional spectra we check for systematic effects in the sky subtraction. For some galaxies located near the edge of their slit, or where multiple objects lie in the same slit, the sky subtraction was imperfect. In those cases, as for DR2, a new sky subtraction was performed by fitting a new Moffat profile, producing a new sky and a new object extraction.

\subsubsection{Telluric Absorption}\label{sec:telluric}
For DR2 we created a telluric absorption spectrum for each of the 32 masks based on the spectrum of a blue star included in those masks. This left two unaddressed problems. First, for some blue stars the wavelength coverage was insufficient to probe the strong telluric features around 9000\AA. Second, a fixed telluric spectrum for each of the $>100$ objects in a mask leaves very significant high-frequency residuals (see Fig.~\ref{fig:dr23_spec}) due to 2 effects: errors in the wavelength calibration at the level of $\sim0.1-0.2$\AA~due to small, but unavoidable variations in object-slit alignments (if the blue star has a different alignment than a galaxy, then there will be a wavelength shift in the spectrum and, consequently, a difference in the telluric absorption spectrum); galaxies are not point sources, such that the line-spread function is broader for objects with larger angular sizes, leading to a smoother telluric absorption spectra for galaxies than for stars -- this effect varies from galaxy to galaxy.

To remedy these effects we revised the telluric absorption correction for DR3 compared to DR1 and DR2:
\begin{itemize}
    \item All telluric absorption features in the $6000-9300$\AA~wavelength range arise due to only two species:  $H_2O$ and $O_2$. We choose a well-calibrated white dwarf spectrum taken with ESO/X-SHOOTER (WD0839-327, Programme ID 098.D-0392(A)) and create $H_2O$ and $O_2$ telluric absorption template spectra using the wavelength ranges $6860 < \lambda/$\AA$~< 6925$ and $7580 < \lambda/$\AA$~< 7710$ (the A and B bands) for $O_2$ and the wavelength ranges $6925 < \lambda/$\AA$~< 7580$ and $\lambda/$\AA$~> 7710$ for $H_2O$.  Note that the combination covers the entire wavelength range $\lambda > 6860$\AA, that is, most of the LEGA-C spectra.
    \item We fit the $H_2O$ and $O_2$ templates to the blue star spectrum, varying the strength of the $H_2O$ and $O_2$ features. The key here is all features arising due to $O_2$ vary together (primarily as a function of airmass) as do the $H_2O$ features (mostly as a function of humidity). This produces a master telluric absorption spectrum for each mask. This approach remedies incomplete wavelength coverage for the blue stars in some of our masks.
    \item For each galaxy the co-added 1D spectrum is divided by the master absorption spectrum for its mask, but allowing two parameters to vary: a shift in wavelength (to account for slit misalignments) and a smoothing factor (to account for the fact that galaxies have a broader line-spread functions than stars). Note that the strengths of the telluric absorption features are kept fixed for this step.
\end{itemize}

The resulting telluric absorption correction spectra are divided into the galaxy spectra and propagated into the variance spectra.

\subsubsection{Flux Calibration}\label{section:fluxcal}
The standard procedure for flux calibration, based on spectroscopic observations of standard stars with the same instrumental setup, is not feasible for LEGA-C.  Slit/galaxy alignment varies from exposure to exposure, and throughout the field of view, so that it is impossible to construct a calibration spectrum that can be applied to all galaxy spectra. Instead, we calibrate our galaxy spectra with the aid of well-calibrated photometric spectral energy distributions.

The approach used for DR3 is conceptually the same as described in the DR2 paper, but several improvements were made. As demonstrated in Appendix \ref{AppendixA}, the photometric zero points are improved compared to the original photometry from \citet{muzzin13}, and we only use a subset of photometric bands ($BVrizYJ$) for the SED fit. For consistency with full-spectral fitting results that will be described in a forthcoming paper, we perform the photometric SED fits with the {\tt bagpipes} code \citep{carnall18} instead of {\tt FAST} \citep{kriek09}, which was used for DR2. We use {\tt bagpipes} to fit the new photometric SED with double power-law star-formation histories (instead of exponentially declining star-formation histories which were used for DR2); recent work \citep[e.g.,][]{carnall19} shows that, especially for star-forming galaxies, rising-then-declining star-formation rates are a necessary choice to avoid biases in, e.g., stellar mass estimates. But for our purpose of flux calibration this change is not particularly relevant, and only made for consistency.  The best-fitting templates from the SED fits are compared with the uncalibrated LEGA-C spectra by fitting a 5th-order Legendre polynomial to the ratio of the two. The calibrated spectra (as well as the variance spectra) are then obtained by multiplying the polynomials with the uncalibrated spectra.  The resulting, calibrated DR3 spectra differ slightly in overall shape and normalization compared to DR2, as illustrated for a typical example in Figure \ref{fig:dr23_spec}.

\subsection{Hubble Space Telescope Imaging}
\label{sec:hst}

We measure structural parameters as described by \citep{van-der-wel16} from the relatively shallow (but wide area) COSMOS HST/ACS imaging \citep{scoville07}. The formal uncertainties on the structural parameters are assigned on the basis of the total $S/N$ of the object, as described by \citet{van-der-wel12}, but as described below they are adapted based the comparison with the structural parameters from CANDELS \citep{grogin11, koekemoer11, van-der-wel12} as determined with the MegaMORPH software package \citep{haussler13}. The latter are obtained from deeper data, and with differences in methodology, most notably distortion correction, PSF construction and background subtraction. It is beyond the scope of this paper to present an in-depth analysis of the differences, and here we simply compare the results and interpret these as indicative of the systematic uncertainties. For the 151 overlapping galaxies in our catalog for which the S\'ersic parameters did not reach a limiting value (such as S\'ersic index $n=6$) in {\tt galfit} \citep{peng10} we find a median effective radius difference of $-2\%$ (our estimates are smaller), with a bi-weight scatter of 11\%. For the S\'ersic index, shape $b/a$, position angle, and total magnitude these statistics are, respectively: $+3 \pm 17\%$,  $-1\pm3\%$, $-1.8\pm2.8\deg$, and $0.06\pm0.18$~mag (ours are fainter). We apply these scatter values (but not the systematic offsets) as minimum uncertainties in our catalog.  Correlations of this scatter with object brightness, size, and S\'ersic index are very weak; the scatter is, apparently, not due to limiting $S/N$ in the COSMOS imaging but must be mainly due to a series of systematic effects resulting from differences in methodology.  We can therefore interpret the scatter values as the random uncertainty, and we assign those values to all objects for which the formal random uncertainties are smaller. This is the case for more than $90\%$ of the galaxies in our sample.

\begin{figure*}[!t]
\epsscale{1}
\plotone{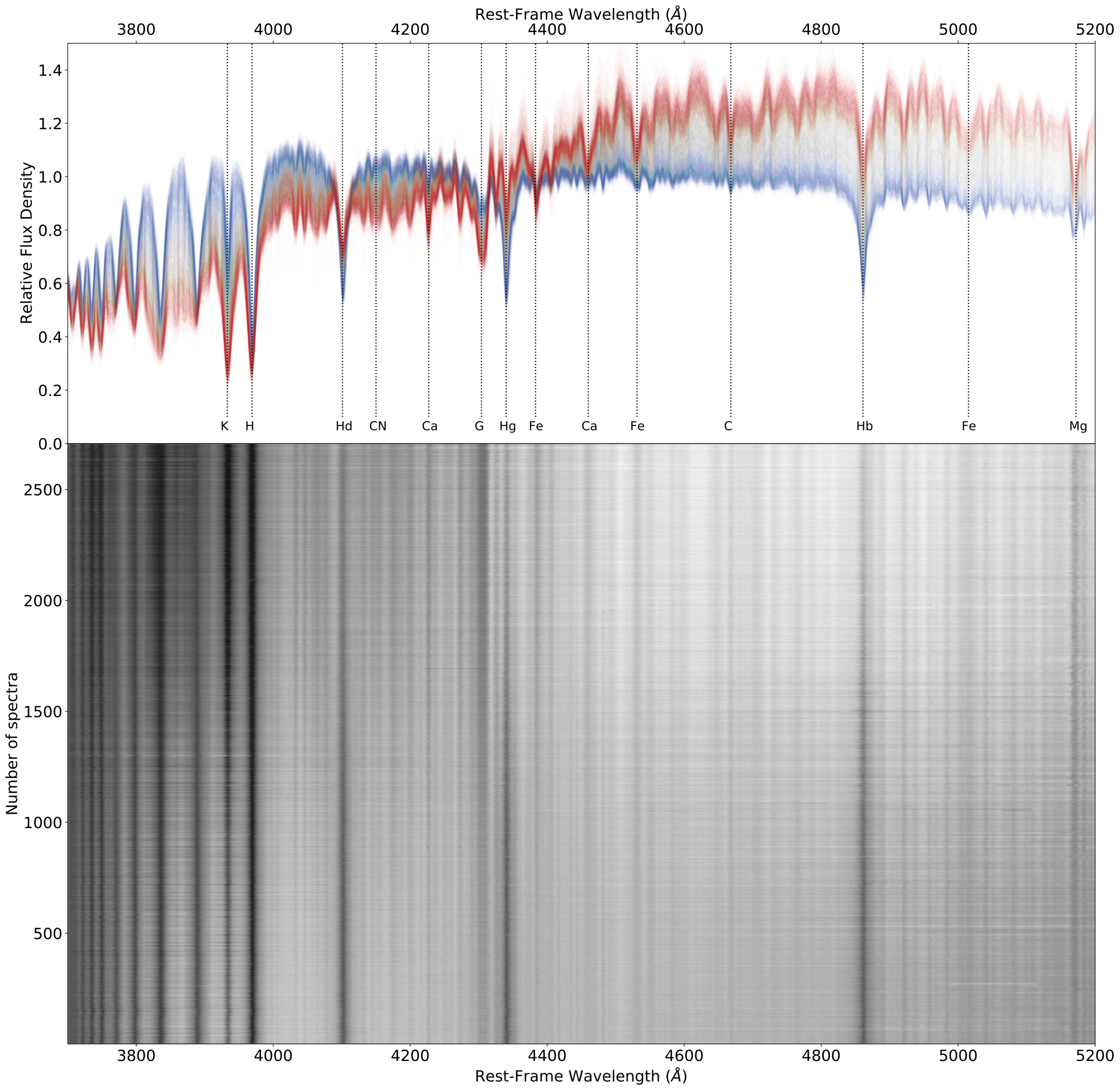}
    \caption{Compilation of 2707 normalized spectra with measured stellar velocity dispersions and H$\gamma_A$ indices. Top: The color coding is based on the order of H$\gamma_A$ absorption (strongest/youngest: blue; weakest/oldest: red) . Bottom: Horizontally aligned, vertically sorted by H$\gamma_A$ absorption. For visualization purposes outlying wavelength elements were replaced by the continuum model described in Section \ref{section:emission} (outliers are defined as wavelength elements with fluxes more than 10\% different from the model), and boxcar smoothed by 200 km s$^{-1}$. In addition, the additive and multiplicative polynomials used to match the model spectra and the observed spectra were removed: this way variations in spectral slope that are not associated with the stellar content (e.g., dust attenuation and errors in the relative flux calibration) are removed. The color intensity (top panel) declines at longer wavelengths due to the smaller number of spectra with wavelength coverage.
    \label{fig:spec_blue}}
\end{figure*}

\begin{figure*}[!th]
\epsscale{.8}
\plotone{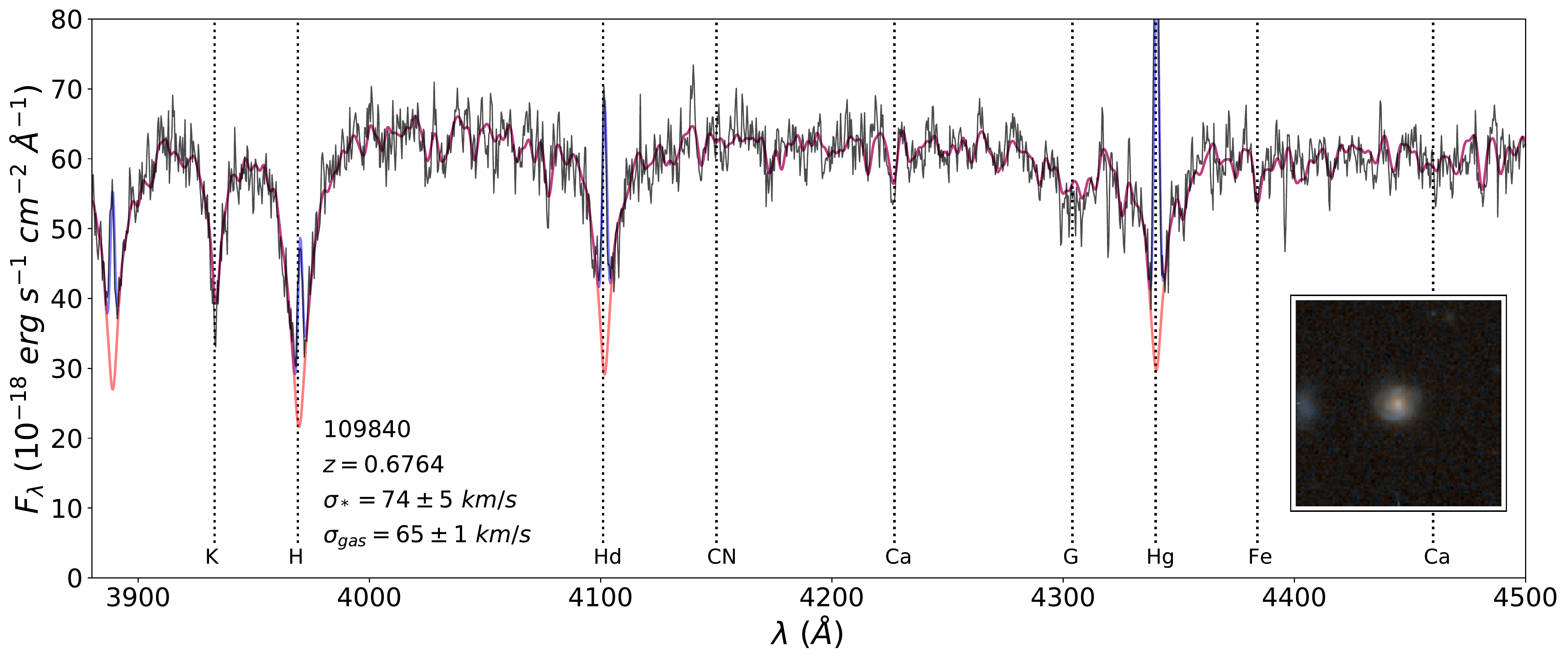} 
\plotone{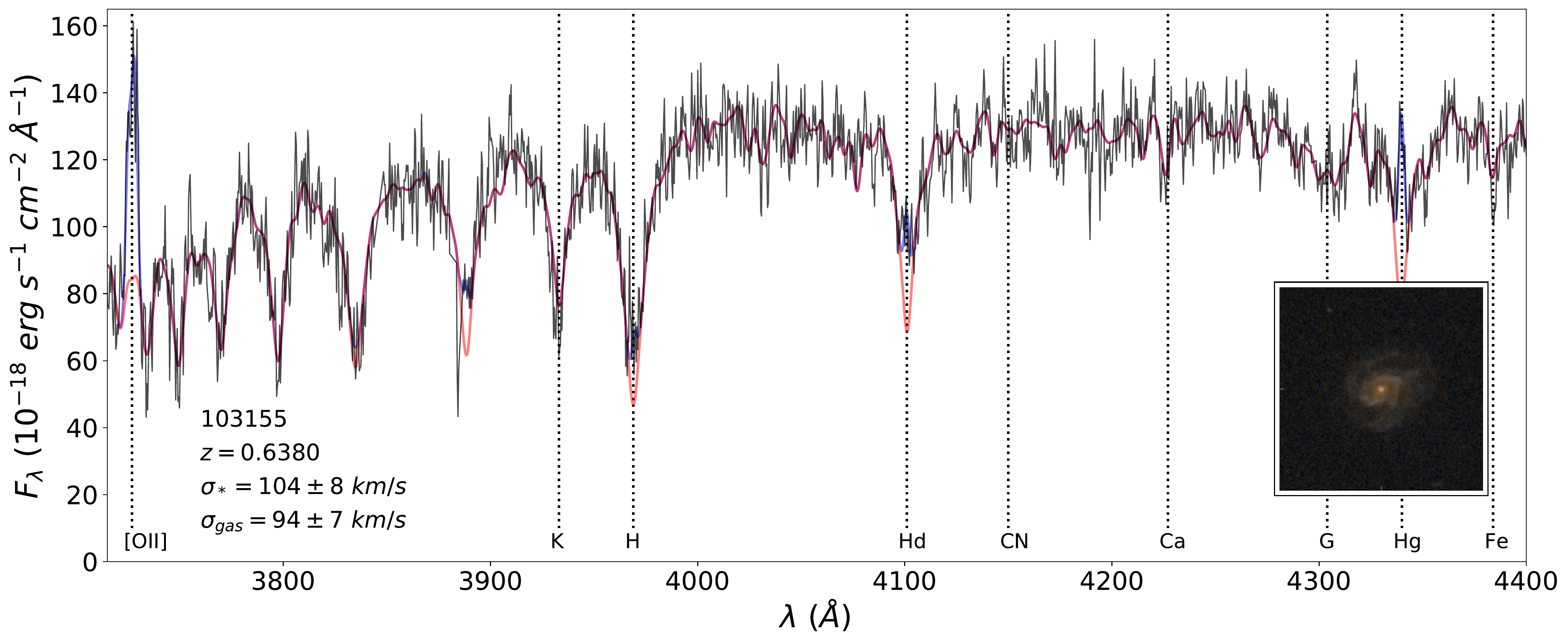} 
\plotone{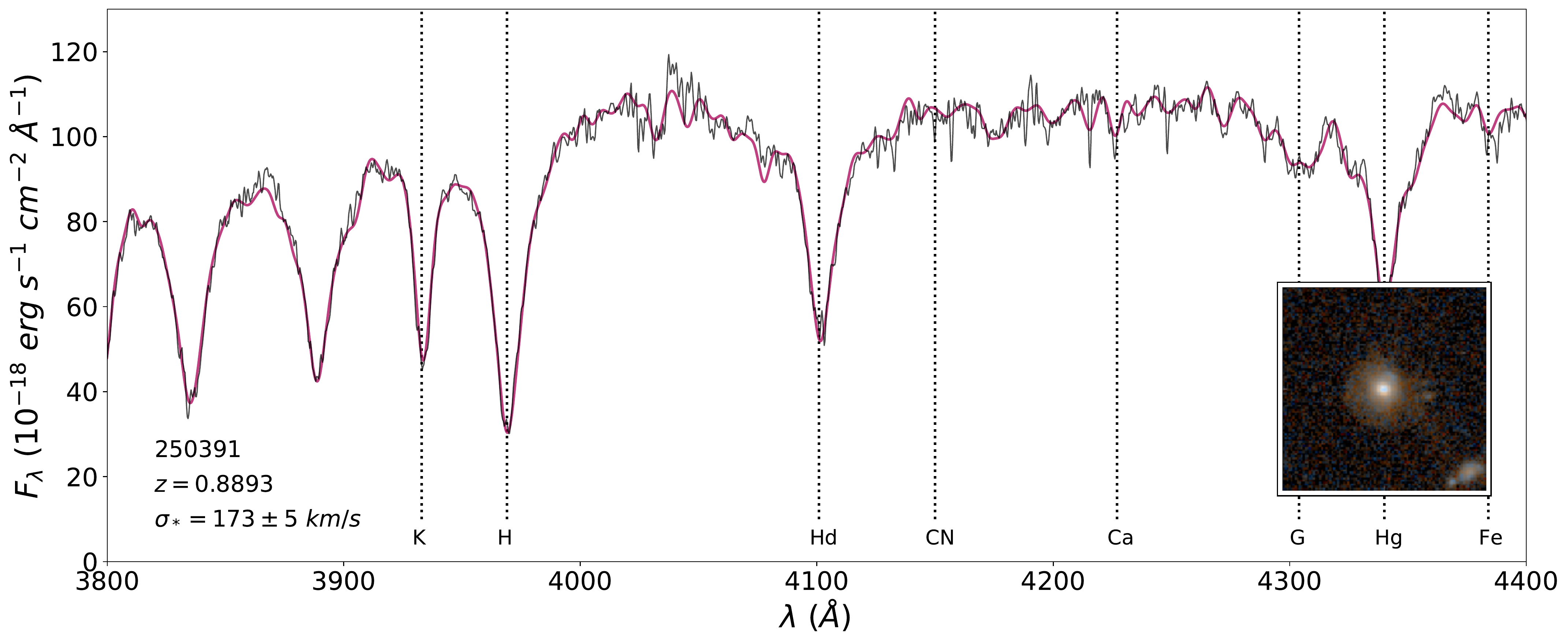} 
\plotone{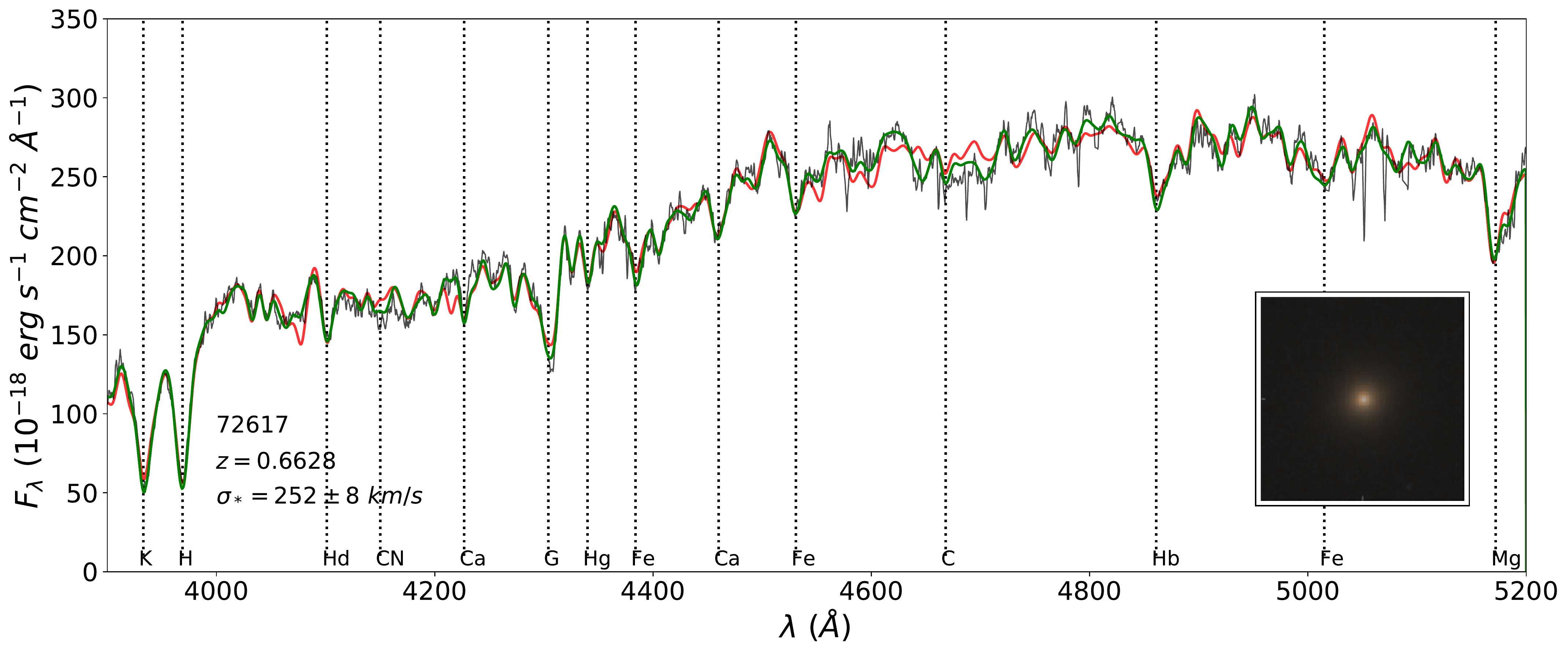} 
    \caption{Four spectra and HST images of galaxies with typical $S/N$ for their respective global properties such as luminosity and redshift. From top to bottom, a blue star-forming galaxy, a red star-forming galaxy, a post-starburst galaxy / merger remnant, and a massive elliptical galaxy.  The templates indicated with red (blue) lines represent the stellar continuum excluding (including) nebular emission lines. The stellar templates are the theoretical Conroy templates used for measuring stellar velocity dispersions. For the star-forming galaxies the broad Balmer absorption lines are filled with narrow Balmer line emission. In the bottom panel no emission lines are present; the green line represents the stellar continuum model using empirical stellar spectra rather than theoretical stellar spectra (see Sections \ref{section:kinematics} and \ref{section:emission} for details).}\label{fig:spectra1}
\end{figure*}

\begin{figure*}[!th]
\epsscale{.8}
\plotone{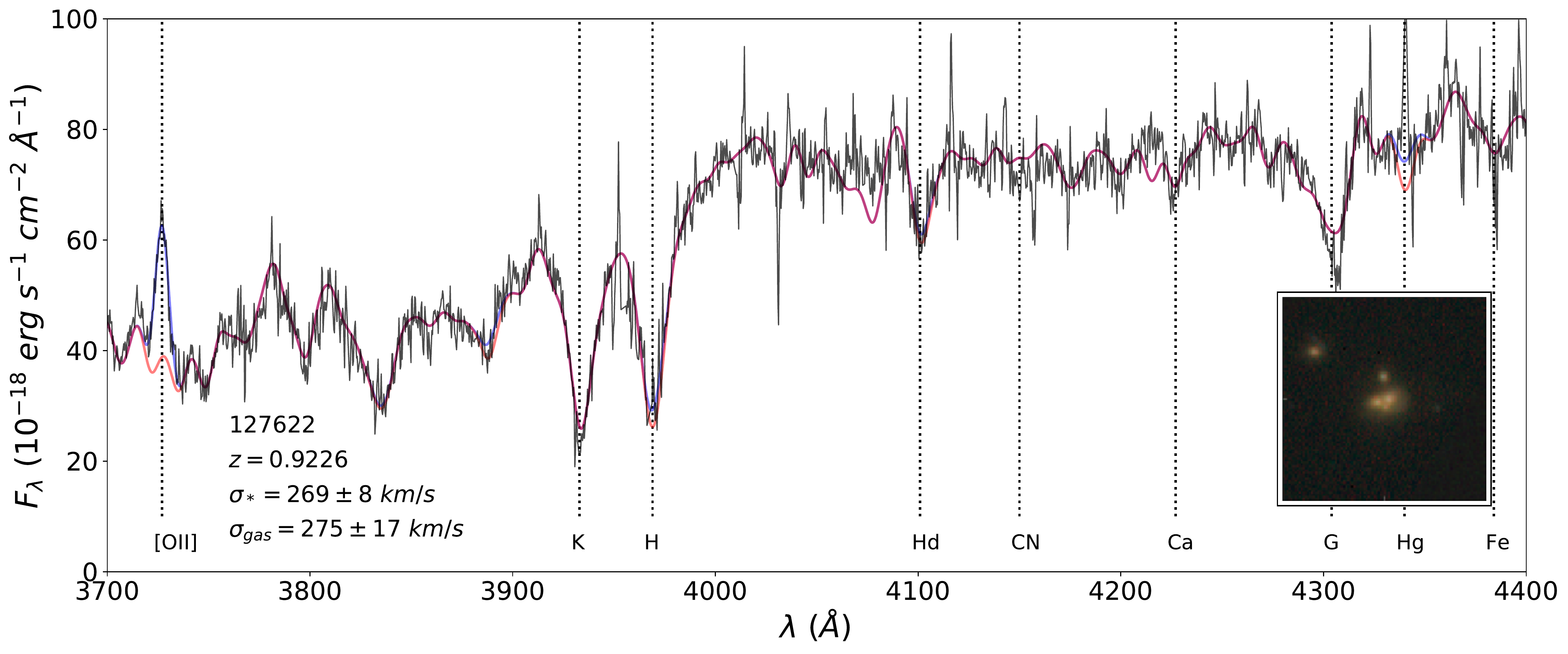} 
\plotone{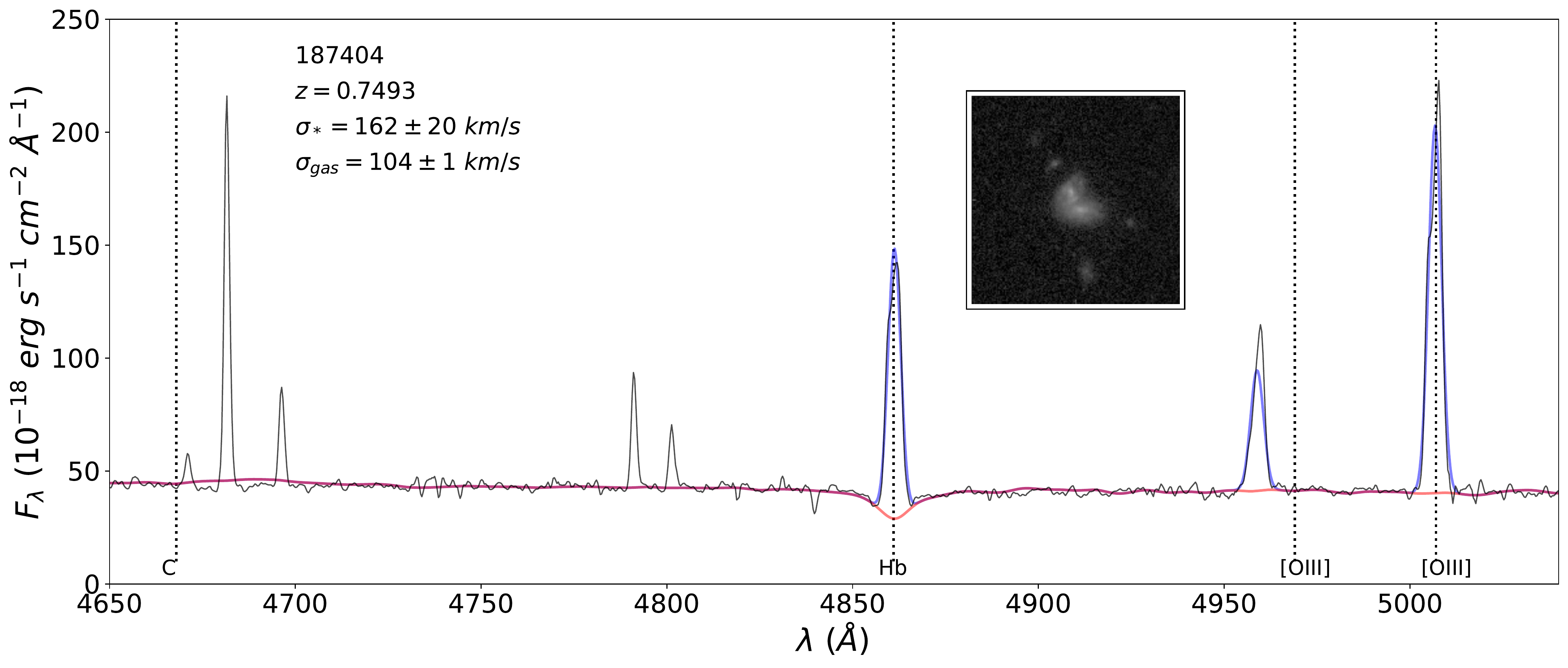} 
\plotone{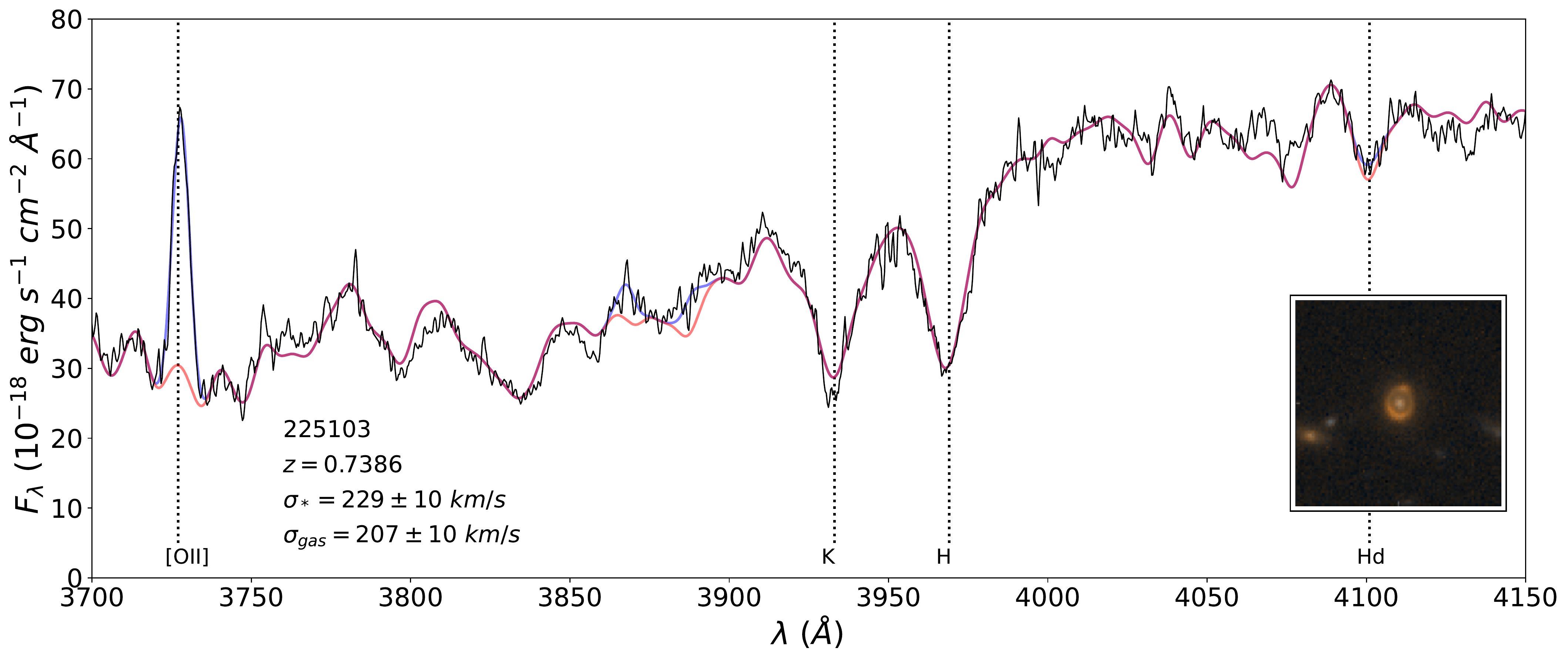} 
\plotone{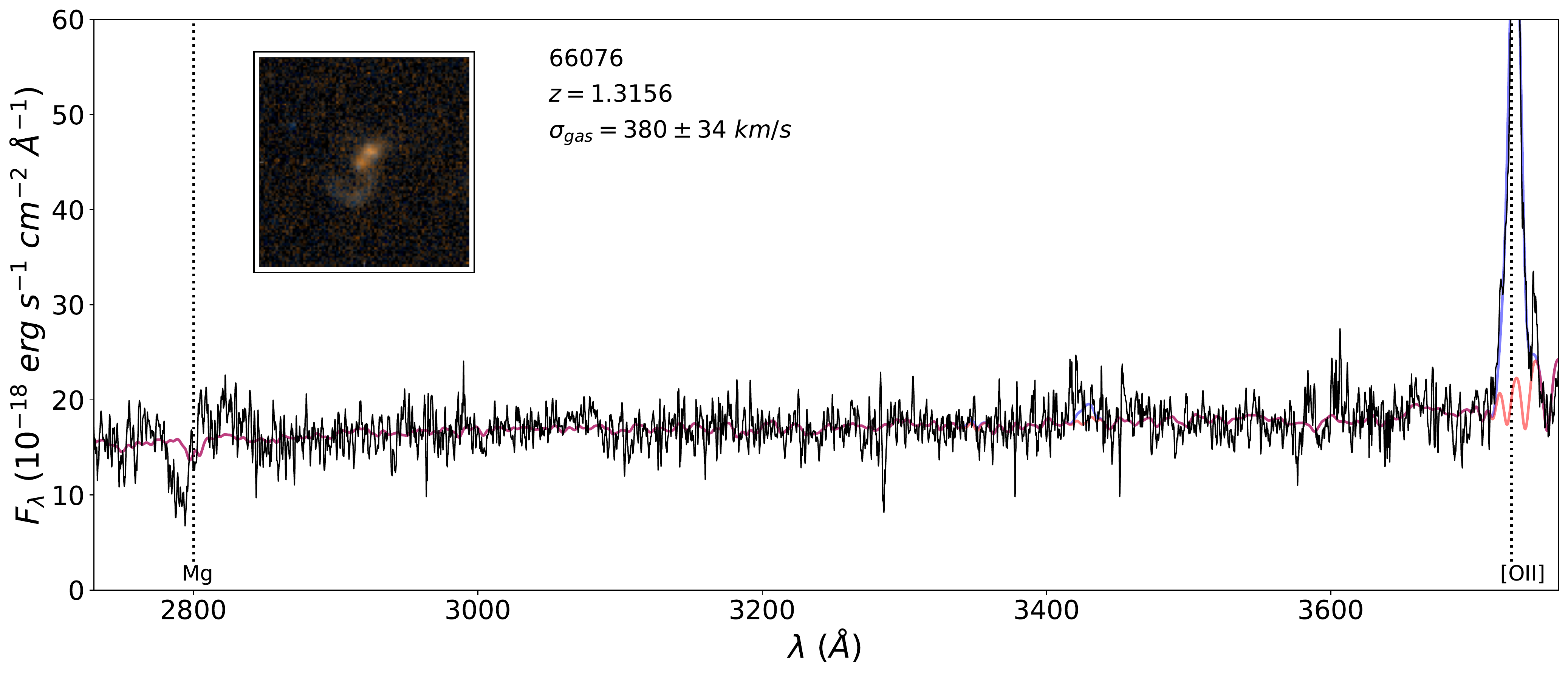} 
    \caption{See caption of Figure \ref{fig:spectra1} for a full description. The top three panels show spectra of objects that were flagged on the basis of their morphology (FLAG\_MORPH -- see text for details). 127622 shows 4 closely separated galaxies in the HST image, most likely all at the same redshift.   The combined light in the spectrum is unambiguous in its interpretation, but the structural properties and stellar kinematics are difficult to assess. 187404 is flagged due to the ambiguous interpretation of the spectrum: this spectrum (as well as the HST image) shows two galaxies at different redshifts.  225103 shows an almost perfect Einstein ring of a galaxy at an unknown redshift, lensed by an elliptical galaxy. The fourth panel (66076) shows an example of the several dozen $z>1.2$ Mg absorbers in the LEGA-C sample: this merger galaxy shows outflowing material with velocities up to $\sim2000$~km s$^{-1}$. }\label{fig:spectra2}
\end{figure*}

\begin{figure*}[!th]
\epsscale{1}
\plotone{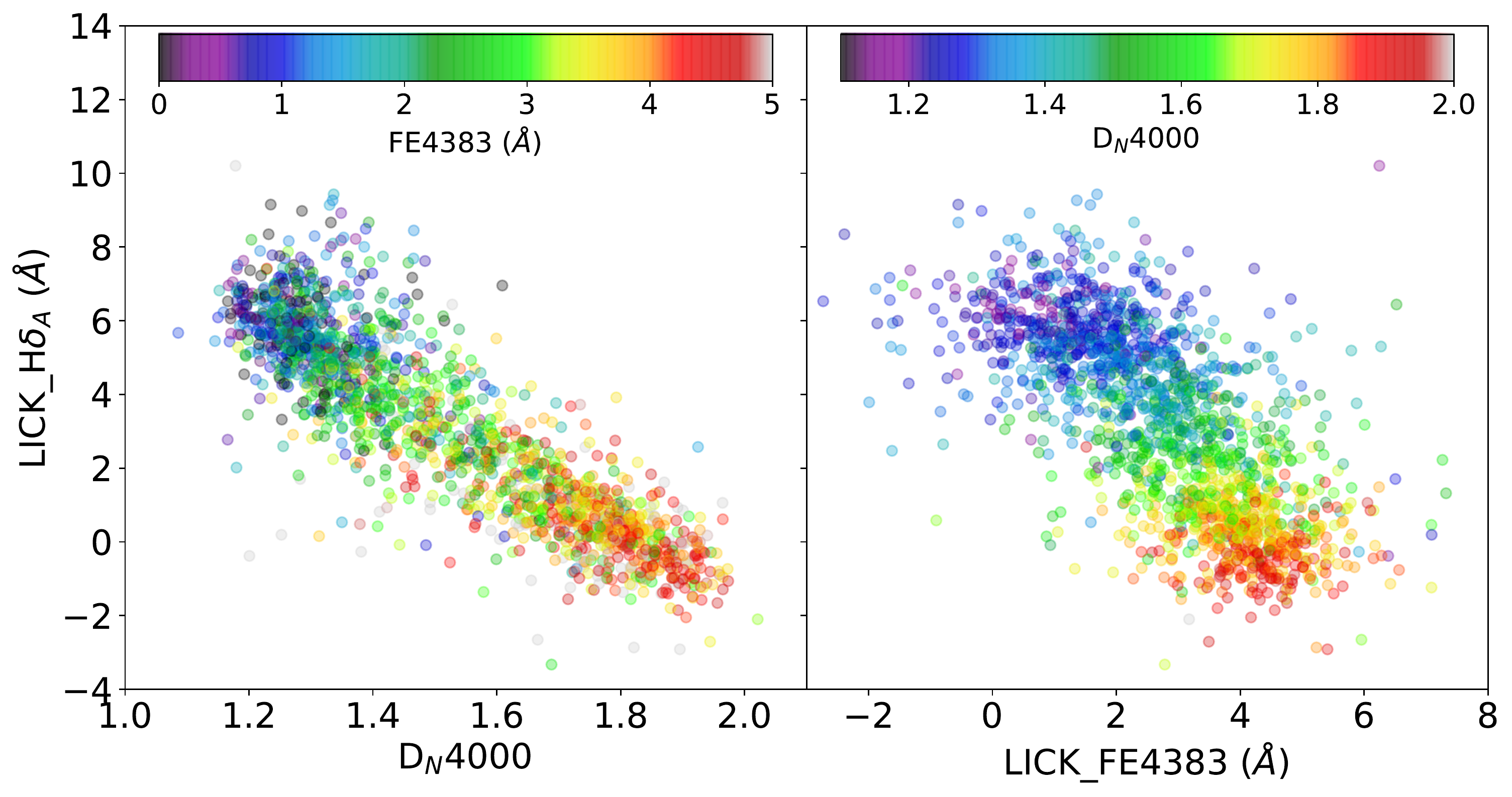} 
    \caption{ Left: H$\delta_A$~vs.~D$_N$4000, color-coded with FE4383. Right: H$\delta_A$ vs.~FE4383, color-coded with D$_N$4000.\label{fig:hd_d4000}}
\end{figure*}

\begin{figure*}[!th]
\epsscale{1}
\plotone{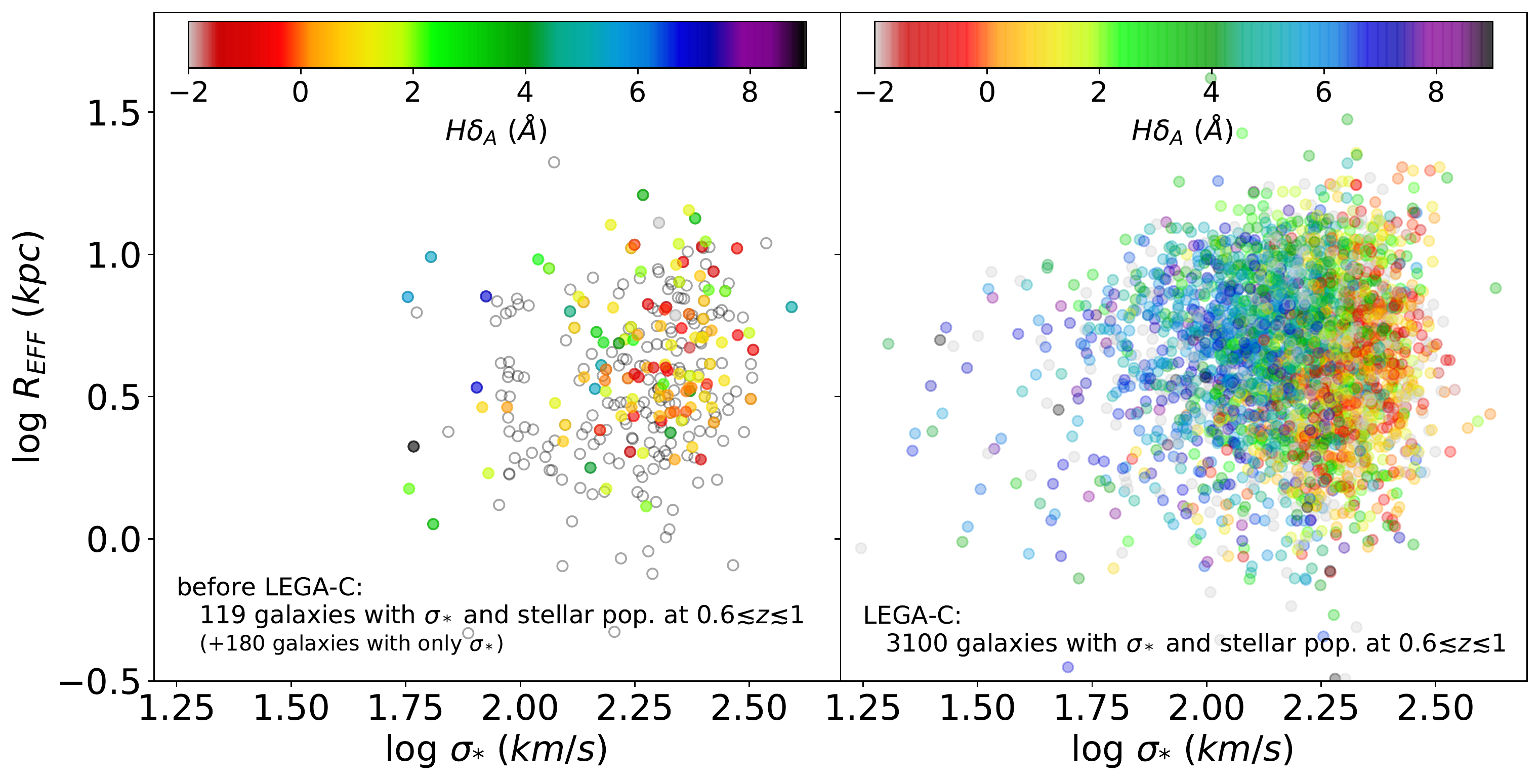}
\plotone{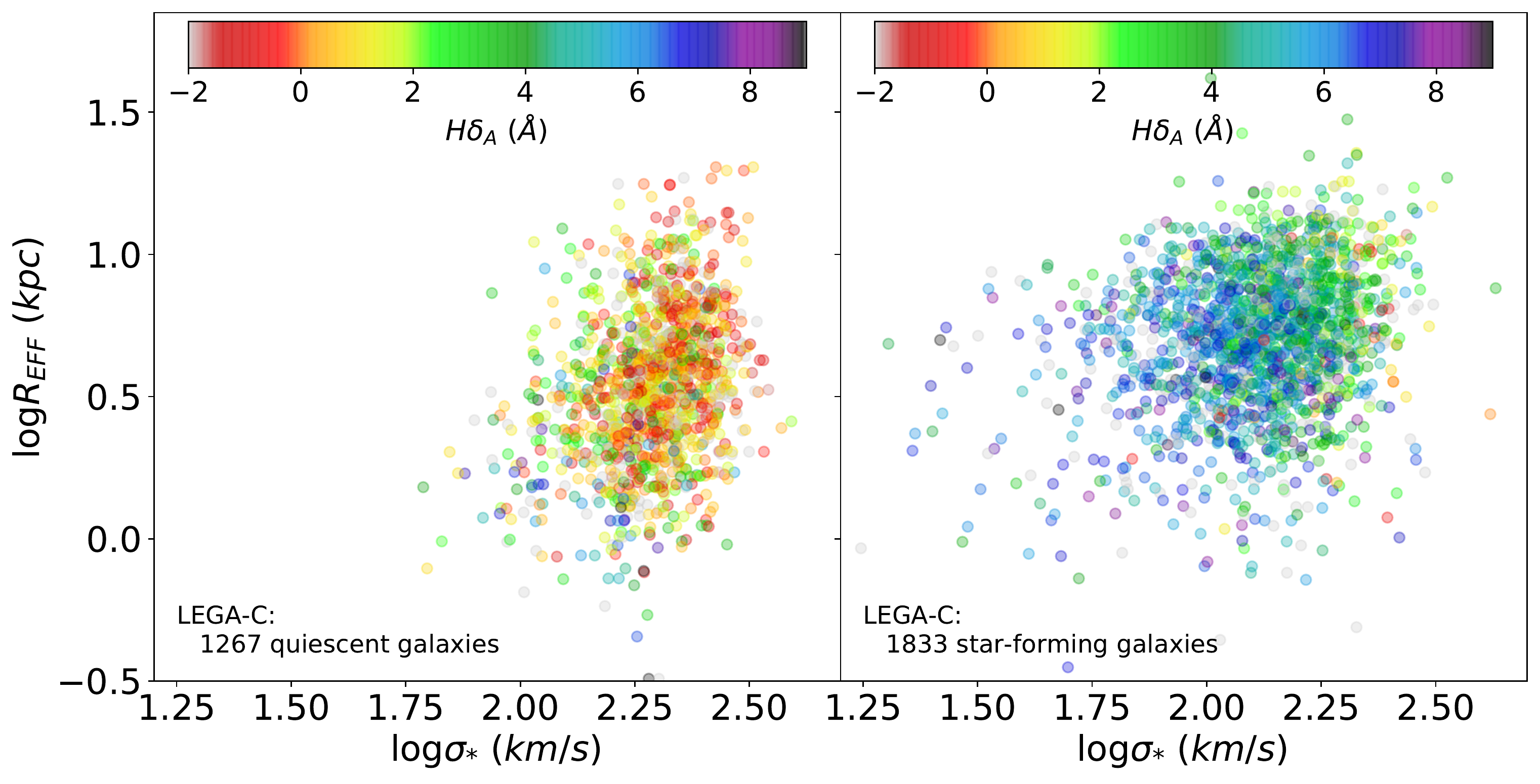}
    \caption{Effective radius vs.~stellar velocity dispersion for galaxies in the redshift range $0.6\lesssim z \lesssim 1$, color-coded $H\delta_A$ absorption equivalent width. From left to bottom right: literature compilation before LEGA-C (see text for references); the full LEGA-C sample; quiescent galaxies in LEGA-C; star-forming galaxies in LEGA-C. A trend between $H\delta_A$ and \sigs~is evident; older galaxies have higher velocity dispersions. A similar trend with size is not as clear.
    \label{fig:sig_re}}
\end{figure*}

\begin{figure*}[!th]
\epsscale{1.15}
\plotone{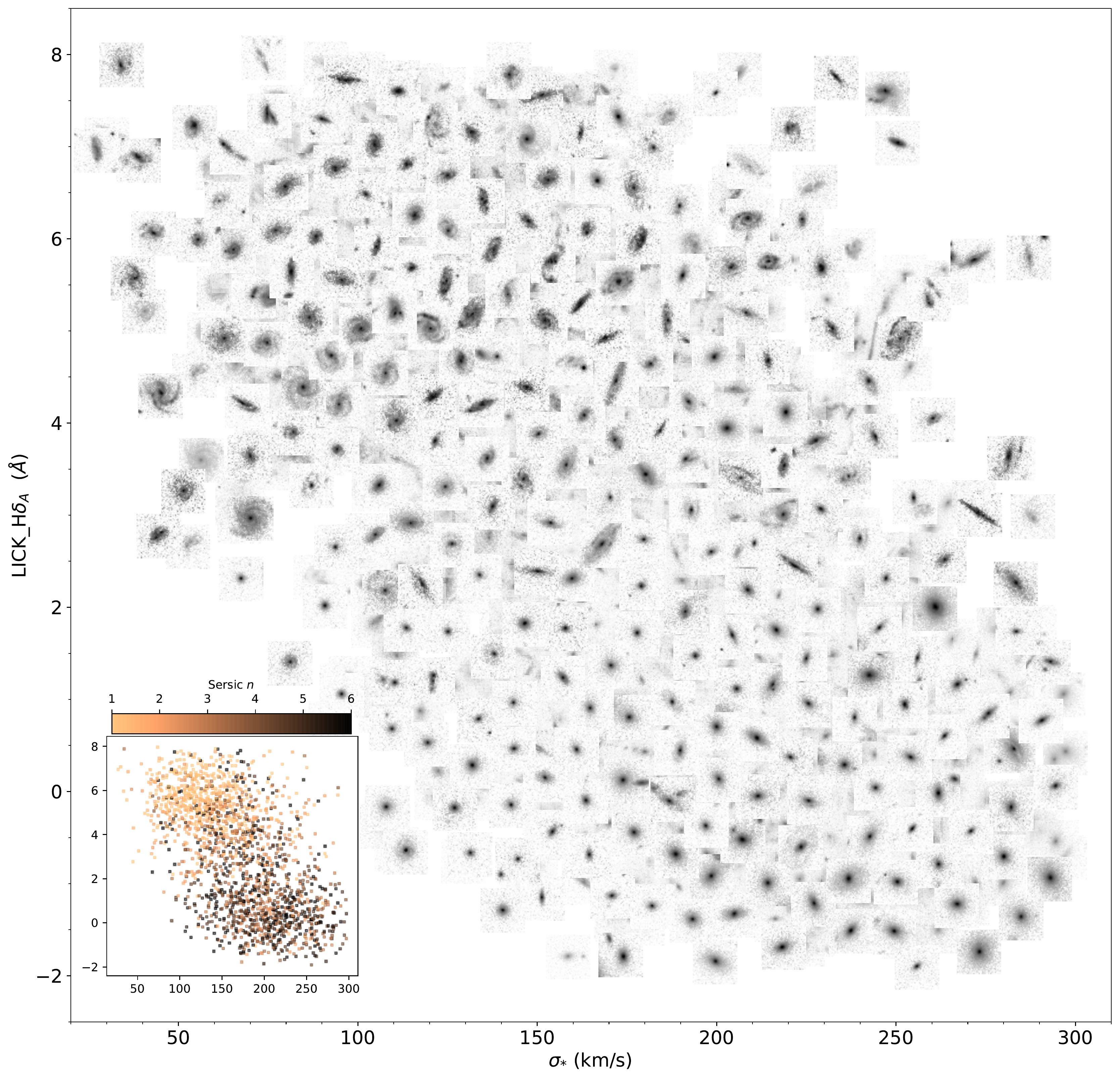}
    \caption{HST/ACS F814W images ($\sim$30 physical kpc on the side) as a function of $H\delta_A$ absorption equivalent width and stellar velocity dispersion \sigs. All galaxies with HST images, and measured \sigs, $H\delta_A$ and S\'ersic $n$~are included in the figure, with a non-overlapping foreground layer of 257 galaxies. The inset panel shows the full distribution, now color-coded with S\'ersic index. From top (young) to bottom (old) we see a gradual transition from late- to early-type visual morphologies. At high $H\delta_A>6$~we also often see irregular morphologies, indicative of interactions. The key observation is that at fixed \sigs~ galaxies show a large variety in visual morphology and S\'ersic index, while H$\delta_A$~(an age indicator) predicts these properties rather well. And conversely, morphology and S\'ersic index broadly predict age.
    \label{fig:sig_hd}}
\end{figure*}

\section{Spectral Measurements}
\label{section:measurements}

\subsection{Redshifts}\label{section:redshifts}
Given the depth of the spectra a redshift measurement is almost always possible, especially for the primary targets.  The 4081 spectra yield 3937 redshift measurements ($99.6\%$~for primary targets; $87.4\%$~for fillers -- see Table \ref{tab:sample}), determined by the XCSAO cross-correlation software package \citep{kurtz92} with the galaxy templates from \citet{hinton16} amended with four high-$S/N$ LEGA-C spectra DR2 for galaxies with different properties (quiescent, star-forming, dusty and emission-line dominated). The redshift measurements are further refined below (Section \ref{section:kinematics}) when modeling the stellar and gas kinematics.

\subsection{Stellar and Ionized Gas Kinematics}\label{section:kinematics}
We use the recent Python implementation of pPXF \citep{cappellari17} to estimate stellar and ionized gas velocity dispersions. The procedure is explained in detail by \citet{bezanson18a} and in the DR2 paper. Briefly, we use a set of single stellar population models based on high-resolution ($R=10,000$) theoretical templates from C.~Conroy (priv.~communication); this high resolution allows us to take maximum advantage of the high effective resolution of the LEGA-C spectra (see DR2 paper: $R\sim3500$; FWHM=$86$km$/$s~or~$\sigma=36$km$/$s). In addition to the stellar continuum model we add a set of emission lines (see Table \ref{tab:cat}). The pPXF fit consists of optimizing the combination of the templates, emission lines, a 3rd-order multiplicative polynomials and an additive polynomial to reproduce the integrated, 1-dimensional spectra. The integrated stellar velocity dispersions \sigs~and\sigg\footnote{The notation $\sigma$ `prime' and `integrated', introduced by \citet{bezanson18a}, refers to the fact that these are line-of-sight dispersions, integrated over the entire galaxy (modulo slit losses).}~are estimated by broadening the templates with Gaussians (no higher-order moments are fit); the stellar and emission line dispersions are allowed to be different, but emission lines all have the same velocity dispersion in a given spectrum. This produces a stellar velocity dispersion for 3528 of the 3937 spectra with redshift measurements ($96.8\%$~for primary targets; $56.7\%$~for fillers -- see Table \ref{tab:sample}). The precision of the resulting uncertainties is assessed in Appendix \ref{AppendixB}. The pPXF fit also refines the initial redshift measurement described in Section \ref{section:redshifts}. If no successful pPXF fit was performed, the initial redshift is adopted as the final redshift estimate.

\subsection{Emission Line Fluxes and Equivalent Widths}\label{section:emission}
As a first step we rerun pPXF as described above in Section \ref{section:kinematics} but replacing the theoretical Conroy templates with the lower-resolution empirical templates based on the MILES library \citep{sanchez-blazquez06, falcon-barroso11}.  The reason for this choice is that the empirical templates better reproduce the stellar continuum than the theoretical templates (see Figure \ref{fig:spectra1}, bottom panel for an illustration), perhaps due to missing species in the theoretical spectra (C.~Conroy, priv.~comm.). We prefer the theoretical templates for the kinematic measurements due to their superior spectral resolution; $R=10000$ (or FWHM$=30~$km$/$s)~vs.~$R\sim1800$ (or FWHM=$167~$km$/$s). We also note that these theoretical templates are limited to fixed, solar abundance ratios, and do not use variable abundance ratios as in, e.g., \citet{conroy18}. Importantly, there is no systematic difference between the inferred stellar velocity dispersions and the typical difference is about half of the measurement uncertainty. This implies that template mismatch does not play a dominant role in our velocity dispersion measurements.

Since the empirical templates produce a more accurate continuum subtraction they provide a better representation of the emission line velocity profiles and strengths.  This produces an ionized gas velocity dispersion for 2674 of the 3937 spectra with redshift measurements ($67.9\%$) -- many of the passive galaxies do not show sufficiently bright emission lines to allow for a dispersion measurement. To determine emission line fluxes and equivalent widths we use {\tt platefit} \citep{brinchmann04} on the stellar continuum subtracted spectra. All lines are fit simultaneously, allowing for a 300 km s$^{-1}$ offset between the Balmer and forbidden lines.  The equivalent widths are then computed with respect to the stellar continuum model. The procedure (and first scientific application) are described in full detail by Maseda et al.~(in prep.). The reason for not using {\tt pPXF} to measure the emission-line fluxes is that it is not advisable to include a large number of (mostly faint) emission lines, which would interfere with the continuum modeling;  only the Balmer lines and bright Oxygen lines are therefore modeled with {\tt pPXF}. {\tt Platefit} is custom-made to measure a large number of emission lines in a flexible manner.

The uncertainties are described in full in Appendix \ref{AppendixB}, but here it is important to emphasize that the irregular spatial distribution of line-emitting regions within galaxies leads to unpredictable slit losses. Our duplicate observations regularly show differences in emission line characteristics -- kinematic as well as line strength -- that far exceed the measurement uncertainties. Visual inspection reveals that such cases arise due to particularly bright, centrally offset emission-line regions, which may fall inside the slit for one of the duplicates but outside for the other. These differences arise through variations in slit alignment.  This implies a systematic uncertainty that cannot be quantified on a case by case basis, as we do not have a good description of the spatial distribution of emission lines.\footnote{For stellar light we rely on the HST image, and variations in slit losses play a very minor role for the more symmetric stellar light in the first place.} We estimate that this uncertainty on the emission line fluxes and gas velocity dispersions is of the order of 15\%, but we do not propagate this into the generally much more precise formal measurement uncertainties in the catalog.

\subsection{Absorption Line Indices}
Absorption lines are often weak and variations among galaxies small. Controlling for systematic effects in the sky subtraction, the noise model, and the wavelength calibration is challenging, especially in the red part of the optical spectrum with large and rapid changes in variance with wavelength. Absorption line indices are defined to be measured from constant-variance spectra -- which is a good assumption in the blue part of the optical spectrum -- but weighing wavelength elements differently across the index wavelength interval introduces a bias in the equivalent width.  Yet ignoring the variance spectrum can also introduce a bias, and always increases the random uncertainty due to the equal weights given to high-noise wavelength elements. These issues do not have a perfect solution and motivate the full spectral fitting approach, but since those are strongly model dependent, index measurements still play an indispensable role in empirically presenting stellar population properties in a manner that will not change over time as our knowledge of the spectral evolution of stellar populations further improves.

We implement a solution for obtaining approximately bias-free index measurements that do not strongly suffer from high-noise wavelength elements as follows. We create a model spectrum consisting of a linear combination of the stellar spectra in the MILES library\footnote{We do not use the continuum models constructed in Secs.~\ref{section:kinematics} and \ref{section:emission} in order to reduce as much as possible any stellar population model dependency in our index measurements} and a 15th order, multiplicative Legendre polynomial\footnote{The order of the polynomial was based on a convergence test with increasingly larger orders. Higher orders do not change the statistical properties of the residuals, while lower orders leave systematic residuals.}. First, we scale the error spectrum such that the reduced $\chi^2$ value is unity, resulting in slightly (typically, 20\%) larger formal measurement uncertainties. Second, we linearly interpolate over spectral elements that deviate by more than 2$\sigma$ (and with a minimum of 10\%) from the model.  About 7\% of all wavelength elements used in index measurements are flagged and interpolated over in this manner.  If the fraction of deviating pixels exceeds 30\% within the wavelength bandpasses of the index then the index is considered to be not measured. To reiterate: the model is only used to identif deviant wavelength elements, not to replace them; the models do not directly enter into the index measurements. In Appendix \ref{AppendixB} we assess the formal measurement uncertainties.  Note that the indices are measured without wavelength convolution to a fixed spectra resolution; a $\sigma_*$ dependence therefore exists. This is usually accounted for when comparing with models \citep[e.g.,][]{gallazzi14}.

The catalog (Section \ref{sec:cat}) includes all measured indices, but we recommend using only those for which the galaxy has a measured stellar velocity dispersion. But given that $>95\%$ of primary targets have a \sigs~measurement the subsample with stellar population information is highly complete. Naturally, which absorption features are actually measured depends on the wavelength coverage. For example, 81\% of primary targets with \sigs~measurements have a H$\gamma$ absorption measurement, but only 21\% have a Mg$b$ absorption measurement. The wavelength coverage of each spectrum is provided in the catalog.

To illustrate the bulk stellar population properties and trends for the galaxy population as a whole we show in Figure \ref{fig:hd_d4000} the distribution of D$_N$4000, H$\delta_A$, and FE4383. The age and metallicity indicators are strongly correlated, but unfortunately this does not translate in a straightforward manner into a correlation between age and metallicity: metal features such as FE4383 are also primarily sensitive to age, as is shown by the age-metallicity grid. In forthcoming papers the stellar population properties will be analyzed in detail to provide age and metallicity estimates, and we will build on our previous full-spectral fitting efforts \citep{chauke18, chauke19, barisic20} to explore the full star-formation and chemical enrichment histories.

\section{Data Release Contents}\label{sec:dr}
\subsection{Spectra}\label{sec:spectra}

The primary data product consists of the spectra and corresponding weights (inverse variance spectra)\footnote{The spectra and catalog have been released by ESO: \url{http://archive.eso.org/cms/eso-archive-news/Third-and-final-release-of-the-Large-Early-Galaxy-Census-LEGA-C-Spectroscopic-Public-Survey-published.html}}$^{,}$\footnote{The spectra and catalog are available on this website: \url{https://users.ugent.be/\string~avdrwel/research.html\#legac}}. Figure \ref{fig:spec_blue} compiles 2707 emission-line subtracted spectra with measured stellar velocity dispersions and H$\gamma_A$~absorption indices, visualizing the rich content of the dataset. First, we see how the Balmer break seen for young galaxies gradually transitions into the 4000\rm{\AA} break for old galaxies. Second, although the old galaxies show stronger metal features, the spectra of individual young galaxies show the same features, enabling the measurement of their star-formation histories and metallicities.

For a more detailed look into the characteristics of the data we examine a small subset of galaxies in Figure \ref{fig:spectra1}, which shows four very different types of galaxies, from blue/late-type to red/early-type. All steps in the analysis described above are illustrated here. The Doppler broadening of the absorption and emission line features -- decomposed into two components: continuum and emission lines --  produces velocity dispersion measurements (for the two examples in the top two panels the gas and stellar dispersions agree well). The bottom panel demonstrates the difficulty of finding a single-size-fits-all analysis method: the high-spectral resolution theoretical stellar population models used for stellar velocity dispersion measurements do not fit the spectra of old galaxies as well as models based on empirical stellar libraries. This mismatch, which may be the result of missing species in the theortical templates and/or non-solar abundance ratios, will be explored in future work.

Figure \ref{fig:spectra2} shows four examples of non-standard objects. We see an early-stage merger, a projected pair (that would have been classified as a merger without a spectrum), and a strong gravitational lens -- these 3 objects are flagged in the catalog as described in Section \ref{sec:cat}. Finally, we show an outflow seen in Mg absorption at $z=1.3$. The LEGA-C sample size guarantees the frequent occurrence of such cases, and makes for a fascinating treasure trove.

\newpage
\subsection{Catalog}\label{sec:cat}
In Table \ref{tab:cat} we list the contents of the electronic public data release. For the 4081 spectra we provide 61 quantities and, where relevant, their uncertainties. For each quantity we indicate for how many of the spectra it has been measured and with what range in values and uncertainties. For example, we have a median redshift ({\tt Z\_SPEC}) of $z=0.808$ and 3528 measured stellar velocity dispersions \sigs~({\tt SIGMA\_STARS\_PRIME}) with a median value of 161 km s$^{-1}$ (the 16-84\%-ile range is 106 to 222 km s$^{-1}$) and median uncertainty of 13 km s$^{-1}$ (the 16-84\%-ile range is 7 to 24 km s$^{-1}$). 

The user should be aware of various meta parameters and peculiarities of the dataset. {\tt ID\_LEGAC} is a running integer number to identify unique spectra, but because of the duplicate observations, there are cases with multiple spectra of the same objects and some {\tt ID} entries from the \citet{muzzin13a} catalog appear more than once. The {\tt PRIMARY} flag indicates whether the galaxy was selected from the $Ks$-band selected parent sample (1) or as a filler (0).

We have created two flags that should be kept in mind when assigning meaning to the measured quantities. {\tt FLAG\_SPEC:} A subset of spectra show clear evidence of AGN affecting the continuum shape, compromising the interpretation of the index measurements.  Upon visual inspection we decided to flag all 107 galaxies with mid-infrared and/or X-ray AGN even if the majority do not show an obvious issue; the catalog entry FLAG\_SPEC$=$1 indicates such AGN. Narrow-line and radio AGN do not present a problem for our spectral analysis and are not flagged. However, in 25 cases we found that the photometry-based flux calibration showed significant imperfections, potentially compromising the measurement of absorption and emission indices. These are indicated by {\tt FLAG\_SPEC}$=$2.

{\tt FLAG\_MORPH:} In most cases the light coming through the slit is from a single galaxy with a regular morphology, but in a significant minority of cases this is not the case.  In order to address this we devise the catalog parameter {\tt FLAG\_MORPH} with value 0, 1 or 2. Single galaxies with regular morphologies have a value 0.  Spectra for which the light coming through the slit does not come from single, regular galaxies get a flag value 1; these are often irregular galaxies such as merger remnants, but also multiple galaxies that are separated in the HST image, but not in the spectrum. The guiding principle when assigning flag values of 1 is that the combination of the stellar velocity dispersion (from the spectrum) and structural parameters (from the HST image) cannot be included in an analysis of the Fundamental Plane or other scaling relations. Spectra that have, on top of this, the problem that the light comes from galaxies at different redshifts receive a flag value 2. Again, the guiding principle is that the presence of a secondary object at the different redshift prevents us from using the stellar velocity dispersion and the structural parameters for a scaling relation analysis. In total, 257 are flagged, 14 of which have flag value 2. Many more galaxies have low-level `contamination' from secondary sources in the slit, but not to the extent that the measurements presented here are ambiguous in their interpretation.

Note that the measurements as presented in the catalog are not wrong for spectra with FLAG\_MORPH value 1 or 2 -- the flag simply alerts the user to a possible issue with the interpretation of the measurement, not an issue with the measurements themselves. For example, we have a spectrum with a correctly measured velocity dispersion of 623 km s$^{-1}$, and close inspection of the HST image reveals a double nucleus. This combination indicates that we are seeing an early-stage merger where the two components happen to be aligned almost perfectly or closely separated in real space. Obviously, the high stellar velocity dispersion does not trace the internal dynamical structure of either component, but rather the orbital motion of the pair.

The structural parameters (Section \ref{sec:hst}) for 3800 objects are also included in the catalog. Most missing values are the result of imperfect overlap between the LEGA-C and COSMOS-ACS footprints. We give a {\tt SERSIC\_LIMIT} flag to indicate for which objects the parameters reached the {\tt galfit} limiting value (e.g., $n=6$). 

Finally, we publish the virial mass ({\tt LOG\_MVIR}) and its uncertainty, derived as explained in an accompanying paper (van Houdt et al., in prep.). In short, it is given by 
\begin{equation}
     M_{\rm{vir}} = K(n)\frac{\sigma^{2}_{\star,\rm{vir}}R_{\rm{sma}}}{G},
\end{equation}
where $n$ is {\tt SERSIC\_N}, $K(n) = 8.87 - 0.831 n + 0.0241 n^{2}$, $R_{\rm{sma}}$ is the semi-major axis length {\tt SERSIC\_RE} in kpc, $G$ the gravitational constant, and $\sigma_{\star,\rm{vir}}$ ({\tt SIGMA\_STARS\_VIR}) is the inclination- and aperture-corrected velocity dispersion:

\begin{equation}
    \sigma_{\star,\rm{vir}} = \sigma'^2_{\star,\rm{int}}\left(0.87+0.39\cdot e^{-3.78(1-q)}\right),
    \label{eq:sigma_correction}
\end{equation}
with $q$ the projected axis ratio ({\tt SERSIC\_Q}).  In principle, at least for flat galaxies, one might expect a difference in $\sigma_{\star,\rm{vir}}$ (and \sigs), depending on the angle between the major axis of the galaxy and the slit, but we find no significant evidence.  This might be due to the small sample size of just ten duplicate observations for flat galaxies ($b/a<0.5$) and for which the position angle differs by more than 45 degrees, and for those we can only constrain the effect on $\sigma_{\star,\rm{vir}}$ (and \sigs) to be less than 10\%. Larger samples are needed to provide a more precise constraint.

Our new definition of the virial mass is different from (and up to 50\% higher than) what has been the norm in the high-redshift community: instead of circularized radius we use the major-axis length, and we use an inclination-corrected velocity dispersion relying on the observational result that most galaxies are oblate rotators \citep[e.g.,][]{weijmans14, foster17}. (Most intrinsically round ellipticals have $q\sim 0.8$, and therefore have an inclination correction near unity.) These masses were calibrated against dynamical models applied to spatially resolved stellar kinematic measurements.

\section{First-Look Results \& Outlook}
\label{section:results}

In order to visualize the increase in sample size and parameter space compared to previous work that LEGA-C represents (not counting previous LEGA-C data releases), we show the size-\sigs~distribution in Figure \ref{fig:sig_re}. Before LEGA-C the combined efforts of observing programs with the goal to collect spectra with sufficient quality to determine the stellar velocity dispersions  for individual galaxies with HST imaging in the $0.6 \lesssim z \lesssim 1$ redshift range yielded 231 such measurements \citep{van-der-wel05, treu05a, gallazzi14, bezanson15}, of which the vast majority are quiescent and very massive due to sample selection. A subset of 119 of these have stellar population measurements from \citet{gallazzi14} and \citet{shetty15}.\footnote{For the galaxies from \citet{shetty15} we crudely estimated $H\delta_A$ from their published ages. While not very precise these estimates serve the purpose of illustrating the existing samples before LEGA-C.} We note that the SHELS survey \citet{geller16} that mostly targets galaxies at $z\sim0.3$ has a high-redshift tail of $\sim$60 galaxies at $z\sim0.6$. Their $\sigma_*$ and stellar population properties have been used in publications \citep{zahid16} but not published in tabulated form.

LEGA-C expands the sample of galaxies with measured and stellar population properties and velocity dispersions to over 3000, while also greatly expanding parameter space of galaxy mass and type. Note that we only counted `field' galaxies in this comparison -- dozens of cluster galaxies  in this redshift range also have $\sigma_*$ measurements \citep{van-dokkum98, wuyts04, holden10, jorgensen13}. In addition, field galaxy measurements at $z>1$ now also number in the dozens \citep{van-dokkum09, newman10, toft12a, van-de-sande13, van-de-sande14, belli14, longhetti14, newman15, hill16, belli17, toft17, saracco19}.

The near-absence of a correlation between galaxy size and velocity dispersion -- this is an almost face-on projection of the fundamental plane -- vividly reveals the connection between galaxy structure and stellar population content.  High-$\sigma$ galaxies have weaker H$\delta$ absorption (that is, older ages), a trend that is seen for the population as a whole as well as for UVJ-selected star-forming galaxies. Within the quiescent population there is a smattering for compact galaxies with strong H$\delta$ absorption, which we can readily identify as post-starburst galaxies \citep{wu18b,wu20}. It goes far beyond the scope of this paper, but Figure \ref{fig:sig_re} clearly invites an analysis as presented for present-day galaxies by \citet{graves09a} with the added advantage of being able to measure evolution in the size-\sigs~plane. 

Figure \ref{fig:sig_hd} makes the explicit connection between galaxy morphology, structure, kinematics and stellar population age. Morphologically late-type galaxies generally have young ages and low stellar velocity dispersion, whereas early-type galaxies have old ages and high $\sigma_*$. At fixed $\sigma_*$ we see, from young to old age, a change from late- to early-type morphology, and an increase in S\'ersic index. Galaxy morphology at fixed H$\delta$ shows less variety or trend than at fixed $\sigma_*$, although for late types we do see an increased S\'ersic index at high $\sigma_*$, that is, increased bulge prominence. In short, the visual (morphological) appearance of a galaxy in this redshift range first and foremost reflects its age, rather than its dynamical structure. A more in-depth analysis is beyond the scope of this data release paper, but this Figure shows the potential of LEGA-C to produce a significant leap forward in our understanding of galaxy formation and evolution by providing for the first time SDSS-quality spectroscopy for the general galaxy population 7 Gyr ago. 
We encourage the community to take advantage of this unique, publicly available dataset. 

\acknowledgments
  Based on observations made with ESO Telescopes at the La Silla Paranal Observatory under programme IDs 194-A.2005 and 1100.A-0949 (The LEGA-C Public Spectroscopy Survey). This project has received funding from the European Research Council (ERC) under the European Union’s Horizon 2020 research and innovation programme (grant agreement No. 683184). CMSS acknowledges support from Research Foundation - Flanders (FWO) through Fellowship 12ZC120N. CP is supported by the Canadian Space Agency under a contract with NRC Herzberg Astronomy and Astrophysics.

\software{{\tt Astropy} \citep{astropy-collaboration13},
{\tt Matplotlib} \citep{hunter07}, {\tt galfit} \citep{peng10}, {\tt XCSAO}, \citep{kurtz92}}

\startlongtable
\begin{deluxetable*}{|l|c|c|c|c|l|}
\tabletypesize{\scriptsize}
\tablecolumns{6}
\tablecaption{Catalog Contents \label{tab:cat}}
\tablehead{
\colhead{Catalog Entry} &  \colhead{Unit} & \colhead{N objects} &  \colhead{Value Range} & \colhead{Uncertainty Range} & \colhead{Comment} \\ 
\colhead{} &  \colhead{} &  \colhead{} & \colhead{[min,max] or [16, 50, 84\%-ile]} & \colhead{[16, 50, 84\%-ile]} & \colhead{}}
\startdata
ID\_LEGAC & INT & 4081 & [1, 4081] &   &  \\ \hline
ID & INT & 4081 &  [3945, 262197] &   & \citet{muzzin13} \\ \hline 
MASK & INT & 4081 & [1, 102] &   & LEGA-C mask id \\ \hline 
RAJ2000A & degrees & 4081 & [149.379760, 150.774770] &   & J2000 \\ \hline 
DECJ2000 & degrees & 4081 & [1.747992, 2.804280] &   & J2000 \\ \hline 
PRIMARY & INT & [1053, 3028]& [0, 1] &   & primary (1) / filler (0) \\ \hline 
Z & & 3937 & [0.672, 0.808, 0.964] &  & LEGA-C spec.-$z$ \\ \hline 
SIGMA\_STARS\_PRIME & km s$^{-1}$ & 3528 & [106, 161, 222] &  [7, 13, 24] & \sigs \\ \hline 
SIGMA\_GAS\_PRIME & km s$^{-1}$ & 2674 & [68, 128, 201] &  [2, 7, 21]$^2$ &  \sigg \\ \hline 
LICK\_HD\_A & \AA & 3147 & [0.3, 3.8, 6.3] &  [0.3, 0.6, 1.4] & Absorption line \\ \hline 
LICK\_HD\_F &  & 3175 & [1.5, 3.3, 4.9] &  [0.2, 0.4, 0.9] & ""  \\ \hline 
LICK\_HG\_A & \AA & 2899 & [-3.2, 1.4, 4.9] &  [0.4, 0.7, 1.5] & "" \\ \hline 
LICK\_HG\_F & \AA & 2913 & [-0.1, 2.5, 4.5] &  [0.2, 0.4, 0.9] & "" \\ \hline 
LICK\_HB & \AA & 1579 & [2.0, 3.3, 4.9] &  [0.4, 0.7, 1.6] & "" \\ \hline 
LICK\_CN1 & mag & 3029 & [-0.12, -0.06, 0.03] &  [0.01, 0.02, 0.04] & "" \\ \hline 
LICK\_CN2 & mag & 3033 & [-0.07, -0.01, 0.08] &  [0.01, 0.02, 0.05] & "" \\ \hline 
LICK\_CA4227 & \AA & 3115 & [0.1, 0.7, 1.1] &  [0.2, 0.4, 0.8] & "" \\ \hline 
LICK\_G4300 & \AA & 3032 & [0.2, 2.5, 4.8] &  [0.4, 0.7, 1.5] & "" \\ \hline 
LICK\_FE4383 & \AA & 2800 & [0.8, 2.7, 4.3] &  [0.5, 1.0, 2.2] & "" \\ \hline 
LICK\_CA4455 & \AA & 2701 & [0.1, 0.9, 1.4] &  [0.3, 0.5, 1.1] & "" \\ \hline 
LICK\_FE4531 & \AA & 2470 & [1.2, 2.5, 3.5] &  [0.4, 0.8, 1.9] & "" \\ \hline 
LICK\_C4668 & \AA & 2026 & [0.5, 3.6, 6.1] &  [0.7, 1.3, 2.9] & "" \\ \hline 
LICK\_FE5015 & \AA & 1120 & [1.6, 3.8, 5.2] &  [0.6, 1.2, 2.8] & "" \\ \hline 
LICK\_MG1 & mag & 438 & [0.00, 0.03, 0.08] &  [0.01, 0.01, 0.03] & "" \\ \hline 
LICK\_MG2 & mag & 432 & [0.046, 0.11, 0.20] &  [0.01, 0.02, 0.04] & "" \\ \hline 
LICK\_MGB & \AA & 761 & [1.0, 2.3, 3.7] &  [0.3, 0.6, 1.5] & "" \\ \hline 
LICK\_FE5270 & \AA & 525 & [0.8, 2.0, 2.8] &  [0.4, 0.7, 1.7] & "" \\ \hline 
LICK\_FE5335 & \AA & 435 & [0.9, 1.9, 2.7] &  [0.4, 0.7, 2.0] & "" \\ \hline 
LICK\_FE5406 & \AA & 320 & [0.4, 1.1, 1.8] &  [0.3, 0.6, 1.4] & "" \\ \hline 
D4000\_N & & 2838 & [1.23, 1.40, 1.76] &  [0.02, 0.03, 0.06] &  \\ \hline 
OII\_3727\_FLUX & $10^{-19}$ erg/s/cm$^{2}/$\AA & 2777 & [53, 316, 863] &  [12, 21, 41]$^2$  & Emission line \\ \hline 
OII\_3727\_EW & \AA & 2765 & [1.51, 8.74, 25.9] &  [0.32, 0.6, 1.2]$^2$  & "" \\ \hline 
NEIII\_3869\_FLUX & $10^{-19}$ erg/s/cm$^{2}/$\AA & 3139 & [-5, 14, 47] &  [8, 13, 24]$^2$  & "" \\ \hline 
NEIII\_3869\_EW & \AA & 3130 & [-0.11, 0.28, 0.98] &  [0.14, 0.28, 0.55]$^2$  & "" \\ \hline 
HEI\_3890\_FLUX & $10^{-19}$ erg/s/cm$^{2}/$\AA & 3190 & [3, 27, 71] &  [8, 14, 25]$^2$  & "" \\ \hline 
HEI\_3890\_EW & \AA & 3181 & [0.09, 0.62, 1.55] &  [0.18, 0.33, 0.66]$^2$  & "" \\ \hline 
Hd\_FLUX & $10^{-19}$ erg/s/cm$^{2}/$\AA & 3392 & [-3, 29, 103] &  [8, 13, 25]$^2$  & "" \\ \hline 
Hd\_EW & \AA & 3368 & [-0.04, 0.59, 2.13] &  [0.13, 0.25, 0.52]$^2$  & "" \\ \hline 
Hg\_FLUX & $10^{-19}$ erg/s/cm$^{2}/$\AA & 3195 & [1, 61, 212] &  [8, 16, 28]$^2$  & "" \\ \hline 
Hg\_EW & \AA & 3163 & [0.04, 1.25, 4.29] &  [0.13, 0.26, 0.56]$^2$  & "" \\ \hline 
OIII\_4363\_FLUX & $10^{-19}$ erg/s/cm$^{2}/$\AA & 3167 & [-15, 2, 19] &  [8, 14, 26]$^2$  & "" \\ \hline 
OIII\_4363\_EW & \AA & 3205 & [-0.21, 0.03, 0.29] &  [0.11, 0.20, 0.42]$^2$  & "" \\ \hline 
Hb\_FLUX & $10^{-19}$ erg/s/cm$^{2}/$\AA & 1858 & [7, 164, 580] &  [11, 19, 34]$^2$  & "" \\ \hline 
Hb\_EW & \AA & 1830 & [0.09, 3.26, 10.8] &  [0.13, 0.28, 0.58]$^2$  & "" \\ \hline 
OIII\_4959\_FLUX & $10^{-19}$ erg/s/cm$^{2}/$\AA & 1501 & [4.2, 44, 149] &  [9, 17, 30]$^2$  & "" \\ \hline 
OIII\_4959\_EW & \AA & 1529 & [0.03, 0.43, 2.0] &  [0.06, 0.11, 0.24]$^2$  & "" \\ \hline 
OIII\_5007\_FLUX & $10^{-19}$ erg/s/cm$^{2}/$\AA &  1372 & [28, 127, 429] &  [12, 22, 42]$^2$  & "" \\ \hline 
OIII\_5007\_EW & \AA & 1348 & [0.36, 1.5, 6.7] &  [0.13, 0.26, 0.61]$^2$  & "" \\ \hline 
SERSIC\_RE & arcsec & 3800 & [0.29, 0.61, 1.13] &  [0.03, 0.07, 0.13] & Eff.~radius (maj.axis) \\ \hline 
SERSIC\_N & & 3800 & [0.82, 2.53, 5.32] &  [0.14, 0.44, 0.92] & S\'ersic index \\ \hline 
SERSIC\_Q & & 3800 & [0.38, 0.64, 0.85] &  [0.03, 0.03, 0.03] & Axis ratio \\ \hline 
SERSIC\_PA & degrees & 3800 & [-62, 2, 61] &  [2.8, 2.8, 2.8] & From North to East \\ \hline 
SERSIC\_MAG & F814W AB mag & 3800 & [20.7, 21.5, 22.7] &  [0.18, 0.18, 0.18] & Total mag\\ \hline 
SERSIC\_LIMIT & INT & [3341, 459] & [0, 1] &  & Galfit constraint flag\\ \hline 
SIGMA\_STARS\_VIR & km s$^{-1}$ & 3100 & [106, 160, 222] &  [7, 12, 24] & Virial vel.~disp. \\ \hline 
LOG\_MVIR & $M_\odot$ & 3100 & [10.80, 11.22, 11.59] &  [0.06, 0.08, 0.14] & Virial mass \\ \hline
SN & \AA$^{-1}$ & 3933 & [5.3, 12, 24.8]$^1$ &   & Avg. $S/N$\\ \hline 
SN\_RF\_4000 & \AA$^{-1}$ & 3764 & [7, 14.8, 26.4]$^1$ &   & $S/N$ at rest 4000\AA \\ \hline
SN\_OBS\_8030 & \AA$^{-1}$ & 3904 & [6.8, 15.8, 30.9]$^1$ &  & $S/N$ at obs 8030\AA\\ \hline 
TCOR & & 2830 & [1.5, 2.4, 6.3] &  & Completeness corr. \\ \hline
SCOR & & 2830 & [1.7, 2.6, 4.4] &  & Selection corr. \\ \hline
VCOR & & 2830 & [1.0, 1.0, 1.5] &  & Volume corr. \\ \hline
FLAG\_MORPH & INT & [3824, 243, 14] & [0, 1, 2] &  & Morphology flag \\ \hline 
FLAG\_SPEC & INT & [3949, 107, 25] & [0, 1, 2] &  & Spectroscopy flag
\enddata
\tablecomments{Data release catalog. Percentile value ranges are given for those objects for which the quantity was measured, that is, non-detections and failed measurements are not included. $^1$The $S/N$ values are given for the full sample, not just the primary targets that have higher $S/N$ than non-primary targets. $^2$ These are the formal measurement uncertainties. As described in Section \ref{section:emission} there is a systematic uncertainty of 15\% due to unknown variations in slit losses.} 
\end{deluxetable*}

\clearpage
\bibliography{mypapers}{}

\begin{thebibliography}{}
\expandafter\ifx\csname natexlab\endcsname\relax\def\natexlab#1{#1}\fi
\providecommand{\url}[1]{\href{#1}{#1}}
\providecommand{\dodoi}[1]{doi:~\href{http://doi.org/#1}{\nolinkurl{#1}}}
\providecommand{\doeprint}[1]{\href{http://ascl.net/#1}{\nolinkurl{http://ascl.net/#1}}}
\providecommand{\doarXiv}[1]{\href{https://arxiv.org/abs/#1}{\nolinkurl{https://arxiv.org/abs/#1}}}

\bibitem[{{Astropy Collaboration} {et~al.}(2013){Astropy Collaboration},
  {Robitaille}, {Tollerud}, {Greenfield}, {Droettboom}, {Bray}, {Aldcroft},
  {Davis}, {Ginsburg}, {Price-Whelan}, {Kerzendorf}, {Conley}, {Crighton},
  {Barbary}, {Muna}, {Ferguson}, {Grollier}, {Parikh}, {Nair}, {Unther},
  {Deil}, {Woillez}, {Conseil}, {Kramer}, {Turner}, {Singer}, {Fox}, {Weaver},
  {Zabalza}, {Edwards}, {Azalee Bostroem}, {Burke}, {Casey}, {Crawford},
  {Dencheva}, {Ely}, {Jenness}, {Labrie}, {Lim}, {Pierfederici}, {Pontzen},
  {Ptak}, {Refsdal}, {Servillat}, \& {Streicher}}]{astropy-collaboration13}
{Astropy Collaboration}, {Robitaille}, T.~P., {Tollerud}, E.~J., {et~al.} 2013,
  \aap, 558, A33, \dodoi{10.1051/0004-6361/201322068}

\bibitem[{{Bari{\v s}i{\'c}} {et~al.}(2017){Bari{\v s}i{\'c}}, {van der Wel},
  {Bezanson}, {Pacifici}, {Noeske}, {Mu{\~n}oz-Mateos}, {Franx}, {Smol{\v
  c}i{\'c}}, {Bell}, {Brammer}, {Calhau}, {Chauk{\'e}}, {van Dokkum}, {van
  Houdt}, {Gallazzi}, {Labb{\'e}}, {Maseda}, {Muzzin}, {Sobral}, {Straatman},
  \& {Wu}}]{barisic17}
{Bari{\v s}i{\'c}}, I., {van der Wel}, A., {Bezanson}, R., {et~al.} 2017, \apj,
  847, 72, \dodoi{10.3847/1538-4357/aa8768}

\bibitem[{{Bari{\v{s}}i{\'c}} {et~al.}(2020){Bari{\v{s}}i{\'c}}, {Pacifici},
  {van der Wel}, {Straatman}, {Bell}, {Bezanson}, {Brammer}, {D'Eugenio},
  {Franx}, {van Houdt}, {Maseda}, {Muzzin}, {Sobral}, \& {Wu}}]{barisic20}
{Bari{\v{s}}i{\'c}}, I., {Pacifici}, C., {van der Wel}, A., {et~al.} 2020,
  \apj, 903, 146, \dodoi{10.3847/1538-4357/abba37}

\bibitem[{{Belli} {et~al.}(2014){Belli}, {Newman}, \& {Ellis}}]{belli14}
{Belli}, S., {Newman}, A.~B., \& {Ellis}, R.~S. 2014, \apj, 783, 117,
  \dodoi{10.1088/0004-637X/783/2/117}

\bibitem[{{Belli} {et~al.}(2017){Belli}, {Newman}, \& {Ellis}}]{belli17}
---. 2017, \apj, 834, 18, \dodoi{10.3847/1538-4357/834/1/18}

\bibitem[{{Bertelli} {et~al.}(1994){Bertelli}, {Bressan}, {Chiosi}, {Fagotto},
  \& {Nasi}}]{bertelli94}
{Bertelli}, G., {Bressan}, A., {Chiosi}, C., {Fagotto}, F., \& {Nasi}, E. 1994,
  \aaps, 106, 275

\bibitem[{{Bezanson} {et~al.}(2015){Bezanson}, {Franx}, \& {van
  Dokkum}}]{bezanson15}
{Bezanson}, R., {Franx}, M., \& {van Dokkum}, P.~G. 2015, \apj, 799, 148,
  \dodoi{10.1088/0004-637X/799/2/148}

\bibitem[{{Bezanson} {et~al.}(2018{\natexlab{a}}){Bezanson}, {van der Wel},
  {Pacifici}, {Noeske}, {Bari{\v s}i{\'c}}, {Bell}, {Brammer}, {Calhau},
  {Chauke}, {van Dokkum}, {Franx}, {Gallazzi}, {van Houdt}, {Labb{\'e}},
  {Maseda}, {Mu{\~n}os-Mateos}, {Muzzin}, {van de Sande}, {Sobral},
  {Straatman}, \& {Wu}}]{bezanson18}
{Bezanson}, R., {van der Wel}, A., {Pacifici}, C., {et~al.} 2018{\natexlab{a}},
  \apj, 858, 60, \dodoi{10.3847/1538-4357/aabc55}

\bibitem[{{Bezanson} {et~al.}(2018{\natexlab{b}}){Bezanson}, {van der Wel},
  {Straatman}, {Pacifici}, {Wu}, {Bari{\v s}i{\'c}}, {Bell}, {Conroy},
  {D'Eugenio}, {Franx}, {Gallazzi}, {van Houdt}, {Maseda}, {Muzzin}, {van de
  Sande}, {Sobral}, \& {Spilker}}]{bezanson18a}
{Bezanson}, R., {van der Wel}, A., {Straatman}, C., {et~al.}
  2018{\natexlab{b}}, \apjl, 868, L36, \dodoi{10.3847/2041-8213/aaf16b}

\bibitem[{{Brinchmann} {et~al.}(2004){Brinchmann}, {Charlot}, {White},
  {Tremonti}, {Kauffmann}, {Heckman}, \& {Brinkmann}}]{brinchmann04}
{Brinchmann}, J., {Charlot}, S., {White}, S.~D.~M., {et~al.} 2004, \mnras, 351,
  1151, \dodoi{10.1111/j.1365-2966.2004.07881.x}

\bibitem[{{Brinchmann} {et~al.}(2008){Brinchmann}, {Kunth}, \&
  {Durret}}]{brinchmann08}
{Brinchmann}, J., {Kunth}, D., \& {Durret}, F. 2008, \aap, 485, 657,
  \dodoi{10.1051/0004-6361:200809783}

\bibitem[{{Cappellari}(2017)}]{cappellari17}
{Cappellari}, M. 2017, \mnras, 466, 798, \dodoi{10.1093/mnras/stw3020}

\bibitem[{{Carnall} {et~al.}(2019){Carnall}, {Leja}, {Johnson}, {McLure},
  {Dunlop}, \& {Conroy}}]{carnall19}
{Carnall}, A.~C., {Leja}, J., {Johnson}, B.~D., {et~al.} 2019, \apj, 873, 44,
  \dodoi{10.3847/1538-4357/ab04a2}

\bibitem[{{Carnall} {et~al.}(2018){Carnall}, {McLure}, {Dunlop}, \&
  {Dav{\'e}}}]{carnall18}
{Carnall}, A.~C., {McLure}, R.~J., {Dunlop}, J.~S., \& {Dav{\'e}}, R. 2018,
  \mnras, 480, 4379, \dodoi{10.1093/mnras/sty2169}

\bibitem[{{Chauke} {et~al.}(2018){Chauke}, {van der Wel}, {Pacifici},
  {Bezanson}, {Wu}, {Gallazzi}, {Noeske}, {Straatman}, {Mu{\~n}os-Mateos},
  {Franx}, {Bari{\v s}i{\'c}}, {Bell}, {Brammer}, {Calhau}, {van Houdt},
  {Labb{\'e}}, {Maseda}, {Muzzin}, {Rix}, \& {Sobral}}]{chauke18}
{Chauke}, P., {van der Wel}, A., {Pacifici}, C., {et~al.} 2018, \apj, 861, 13,
  \dodoi{10.3847/1538-4357/aac324}

\bibitem[{{Chauke} {et~al.}(2019){Chauke}, {van der Wel}, {Pacifici},
  {Bezanson}, {Wu}, {Gallazzi}, {Straatman}, {Franx}, {Bari{\v s}i{\'c}},
  {Bell}, {van Houdt}, {Maseda}, {Muzzin}, {Sobral}, \& {Spilker}}]{chauke19}
---. 2019, \apj, 877, 48, \dodoi{10.3847/1538-4357/ab164d}

\bibitem[{{Cole} {et~al.}(2020){Cole}, {Bezanson}, {van der Wel}, {Bell},
  {D'Eugenio}, {Franx}, {Gallazzi}, {van Houdt}, {Muzzin}, {Pacifici}, {van de
  Sande}, {Sobral}, {Straatman}, \& {Wu}}]{cole20}
{Cole}, J., {Bezanson}, R., {van der Wel}, A., {et~al.} 2020, \apjl, 890, L25,
  \dodoi{10.3847/2041-8213/ab7241}

\bibitem[{{Conroy} {et~al.}(2009){Conroy}, {Gunn}, \& {White}}]{conroy09}
{Conroy}, C., {Gunn}, J.~E., \& {White}, M. 2009, \apj, 699, 486,
  \dodoi{10.1088/0004-637X/699/1/486}

\bibitem[{{Conroy} {et~al.}(2018){Conroy}, {Villaume}, {van Dokkum}, \&
  {Lind}}]{conroy18}
{Conroy}, C., {Villaume}, A., {van Dokkum}, P.~G., \& {Lind}, K. 2018, \apj,
  854, 139, \dodoi{10.3847/1538-4357/aaab49}

\bibitem[{{de Graaff} {et~al.}(2020){de Graaff}, {Bezanson}, {Franx}, {van der
  Wel}, {Bell}, {D'Eugenio}, {Holden}, {Maseda}, {Muzzin}, {Pacifici}, {van de
  Sande}, {Sobral}, {Straatman}, \& {Wu}}]{de-graaff20}
{de Graaff}, A., {Bezanson}, R., {Franx}, M., {et~al.} 2020, \apjl, 903, L30,
  \dodoi{10.3847/2041-8213/abc428}

\bibitem[{{de Graaff} {et~al.}(2021){de Graaff}, {Bezanson}, {Franx}, {van der
  Wel}, {Holden}, {van de Sande}, {Bell}, {D'Eugenio}, {Maseda}, {Muzzin},
  {Sobral}, {Straatman}, \& {Wu}}]{de-graaff21}
---. 2021, arXiv e-prints, arXiv:2103.12753.
\newblock \doarXiv{2103.12753}

\bibitem[{{D'Eugenio} {et~al.}(2020){D'Eugenio}, {van der Wel}, {Wu
  (吳柏锋)}, {Barone}, {van Houdt}, {Bezanson}, {Straatman}, {Pacifici},
  {Muzzin}, {Gallazzi}, {Wild}, {Sobral}, {Bell}, {Zibetti}, {Mowla}, \&
  {Franx}}]{deugenio20}
{D'Eugenio}, F., {van der Wel}, A., {Wu}, P.-F., {et~al.} 2020,
  \mnras, 497, 389, \dodoi{10.1093/mnras/staa1937}

\bibitem[{{Falc{\'o}n-Barroso} {et~al.}(2011){Falc{\'o}n-Barroso}, {van de
  Ven}, {Peletier}, {Bureau}, {Jeong}, {Bacon}, {Cappellari}, {Davies}, {de
  Zeeuw}, {Emsellem}, {Krajnovi{\'c}}, {Kuntschner}, {McDermid}, {Sarzi},
  {Shapiro}, {van den Bosch}, {van der Wolk}, {Weijmans}, \&
  {Yi}}]{falcon-barroso11}
{Falc{\'o}n-Barroso}, J., {van de Ven}, G., {Peletier}, R.~F., {et~al.} 2011,
  \mnras, 417, 1787, \dodoi{10.1111/j.1365-2966.2011.19372.x}

\bibitem[{{Foster} {et~al.}(2017){Foster}, {van de Sande}, {D'Eugenio},
  {Cortese}, {McDermid}, {Bland-Hawthorn}, {Brough}, {Bryant}, {Croom},
  {Goodwin}, {Konstantopoulos}, {Lawrence}, {L{\'o}pez-S{\'a}nchez}, {Medling},
  {Owers}, {Richards}, {Scott}, {Taranu}, {Tonini}, \& {Zafar}}]{foster17}
{Foster}, C., {van de Sande}, J., {D'Eugenio}, F., {et~al.} 2017, \mnras, 472,
  966, \dodoi{10.1093/mnras/stx1869}

\bibitem[{{Gallazzi} {et~al.}(2014){Gallazzi}, {Bell}, {Zibetti}, {Brinchmann},
  \& {Kelson}}]{gallazzi14}
{Gallazzi}, A., {Bell}, E.~F., {Zibetti}, S., {Brinchmann}, J., \& {Kelson},
  D.~D. 2014, \apj, 788, 72, \dodoi{10.1088/0004-637X/788/1/72}

\bibitem[{{Geller} {et~al.}(2016){Geller}, {Hwang}, {Dell'Antonio}, {Zahid},
  {Kurtz}, \& {Fabricant}}]{geller16}
{Geller}, M.~J., {Hwang}, H.~S., {Dell'Antonio}, I.~P., {et~al.} 2016, \apjs,
  224, 11, \dodoi{10.3847/0067-0049/224/1/11}

\bibitem[{{Graves} {et~al.}(2009){Graves}, {Faber}, \& {Schiavon}}]{graves09a}
{Graves}, G.~J., {Faber}, S.~M., \& {Schiavon}, R.~P. 2009, \apj, 698, 1590,
  \dodoi{10.1088/0004-637X/698/2/1590}

\bibitem[{{Grogin} {et~al.}(2011){Grogin}, {Kocevski}, {Faber}, {Ferguson},
  {Koekemoer}, {Riess}, {Acquaviva}, {Alexander}, {Almaini}, {Ashby}, {Barden},
  {Bell}, {Bournaud}, {Brown}, {Caputi}, {Casertano}, {Cassata}, {Castellano},
  {Challis}, {Chary}, {Cheung}, {Cirasuolo}, {Conselice}, {Roshan Cooray},
  {Croton}, {Daddi}, {Dahlen}, {Dav{\'e}}, {de Mello}, {Dekel}, {Dickinson},
  {Dolch}, {Donley}, {Dunlop}, {Dutton}, {Elbaz}, {Fazio}, {Filippenko},
  {Finkelstein}, {Fontana}, {Gardner}, {Garnavich}, {Gawiser}, {Giavalisco},
  {Grazian}, {Guo}, {Hathi}, {H{\"a}ussler}, {Hopkins}, {Huang}, {Huang},
  {Jha}, {Kartaltepe}, {Kirshner}, {Koo}, {Lai}, {Lee}, {Li}, {Lotz}, {Lucas},
  {Madau}, {McCarthy}, {McGrath}, {McIntosh}, {McLure}, {Mobasher},
  {Moustakas}, {Mozena}, {Nandra}, {Newman}, {Niemi}, {Noeske}, {Papovich},
  {Pentericci}, {Pope}, {Primack}, {Rajan}, {Ravindranath}, {Reddy}, {Renzini},
  {Rix}, {Robaina}, {Rodney}, {Rosario}, {Rosati}, {Salimbeni}, {Scarlata},
  {Siana}, {Simard}, {Smidt}, {Somerville}, {Spinrad}, {Straughn}, {Strolger},
  {Telford}, {Teplitz}, {Trump}, {van der Wel}, {Villforth}, {Wechsler},
  {Weiner}, {Wiklind}, {Wild}, {Wilson}, {Wuyts}, {Yan}, \& {Yun}}]{grogin11}
{Grogin}, N.~A., {Kocevski}, D.~D., {Faber}, S.~M., {et~al.} 2011, \apjs, 197,
  35, \dodoi{10.1088/0067-0049/197/2/35}

\bibitem[{{H{\"a}u{\ss}ler} {et~al.}(2013){H{\"a}u{\ss}ler}, {Bamford}, {Vika},
  {Rojas}, {Barden}, {Kelvin}, {Alpaslan}, {Robotham}, {Driver}, {Baldry},
  {Brough}, {Hopkins}, {Liske}, {Nichol}, {Popescu}, \& {Tuffs}}]{haussler13}
{H{\"a}u{\ss}ler}, B., {Bamford}, S.~P., {Vika}, M., {et~al.} 2013, \mnras,
  430, 330, \dodoi{10.1093/mnras/sts633}

\bibitem[{{Hill} {et~al.}(2016){Hill}, {Muzzin}, {Franx}, \& {van de
  Sande}}]{hill16}
{Hill}, A.~R., {Muzzin}, A., {Franx}, M., \& {van de Sande}, J. 2016, \apj,
  819, 74, \dodoi{10.3847/0004-637X/819/1/74}

\bibitem[{{Hinton} {et~al.}(2016){Hinton}, {Davis}, {Lidman}, {Glazebrook}, \&
  {Lewis}}]{hinton16}
{Hinton}, S.~R., {Davis}, T.~M., {Lidman}, C., {Glazebrook}, K., \& {Lewis},
  G.~F. 2016, Astronomy and Computing, 15, 61,
  \dodoi{10.1016/j.ascom.2016.03.001}

\bibitem[{{Holden} {et~al.}(2010){Holden}, {van der Wel}, {Kelson}, {Franx}, \&
  {Illingworth}}]{holden10}
{Holden}, B.~P., {van der Wel}, A., {Kelson}, D.~D., {Franx}, M., \&
  {Illingworth}, G.~D. 2010, \apj, 724, 714,
  \dodoi{10.1088/0004-637X/724/1/714}

\bibitem[{{Hunter}(2007)}]{hunter07}
{Hunter}, J.~D. 2007, Computing in Science and Engineering, 9, 90,
  \dodoi{10.1109/MCSE.2007.55}

\bibitem[{{J{\o}rgensen} \& {Chiboucas}(2013)}]{jorgensen13}
{J{\o}rgensen}, I., \& {Chiboucas}, K. 2013, \aj, 145, 77,
  \dodoi{10.1088/0004-6256/145/3/77}

\bibitem[{{Koekemoer} {et~al.}(2011){Koekemoer}, {Faber}, {Ferguson}, {Grogin},
  {Kocevski}, {Koo}, {Lai}, {Lotz}, {Lucas}, {McGrath}, {Ogaz}, {Rajan},
  {Riess}, {Rodney}, {Strolger}, {Casertano}, {Castellano}, {Dahlen},
  {Dickinson}, {Dolch}, {Fontana}, {Giavalisco}, {Grazian}, {Guo}, {Hathi},
  {Huang}, {van der Wel}, {Yan}, {Acquaviva}, {Alexander}, {Almaini}, {Ashby},
  {Barden}, {Bell}, {Bournaud}, {Brown}, {Caputi}, {Cassata}, {Challis},
  {Chary}, {Cheung}, {Cirasuolo}, {Conselice}, {Roshan Cooray}, {Croton},
  {Daddi}, {Dav{\'e}}, {de Mello}, {de Ravel}, {Dekel}, {Donley}, {Dunlop},
  {Dutton}, {Elbaz}, {Fazio}, {Filippenko}, {Finkelstein}, {Frazer}, {Gardner},
  {Garnavich}, {Gawiser}, {Gruetzbauch}, {Hartley}, {H{\"a}ussler},
  {Herrington}, {Hopkins}, {Huang}, {Jha}, {Johnson}, {Kartaltepe},
  {Khostovan}, {Kirshner}, {Lani}, {Lee}, {Li}, {Madau}, {McCarthy},
  {McIntosh}, {McLure}, {McPartland}, {Mobasher}, {Moreira}, {Mortlock},
  {Moustakas}, {Mozena}, {Nandra}, {Newman}, {Nielsen}, {Niemi}, {Noeske},
  {Papovich}, {Pentericci}, {Pope}, {Primack}, {Ravindranath}, {Reddy},
  {Renzini}, {Rix}, {Robaina}, {Rosario}, {Rosati}, {Salimbeni}, {Scarlata},
  {Siana}, {Simard}, {Smidt}, {Snyder}, {Somerville}, {Spinrad}, {Straughn},
  {Telford}, {Teplitz}, {Trump}, {Vargas}, {Villforth}, {Wagner}, {Wandro},
  {Wechsler}, {Weiner}, {Wiklind}, {Wild}, {Wilson}, {Wuyts}, \&
  {Yun}}]{koekemoer11}
{Koekemoer}, A.~M., {Faber}, S.~M., {Ferguson}, H.~C., {et~al.} 2011, \apjs,
  197, 36, \dodoi{10.1088/0067-0049/197/2/36}

\bibitem[{{Kriek} {et~al.}(2009){Kriek}, {van Dokkum}, {Labb{\'e}}, {Franx},
  {Illingworth}, {Marchesini}, \& {Quadri}}]{kriek09}
{Kriek}, M., {van Dokkum}, P.~G., {Labb{\'e}}, I., {et~al.} 2009, \apj, 700,
  221, \dodoi{10.1088/0004-637X/700/1/221}

\bibitem[{{Kurtz} {et~al.}(1992){Kurtz}, {Mink}, {Wyatt}, {Fabricant},
  {Torres}, {Kriss}, \& {Tonry}}]{kurtz92}
{Kurtz}, M.~J., {Mink}, D.~J., {Wyatt}, W.~F., {et~al.} 1992, in Astronomical
  Society of the Pacific Conference Series, Vol.~25, Astronomical Data Analysis
  Software and Systems I, ed. D.~M. {Worrall}, C.~{Biemesderfer}, \&
  J.~{Barnes}, 432

\bibitem[{{Le F{\`e}vre} {et~al.}(2003){Le F{\`e}vre}, {Saisse}, {Mancini},
  {Brau-Nogue}, {Caputi}, {Castinel}, {D'Odorico}, {Garilli}, {Kissler-Patig},
  {Lucuix}, {Mancini}, {Pauget}, {Sciarretta}, {Scodeggio}, {Tresse}, \&
  {Vettolani}}]{le-fevre03}
{Le F{\`e}vre}, O., {Saisse}, M., {Mancini}, D., {et~al.} 2003, in Society of
  Photo-Optical Instrumentation Engineers (SPIE) Conference Series, Vol. 4841,
  Instrument Design and Performance for Optical/Infrared Ground-based
  Telescopes, ed. M.~{Iye} \& A.~F.~M. {Moorwood}, 1670--1681,
  \dodoi{10.1117/12.460959}

\bibitem[{{Leja} {et~al.}(2020){Leja}, {Speagle}, {Johnson}, {Conroy}, {van
  Dokkum}, \& {Franx}}]{leja20}
{Leja}, J., {Speagle}, J.~S., {Johnson}, B.~D., {et~al.} 2020, \apj, 893, 111,
  \dodoi{10.3847/1538-4357/ab7e27}

\bibitem[{{Leja} {et~al.}(2019){Leja}, {Johnson}, {Conroy}, {van Dokkum},
  {Speagle}, {Brammer}, {Momcheva}, {Skelton}, {Whitaker}, {Franx}, \&
  {Nelson}}]{leja19}
{Leja}, J., {Johnson}, B.~D., {Conroy}, C., {et~al.} 2019, \apj, 877, 140,
  \dodoi{10.3847/1538-4357/ab1d5a}

\bibitem[{{Longhetti} {et~al.}(2014){Longhetti}, {Saracco}, {Gargiulo},
  {Tamburri}, \& {Lonoce}}]{longhetti14}
{Longhetti}, M., {Saracco}, P., {Gargiulo}, A., {Tamburri}, S., \& {Lonoce}, I.
  2014, \mnras, 439, 3962, \dodoi{10.1093/mnras/stu252}

\bibitem[{{Marigo} \& {Girardi}(2007)}]{marigo07}
{Marigo}, P., \& {Girardi}, L. 2007, \aap, 469, 239,
  \dodoi{10.1051/0004-6361:20066772}

\bibitem[{{McLure} {et~al.}(2018){McLure}, {Pentericci}, {Cimatti}, {Dunlop},
  {Elbaz}, {Fontana}, {Nandra}, {Amorin}, {Bolzonella}, {Bongiorno}, {Carnall},
  {Castellano}, {Cirasuolo}, {Cucciati}, {Cullen}, {De Barros}, {Finkelstein},
  {Fontanot}, {Franzetti}, {Fumana}, {Gargiulo}, {Garilli}, {Guaita},
  {Hartley}, {Iovino}, {Jarvis}, {Juneau}, {Karman}, {Maccagni}, {Marchi},
  {M{\'a}rmol-Queralt{\'o}}, {Pompei}, {Pozzetti}, {Scodeggio}, {Sommariva},
  {Talia}, {Almaini}, {Balestra}, {Bardelli}, {Bell}, {Bourne}, {Bowler},
  {Brusa}, {Buitrago}, {Caputi}, {Cassata}, {Charlot}, {Citro}, {Cresci},
  {Cristiani}, {Curtis-Lake}, {Dickinson}, {Fazio}, {Ferguson}, {Fiore},
  {Franco}, {Fynbo}, {Galametz}, {Georgakakis}, {Giavalisco}, {Grazian},
  {Hathi}, {Jung}, {Kim}, {Koekemoer}, {Khusanova}, {Le F{\`e}vre}, {Lotz},
  {Mannucci}, {Maltby}, {Matsuoka}, {McLeod}, {Mendez-Hernandez},
  {Mendez-Abreu}, {Mignoli}, {Moresco}, {Mortlock}, {Nonino}, {Pannella},
  {Papovich}, {Popesso}, {Rosario}, {Salvato}, {Santini}, {Schaerer},
  {Schreiber}, {Stark}, {Tasca}, {Thomas}, {Treu}, {Vanzella}, {Wild},
  {Williams}, {Zamorani}, \& {Zucca}}]{mclure18}
{McLure}, R.~J., {Pentericci}, L., {Cimatti}, A., {et~al.} 2018, \mnras, 479,
  25, \dodoi{10.1093/mnras/sty1213}

\bibitem[{{Muzzin} {et~al.}(2013{\natexlab{a}}){Muzzin}, {Marchesini},
  {Stefanon}, {Franx}, {Milvang-Jensen}, {Dunlop}, {Fynbo}, {Brammer},
  {Labb{\'e}}, \& {van Dokkum}}]{muzzin13a}
{Muzzin}, A., {Marchesini}, D., {Stefanon}, M., {et~al.} 2013{\natexlab{a}},
  \apjs, 206, 8, \dodoi{10.1088/0067-0049/206/1/8}

\bibitem[{{Muzzin} {et~al.}(2013{\natexlab{b}}){Muzzin}, {Marchesini},
  {Stefanon}, {Franx}, {McCracken}, {Milvang-Jensen}, {Dunlop}, {Fynbo},
  {Brammer}, {Labb{\'e}}, \& {van Dokkum}}]{muzzin13}
---. 2013{\natexlab{b}}, \apj, 777, 18, \dodoi{10.1088/0004-637X/777/1/18}

\bibitem[{{Newman} {et~al.}(2015){Newman}, {Belli}, \& {Ellis}}]{newman15}
{Newman}, A.~B., {Belli}, S., \& {Ellis}, R.~S. 2015, \apjl, 813, L7,
  \dodoi{10.1088/2041-8205/813/1/L7}

\bibitem[{{Newman} {et~al.}(2010){Newman}, {Ellis}, {Treu}, \&
  {Bundy}}]{newman10}
{Newman}, A.~B., {Ellis}, R.~S., {Treu}, T., \& {Bundy}, K. 2010, \apjl, 717,
  L103, \dodoi{10.1088/2041-8205/717/2/L103}

\bibitem[{{Peng} {et~al.}(2010){Peng}, {Ho}, {Impey}, \& {Rix}}]{peng10}
{Peng}, C.~Y., {Ho}, L.~C., {Impey}, C.~D., \& {Rix}, H.-W. 2010, \aj, 139,
  2097, \dodoi{10.1088/0004-6256/139/6/2097}

\bibitem[{{Pentericci} {et~al.}(2018){Pentericci}, {McLure}, {Garilli},
  {Cucciati}, {Franzetti}, {Iovino}, {Amorin}, {Bolzonella}, {Bongiorno},
  {Carnall}, {Castellano}, {Cimatti}, {Cirasuolo}, {Cullen}, {De Barros},
  {Dunlop}, {Elbaz}, {Finkelstein}, {Fontana}, {Fontanot}, {Fumana},
  {Gargiulo}, {Guaita}, {Hartley}, {Jarvis}, {Juneau}, {Karman}, {Maccagni},
  {Marchi}, {Marmol-Queralto}, {Nandra}, {Pompei}, {Pozzetti}, {Scodeggio},
  {Sommariva}, {Talia}, {Almaini}, {Balestra}, {Bardelli}, {Bell}, {Bourne},
  {Bowler}, {Brusa}, {Buitrago}, {Caputi}, {Cassata}, {Charlot}, {Citro},
  {Cresci}, {Cristiani}, {Curtis-Lake}, {Dickinson}, {Fazio}, {Ferguson},
  {Fiore}, {Franco}, {Fynbo}, {Galametz}, {Georgakakis}, {Giavalisco},
  {Grazian}, {Hathi}, {Jung}, {Kim}, {Koekemoer}, {Khusanova}, {Le F{\`e}vre},
  {Lotz}, {Mannucci}, {Maltby}, {Matsuoka}, {McLeod}, {Mendez-Hernandez},
  {Mendez-Abreu}, {Mignoli}, {Moresco}, {Mortlock}, {Nonino}, {Pannella},
  {Papovich}, {Popesso}, {Rosario}, {Salvato}, {Santini}, {Schaerer},
  {Schreiber}, {Stark}, {Tasca}, {Thomas}, {Treu}, {Vanzella}, {Wild},
  {Williams}, {Zamorani}, \& {Zucca}}]{pentericci18}
{Pentericci}, L., {McLure}, R.~J., {Garilli}, B., {et~al.} 2018, \aap, 616,
  A174, \dodoi{10.1051/0004-6361/201833047}

\bibitem[{{Pickles}(1998)}]{pickles98}
{Pickles}, A.~J. 1998, \pasp, 110, 863, \dodoi{10.1086/316197}

\bibitem[{{S{\'a}nchez-Bl{\'a}zquez} {et~al.}(2006){S{\'a}nchez-Bl{\'a}zquez},
  {Peletier}, {Jim{\'e}nez-Vicente}, {Cardiel}, {Cenarro},
  {Falc{\'o}n-Barroso}, {Gorgas}, {Selam}, \& {Vazdekis}}]{sanchez-blazquez06}
{S{\'a}nchez-Bl{\'a}zquez}, P., {Peletier}, R.~F., {Jim{\'e}nez-Vicente}, J.,
  {et~al.} 2006, \mnras, 371, 703, \dodoi{10.1111/j.1365-2966.2006.10699.x}

\bibitem[{{Saracco} {et~al.}(2019){Saracco}, {La Barbera}, {Gargiulo},
  {Mannucci}, {Marchesini}, {Nonino}, \& {Ciliegi}}]{saracco19}
{Saracco}, P., {La Barbera}, F., {Gargiulo}, A., {et~al.} 2019, \mnras, 484,
  2281, \dodoi{10.1093/mnras/sty3509}

\bibitem[{{Scoville} {et~al.}(2007){Scoville}, {Aussel}, {Brusa}, {Capak},
  {Carollo}, {Elvis}, {Giavalisco}, {Guzzo}, {Hasinger}, {Impey}, {Kneib},
  {LeFevre}, {Lilly}, {Mobasher}, {Renzini}, {Rich}, {Sanders}, {Schinnerer},
  {Schminovich}, {Shopbell}, {Taniguchi}, \& {Tyson}}]{scoville07}
{Scoville}, N., {Aussel}, H., {Brusa}, M., {et~al.} 2007, \apjs, 172, 1,
  \dodoi{10.1086/516585}

\bibitem[{{Shetty} \& {Cappellari}(2015)}]{shetty15}
{Shetty}, S., \& {Cappellari}, M. 2015, \mnras, 454, 1332,
  \dodoi{10.1093/mnras/stv1948}

\bibitem[{{Straatman} {et~al.}(2018){Straatman}, {van der Wel}, {Bezanson},
  {Pacifici}, {Gallazzi}, {Wu}, {Noeske}, {Bari{\v s}i{\'c}}, {Bell},
  {Brammer}, {Calhau}, {Chauke}, {Franx}, {van Houdt}, {Labb{\'e}}, {Maseda},
  {Mu{\~n}oz-Mateos}, {Muzzin}, {van de Sande}, {Sobral}, \&
  {Spilker}}]{straatman18}
{Straatman}, C.~M.~S., {van der Wel}, A., {Bezanson}, R., {et~al.} 2018, \apjs,
  239, 27, \dodoi{10.3847/1538-4365/aae37a}

\bibitem[{{Toft} {et~al.}(2012){Toft}, {Gallazzi}, {Zirm}, {Wold}, {Zibetti},
  {Grillo}, \& {Man}}]{toft12a}
{Toft}, S., {Gallazzi}, A., {Zirm}, A., {et~al.} 2012, \apj, 754, 3,
  \dodoi{10.1088/0004-637X/754/1/3}

\bibitem[{{Toft} {et~al.}(2017){Toft}, {Zabl}, {Richard}, {Gallazzi},
  {Zibetti}, {Prescott}, {Grillo}, {Man}, {Lee}, {G{\'o}mez-Guijarro},
  {Stockmann}, {Magdis}, \& {Steinhardt}}]{toft17}
{Toft}, S., {Zabl}, J., {Richard}, J., {et~al.} 2017, \nat, 546, 510,
  \dodoi{10.1038/nature22388}

\bibitem[{{Treu} {et~al.}(2005){Treu}, {Ellis}, {Liao}, {van Dokkum}, {Tozzi},
  {Coil}, {Newman}, {Cooper}, \& {Davis}}]{treu05a}
{Treu}, T., {Ellis}, R.~S., {Liao}, T.~X., {et~al.} 2005, \apj, 633, 174,
  \dodoi{10.1086/444585}

\bibitem[{{van de Sande} {et~al.}(2014){van de Sande}, {Kriek}, {Franx},
  {Bezanson}, \& {van Dokkum}}]{van-de-sande14}
{van de Sande}, J., {Kriek}, M., {Franx}, M., {Bezanson}, R., \& {van Dokkum},
  P.~G. 2014, \apjl, 793, L31, \dodoi{10.1088/2041-8205/793/2/L31}

\bibitem[{{van de Sande} {et~al.}(2013){van de Sande}, {Kriek}, {Franx}, {van
  Dokkum}, {Bezanson}, {Bouwens}, {Quadri}, {Rix}, \&
  {Skelton}}]{van-de-sande13}
{van de Sande}, J., {Kriek}, M., {Franx}, M., {et~al.} 2013, \apj, 771, 85,
  \dodoi{10.1088/0004-637X/771/2/85}

\bibitem[{{van der Wel} {et~al.}(2005){van der Wel}, {Franx}, {van Dokkum},
  {Rix}, {Illingworth}, \& {Rosati}}]{van-der-wel05}
{van der Wel}, A., {Franx}, M., {van Dokkum}, P.~G., {et~al.} 2005, \apj, 631,
  145, \dodoi{10.1086/430464}

\bibitem[{{van der Wel} {et~al.}(2012){van der Wel}, {Bell}, {H{\"a}ussler},
  {McGrath}, {Chang}, {Guo}, {McIntosh}, {Rix}, {Barden}, {Cheung}, {Faber},
  {Ferguson}, {Galametz}, {Grogin}, {Hartley}, {Kartaltepe}, {Kocevski},
  {Koekemoer}, {Lotz}, {Mozena}, {Peth}, \& {Peng}}]{van-der-wel12}
{van der Wel}, A., {Bell}, E.~F., {H{\"a}ussler}, B., {et~al.} 2012, \apjs,
  203, 24, \dodoi{10.1088/0067-0049/203/2/24}

\bibitem[{{van der Wel} {et~al.}(2016){van der Wel}, {Noeske}, {Bezanson},
  {Pacifici}, {Gallazzi}, {Franx}, {Mu{\~n}oz-Mateos}, {Bell}, {Brammer},
  {Charlot}, {Chauk{\'e}}, {Labb{\'e}}, {Maseda}, {Muzzin}, {Rix}, {Sobral},
  {van de Sande}, {van Dokkum}, {Wild}, \& {Wolf}}]{van-der-wel16}
{van der Wel}, A., {Noeske}, K., {Bezanson}, R., {et~al.} 2016, \apjs, 223, 29,
  \dodoi{10.3847/0067-0049/223/2/29}

\bibitem[{{van Dokkum} {et~al.}(1998){van Dokkum}, {Franx}, {Kelson}, \&
  {Illingworth}}]{van-dokkum98}
{van Dokkum}, P.~G., {Franx}, M., {Kelson}, D.~D., \& {Illingworth}, G.~D.
  1998, \apjl, 504, L17, \dodoi{10.1086/311567}

\bibitem[{{van Dokkum} {et~al.}(2009){van Dokkum}, {Kriek}, \&
  {Franx}}]{van-dokkum09}
{van Dokkum}, P.~G., {Kriek}, M., \& {Franx}, M. 2009, \nat, 460, 717,
  \dodoi{10.1038/nature08220}

\bibitem[{{Weijmans} {et~al.}(2014){Weijmans}, {de Zeeuw}, {Emsellem},
  {Krajnovi{\'c}}, {Lablanche}, {Alatalo}, {Blitz}, {Bois}, {Bournaud},
  {Bureau}, {Cappellari}, {Crocker}, {Davies}, {Davis}, {Duc}, {Khochfar},
  {Kuntschner}, {McDermid}, {Morganti}, {Naab}, {Oosterloo}, {Sarzi}, {Scott},
  {Serra}, {Verdoes Kleijn}, \& {Young}}]{weijmans14}
{Weijmans}, A.-M., {de Zeeuw}, P.~T., {Emsellem}, E., {et~al.} 2014, \mnras,
  444, 3340, \dodoi{10.1093/mnras/stu1603}

\bibitem[{{Wu} {et~al.}(2018{\natexlab{a}}){Wu}, {van der Wel}, {Bezanson},
  {Gallazzi}, {Pacifici}, {Straatman}, {Bari{\v s}i{\'c}}, {Bell}, {Chauke},
  {van Houdt}, {Franx}, {Muzzin}, {Sobral}, \& {Wild}}]{wu18b}
{Wu}, P.-F., {van der Wel}, A., {Bezanson}, R., {et~al.} 2018{\natexlab{a}},
  \apj, 868, 37, \dodoi{10.3847/1538-4357/aae822}

\bibitem[{{Wu} {et~al.}(2018{\natexlab{b}}){Wu}, {van der Wel}, {Gallazzi},
  {Bezanson}, {Pacifici}, {Straatman}, {Franx}, {Bari{\v s}i{\'c}}, {Bell},
  {Brammer}, {Calhau}, {Chauke}, {van Houdt}, {Maseda}, {Muzzin}, {Rix},
  {Sobral}, {Spilker}, {van de Sande}, {van Dokkum}, \& {Wild}}]{wu18a}
{Wu}, P.-F., {van der Wel}, A., {Gallazzi}, A., {et~al.} 2018{\natexlab{b}},
  \apj, 855, 85, \dodoi{10.3847/1538-4357/aab0a6}

\bibitem[{{Wu} {et~al.}(2020){Wu}, {van der Wel}, {Bezanson}, {Gallazzi},
  {Pacifici}, {Straatman}, {Bari{\v{s}}i{\'c}}, {Bell}, {Chauke},
  {D{\textquoteright}Eugenio}, {Franx}, {Muzzin}, {Sobral}, \& {van
  Houdt}}]{wu20}
{Wu}, P.-F., {van der Wel}, A., {Bezanson}, R., {et~al.} 2020, \apj, 888, 77,
  \dodoi{10.3847/1538-4357/ab5fd9}

\bibitem[{{Wuyts} {et~al.}(2004){Wuyts}, {van Dokkum}, {Kelson}, {Franx}, \&
  {Illingworth}}]{wuyts04}
{Wuyts}, S., {van Dokkum}, P.~G., {Kelson}, D.~D., {Franx}, M., \&
  {Illingworth}, G.~D. 2004, \apj, 605, 677, \dodoi{10.1086/381746}

\bibitem[{{Zahid} {et~al.}(2016){Zahid}, {Geller}, {Fabricant}, \&
  {Hwang}}]{zahid16}
{Zahid}, H.~J., {Geller}, M.~J., {Fabricant}, D.~G., \& {Hwang}, H.~S. 2016,
  \apj, 832, 203, \dodoi{10.3847/0004-637X/832/2/203}

\end{thebibliography}
\bibliographystyle{aasjournal}

\appendix
\restartappendixnumbering
\section{Completeness}
\label{AppendixA1}
\label{sec:compl}

\begin{figure}[t]
\epsscale{0.7}
\plotone{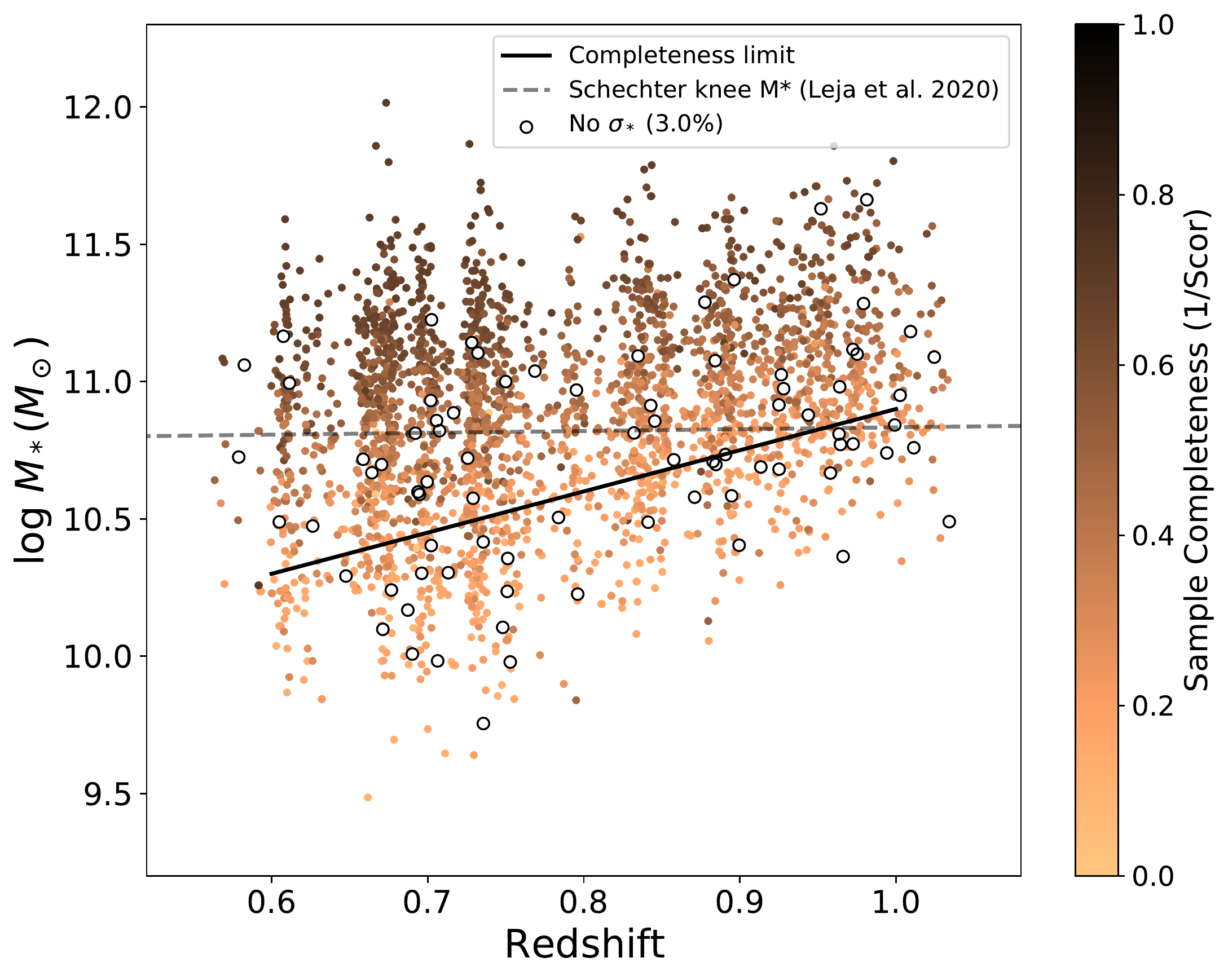} 
    \caption{Stellar mass vs.~redshift for the primary LEGA-C targets, color-coded with the $K_s$-dependent completeness with respect to the parent sample of $K_s$-selected galaxies; this is the inverse of the completeness correction factor {\tt Scor} in the catalog as described in Section \ref{sec:cat}).  The black line indicates the stellar mass completeness limit of the parent sample from which the LEGA-C primary targets have been selected: above this stellar mass limit all galaxies have $K_s$-band magnitudes brighter than $K_{s,LIM}$, whereas below the limit (a subset of dusty) galaxies have fainter magnitudes. The open data points show the few galaxies without velocity dispersion measurements. The dotted line -- the knee of the stellar mass function as determined by \citet{leja20} -- is shown for reference. LEGA-C is representative down to $\sim$0.3$L*$ at $z=0.6$ and $\sim L*$ at $z=1$. The color coding demonstrates that completeness with respect to the parent sample increases with mass, as a result of survey design: probability of observation is determined by $K_s$ magnitude as described in Section \ref{section:data}.
    \label{fig:m_z}}
\end{figure}

\begin{figure}[t]
\epsscale{0.7}
\plotone{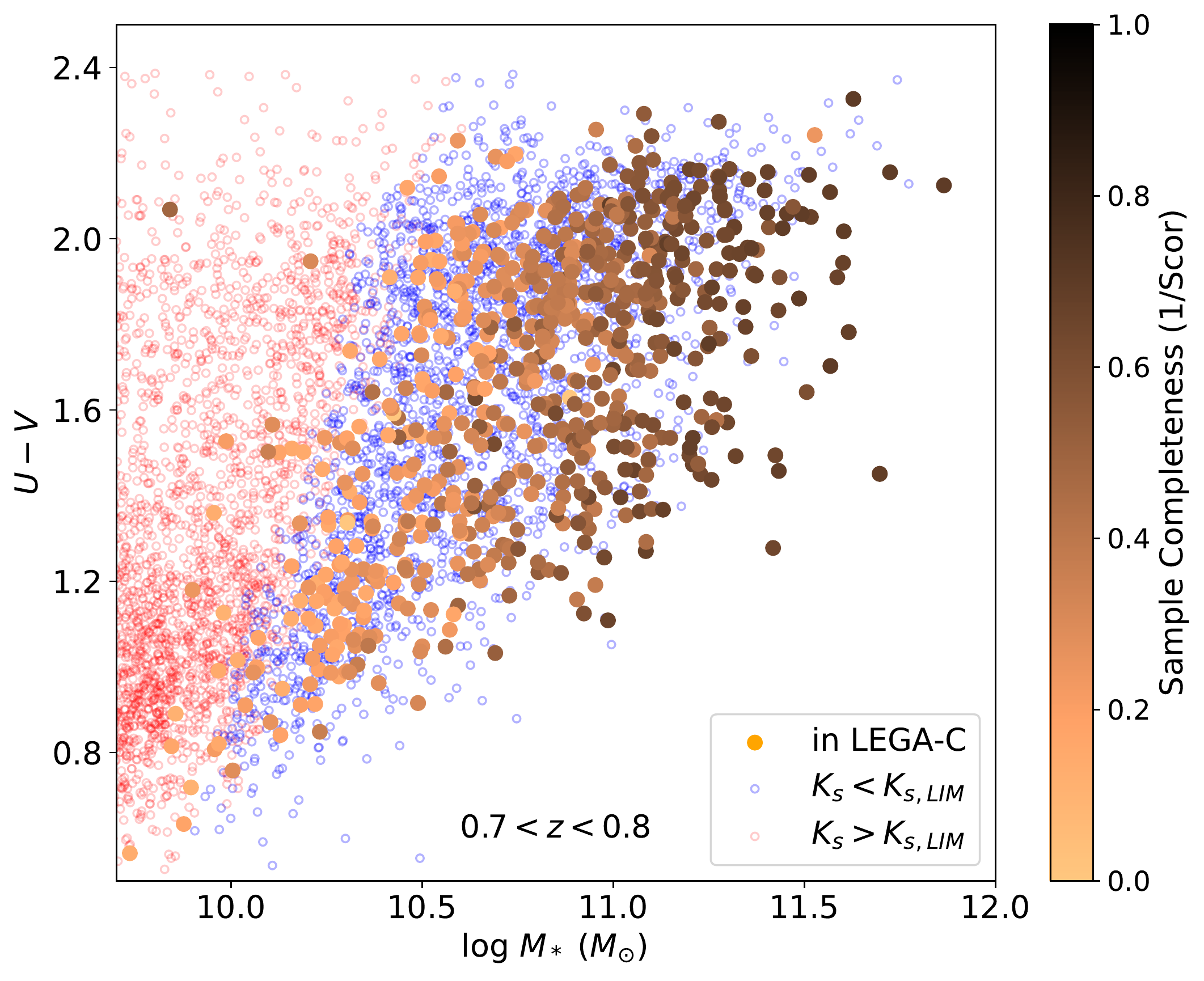} 
    \caption{Rest-frame $U-V$ color vs. stellar mass for galaxies in the redshift range $0.7<z<0.8$. The primary LEGA-C targets is color-coded with the $K_s$-dependent completeness with respect to the parent sample of $K_s$-selected galaxies, as in Figure \ref{fig:m_z}. The blue open circles represent the (unobserved) parent sample of $K_s$-selected galaxies; the red open circles represent all other galaxies in the UltraVISTA photometric catalog with $K_s$-band magnitudes fainer than the selection limit. The observed sample is less complete toward lower masses, but in a well-understood manner, with $Scor$ as the correction factor. The $K_s$ magnitude limit is color-dependent, with a lower stellar mass limit for blue galaxies compared to red galaxies. The LEGA-C sample is representative at $z=0.7-0.8$ down to a stellar mass limit of $\sim 3\times10^{10}~M_{odot}$.
    \label{fig:m_uv}}
\end{figure}

\begin{figure*}[t]
\epsscale{1.1}
\plottwo{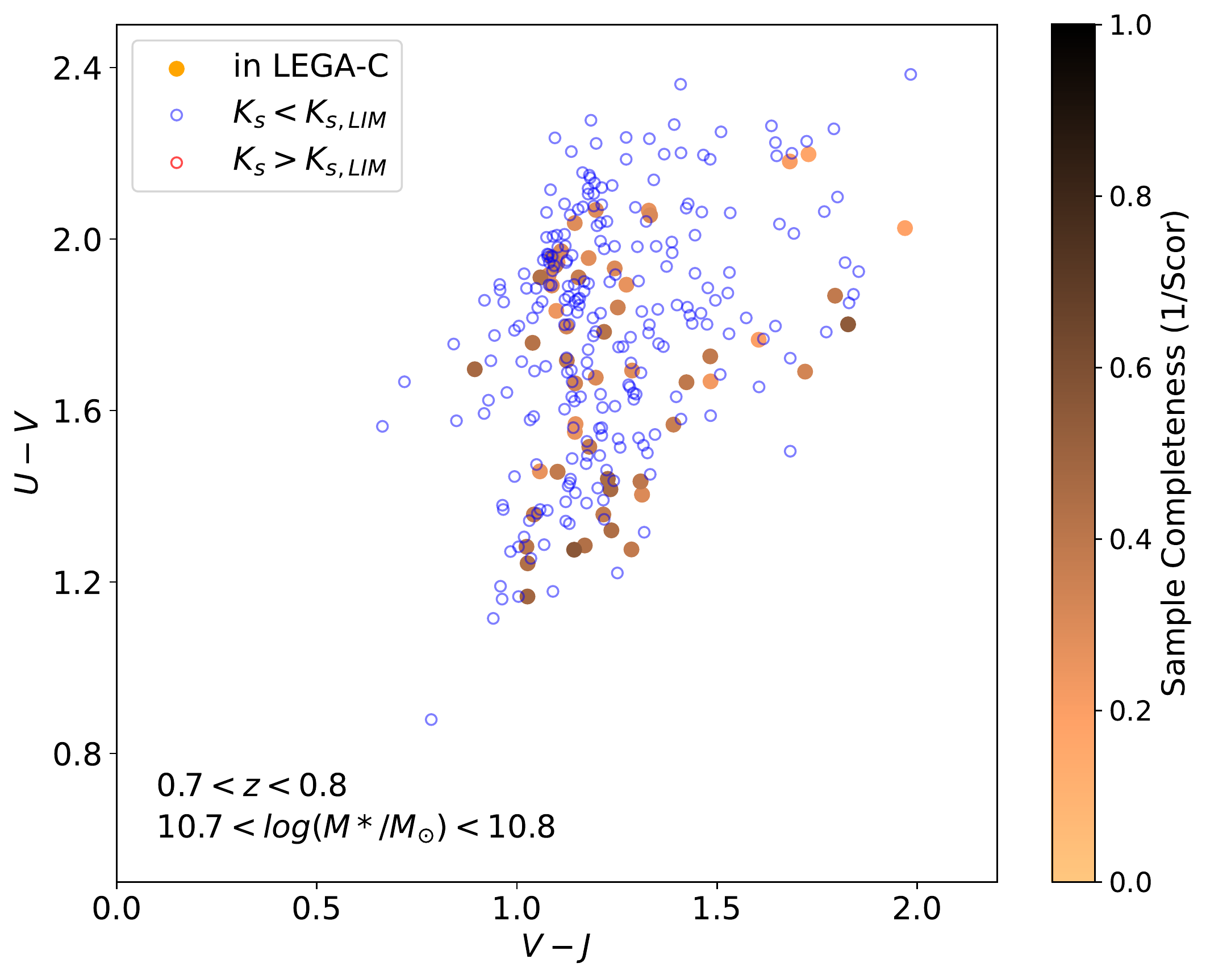}{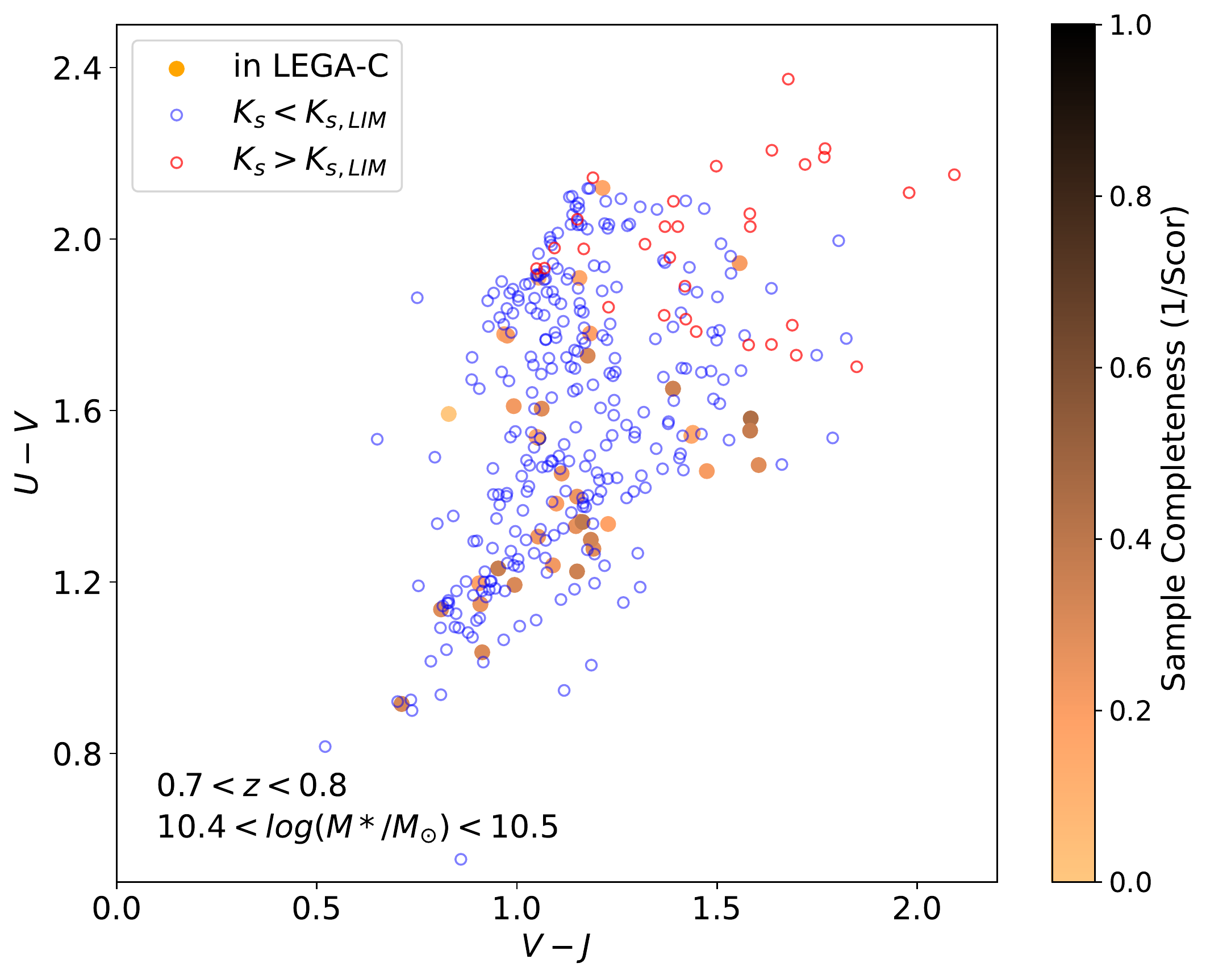} 
    \caption{Rest-frame color-color diagrams for two stellar mass-redshift bins to illustrate the completeness of the $K_s$-selected parent sample.  Left: example of a bin where all galaxies are brighter than the $K_s$ magnitude limit of the parent sample (there are no red open circles). The filled, color-coded circles are the galaxies included in the LEGA-C survey. The completeness  of the observed galaxies (1/Scor in the catalog), with respect to the parent sample, is $\sim35\%$ in this bin, but this number depends on galaxy color: redder galaxies are less likely to be included (due to their fainer $K_s$ magnitude) but since the parent sample itself does not suffer from incompleteness the observed galaxies are representative of the parent sample, removing any biases by assigning weight factors Scor to each individual galaxy. Right: example of a bin below the stellar mass limit. A large fraction of (red, dusty) galaxies has faint $K_s$ magnitudes so that they are not included in the parent sample. The observed galaxies are representative of the parent sample (the blue data points) but not the full population in this stellar mass-redshift bin.
    \label{fig:uvj}}
\end{figure*}

\begin{figure}[t]
\epsscale{0.55}
\plotone{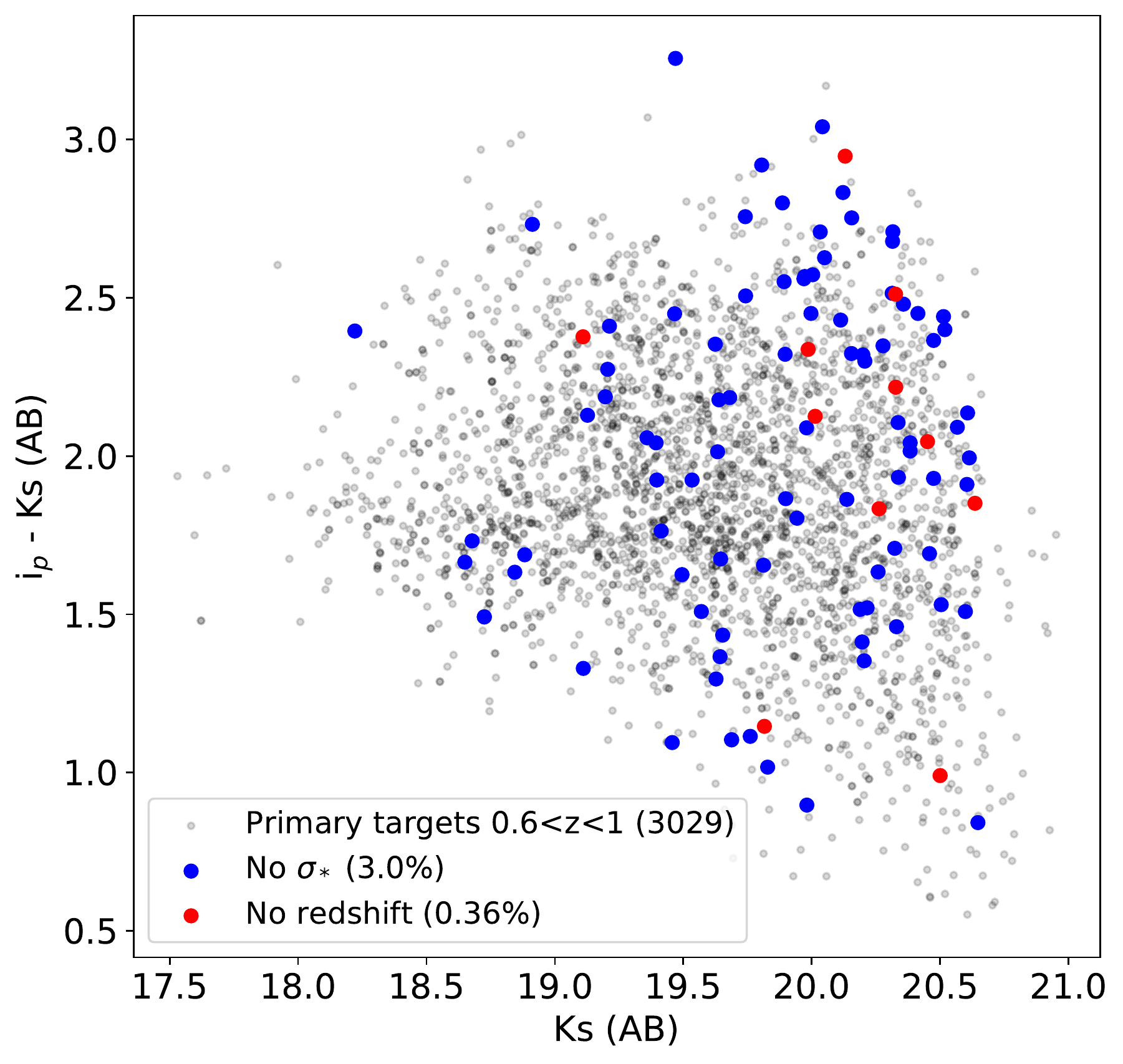} 
    \caption{Observed-frame color-magnitude diagram ($i_p$ - $Ks$ vs. $Ks$) of the primary sample, indicating those galaxies without successful redshift or stellar velocity dispersion measurements (naturally, a measured redshift is a prerequisite for a dispersion measurement). The success rate is extremely high -- even among the tail of optically faint targets ($i_p\sim22.5-23.0$) a velocity dispersion is measured in 85\% of the cases (see text for details).
    \label{fig:k_ik}}
\end{figure}

\begin{figure}[t]
\epsscale{0.6}
\end{figure}

The measurements from high-redshift galaxy spectroscopic surveys are often difficult to connect to the full galaxy population in terms of their completeness. Three factors play a key role in establishing this connection:

\begin{enumerate}
    \item How does the parent sample from which the spectroscopic targets are chosen connect to the full galaxy population in terms of its physical parameters? (For example, for LEGA-C, how does the $K_s$ magnitude limit connect to a stellar mass limit?)
    \item Given the parent sample, how are targets prioritized for observation? (If chosen randomly, this is not an issue; for LEGA-C, brighter targets in $K_s$ have higher priority.)
    \item What is the success rate of measuring the desired quantities? (For LEGA-C, these are stellar velocity dispersions and stellar population properties for which the success rate is very high: 97\%.)
\end{enumerate}

The LEGA-C survey is designed to control all three factors. \textbf{Factor 1:} Our $K_s$-band selection of the parent sample from which the targets are chosen is fairly close (but clearly not identical) to a stellar mass selection, with the added advantage that the stellar mass is a model-dependent quantity that may change over time, while the interpretation of a magnitude is fixed. To illustrate the effect of our parent sample selection -- $K_s < K_{s,LIM}$ as described in Section \ref{section:data} and Figure \ref{fig:K_hist} -- on completeness in terms of stellar mass we show in Figure \ref{fig:m_z} the stellar mass-redshift distribution of the primary targets\footnote{Stellar mass estimates do not play a major role in this data release paper and are used only to illustrate the issue of completeness here. The stellar masses used here, determined with the {\tt Prospector} SED fitting code with the setup used by \citet{leja19} and the $BVrizYJ$ photometry described in Appendix \ref{AppendixA}.}. $K_{s,LIM}$ results in a redshift-dependent stellar mass limit above which \emph{all} galaxies satisfy $K_s < K_{s,LIM}$, and below which this is not the case. 

In Figure \ref{fig:m_uv} we show the color-dependence on the mass completeness limit. This effect is illustrated in more detail in Figure \ref{fig:uvj} where we show, for specific stellar mass and redshift bins, the color-color distribution of the LEGA-C targets, the galaxies in the parent sample from which the targets are chosen, and galaxies that are fainter than $K_{s,LIM}$ (if they exist).  There we see that below the mass limit the parent sample is incomplete in terms of the reddest (dusty) galaxies. This implies that this completeness limit in reality depends on the science question at hand: if a user is interested in face-on blue disks the stellar mass completeness limit will be much lower. Figure \ref{fig:m_z} and Figure \ref{fig:uvj} merely serve to illustrate the mechanism by which incompleteness can be assessed, depending on one's requirements.

\textbf{Factor 2:} Galaxies from the $K_s$-selected parent sample are sorted by $K_s$ flux and included in the mask designs in order of decreasing flux. As a result, a larger fraction of $K_s$-bright galaxies from the parent sample are included in the survey compared to $K_s$-faint galaxies. In fact, the fraction of included galaxies (with respect to the parent sample) is a strictly monotonously increasing function of $K_s$ magnitude, which means that we know the probability of survey inclusion based on (and only on) $K_s$ magnitude. This probability is the inverse of the sample correction factor ({\tt Scor} in the catalog -- see Section \ref{sec:cat}).  In Figure \ref{fig:uvj} the filled data points (included in LEGA-C) are representative of the blue data points (the parent sample) when taking into account the completeness via {\tt Scor}. 

A second correction -- the volume correction factor {\tt Vcor} -- accounts for the fact that a galaxy at $z=0.7$ that is brighter than $K_{s,LIM}$ might not be sufficiently bright if it had been at $z=0.9$.  The product of {\tt Scor} and {\tt Vcor} -- which we call {\tt Tcor} in the catalog -- allows for the calculation of volume-limited quantities such as the stellar velocity dispersion function.  The extent to which this is useful depends on the amount of evolution in the time between $z=1$ and $z=0.6$; the correction {\tt Tcor} implicitly assumes $z=0.6$ and $z=1$ galaxies are the same.

\textbf{Factor 3:} Crucially, the observed primary sample is highly successful in measuring redshifts and stellar velocity dispersion measurements (Figure \ref{fig:k_ik}). Only 11 (0.4\%) of primary targets do not have a redshift measurement, for a variety of technical and astrophysical reasons. Two galaxies had previously measured spectroscopic measurements that turned out to be incorrect; three have poor spectra for technical reasons (e.g., vignetting), four spectra suffer from contamination by lower-redshift interlopers, preventing the useful extraction of spectrum, and two galaxies have good spectra without absorption or emission features. 97 (3\%) do not have a stellar velocity dispersion measurement; a measured velocity dispersion also serves as an indication that the stellar population properties (in the form of absorption line index values) can be measured with reasonable precision. A stellar velocity dispersion measurement most commonly fails among the faintest, reddest galaxies. But even among those 148 galaxies with $i$-band magnitudes fainter than 22.5, 85\% have a successful stellar velocity dispersion measurement, and all have redshift measurements. Other causes that lead to a failed $\sigma_*$ measurement are AGN and (very rarely) flux calibration errors. These issues are flagged as described below. The implication is that measurement success plays a negligible role in assessing the survey completeness. 

In summary, connecting the LEGA-C sample to the full galaxy population at $z=0.6-1$ is possible thanks to the high success rate of relevant measurements and the well-understood selection function. 

One issue has been neglected in the above analysis: we have assumed that the photometric redshift used for target selection is identical to the actual redshift (as measured on the basis of LEGA-C spectra). In reality there is a r.m.s. of 0.02 in $\Delta z = z_{\rm{phot}} - z_{\rm{LEGA-C}}$ for primary targets. For 97\% of the primary sample this scatter does not matter:  $(z_{\rm{phot}}$ and $z_{\rm{LEGA-C}}$ are both in the redshift range $0.6<z<1.0$.

The other 3\% can be divided into two groups. First, the group that arises due to random scatter near the redshift limits $z=0.6$ and $z=1$. We assign to this group all galaxies (2\%) with $0.56 < z_{\rm{LEGA-C}} < 0.60$ and  $1.00 < z_{\rm{LEGA-C}} < 1.04$, that is, within $2\sigma$ of the the redshift limits. These are still treated as primary targets and included in the above completeness and volume corrections. This does not artificially increase the number density, since there is a set of galaxies that is not included in the survey while they should have been: galaxies with $0.6<z<1.0$ but with $0.56 < z_{\rm{phot}} < 0.6$ or $1.00 < z_{\rm{phot}} < 1.04$. Since the scatter in $\Delta z$ is symmetric, the sets of galaxies that were accidentally included and accidentally excluded cancel out in the above volume correction calculation.

The second group consists of galaxies (1\%) with more strongly deviating redshift measurements:  $z_{\rm{LEGA-C}} < 0.56$ or $z_{\rm{LEGA-C}} > 1.04$. These `catastrophic' outliers are excluded from the primary sample and not taken into account when calculating the completeness and volume corrections, which has a negligible effect on the analysis given their small number.

\restartappendixnumbering
\section{Photometry and Fitting the Spectral Energy Distribution}
\label{AppendixA}

We use a subset of the photometry from \citet[][hereafter: \citetalias{muzzin13}]{muzzin13} with revised zero points.  In M13 zero point corrections are applied that remove systematic differences between the initially calibrated photometry and template spectral energy distributions.  While this greatly aids photometric redshift estimates, systematic errors in the colors of galaxies can be propagated into the catalog if the templates themselves -- which are based on a particular flavor of stellar population synthesis models -- contain systematic errors. We will show below that small but significant systematic offsets exist in the rest-frame UV and near-IR parts of the spectrum.  But first we reassess the relative flux calibrations of the M13 UltraVISTA catalog.

\begin{figure}[!h]
\epsscale{1.1}
\plotone{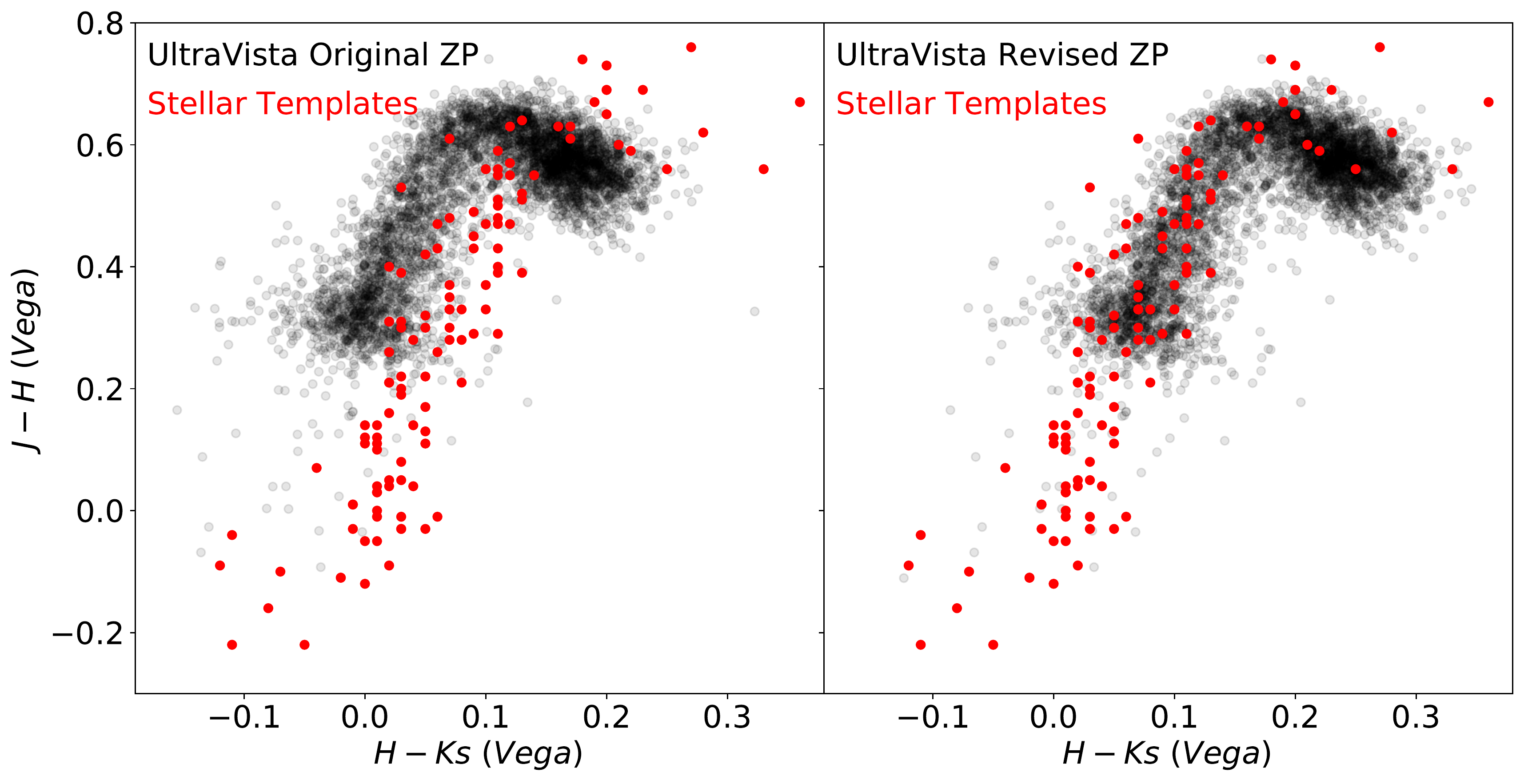} 
    \caption{$J-H$ vs.~$H-K_s$ colors of stars (black) in the UltraVISTA catalog from \citet{muzzin13} compared with synthesized colors of stars (red) from the Pickles library. The left-hand panel shows the star colors with the original zero points; the right-hand stars show the colors with the revised zero points. The red points are idential in both panels. \label{fig:jhk_stars}}
\end{figure}

\begin{figure}[!h]
\epsscale{1.1}
\plotone{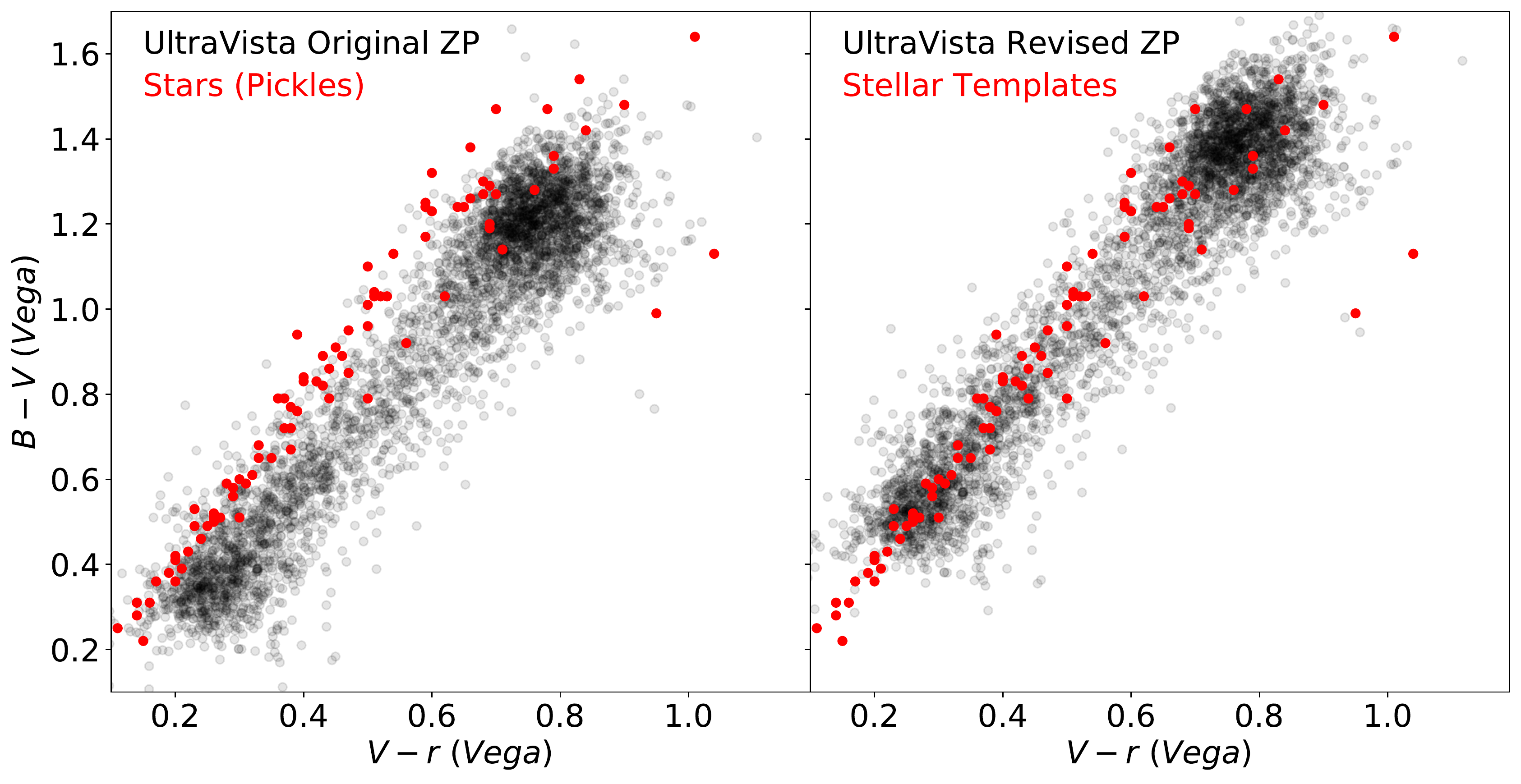} 
    \caption{$B-V$ vs.~$V-r$ colors of stars (black) in the UltraVISTA catalog from \citet{muzzin13} compared with synthesized colors of stars (red) from the Pickles library. The left-hand panel shows the star colors with the original zero points; the right-hand stars show the colors with the revised zero points. The red points are identical in both panels.\label{fig:bvr_stars}}
\end{figure}

In Figure \ref{fig:jhk_stars} (left-hand panel) we show that the near-IR photometry of stars in the original M13 catalog does not coincide with the synthesized photometry of the \citet{pickles98} library of stellar spectra\footnote{M stars or AGB stars in the Pickles library are known to have imperfect model spectra, but only a few such stars (the reddest in Figure \ref{fig:jhk_stars}) are included here, and these do not affect our analysis.}. An examination of the photometry in all nine filters listed in Table \ref{tab:zp} made clear that the $K_s$ photometry is systematically offset from the other UltraVISTA filters ($Y$, $J$, and $H$) by 0.07 mag. Applying this shift aligns the observed and synthesized stellar photometry well. We elected to keep the total $K_s$ band magnitude unchanged in order not to change the absolute calibration of the source detection filter ($K_s$). For the optical filters $V$, $r$, $i$ and $z$ similar -- marginally smaller -- offsets were required to align the photometry (Table \ref{tab:zp}). For the $B$ band a larger shift was required (see Figure \ref{fig:bvr_stars}). All in all, only the $B$ and $K_s$ filters received a relative zero point correction larger than 0.03 mag.  The filters $V$, $r$, $i$, $z$, $Y$, $J$ and $H$ were already consistent with each other to within 3\%.

\begin{figure}[!h]
\epsscale{1.1}
\plottwo{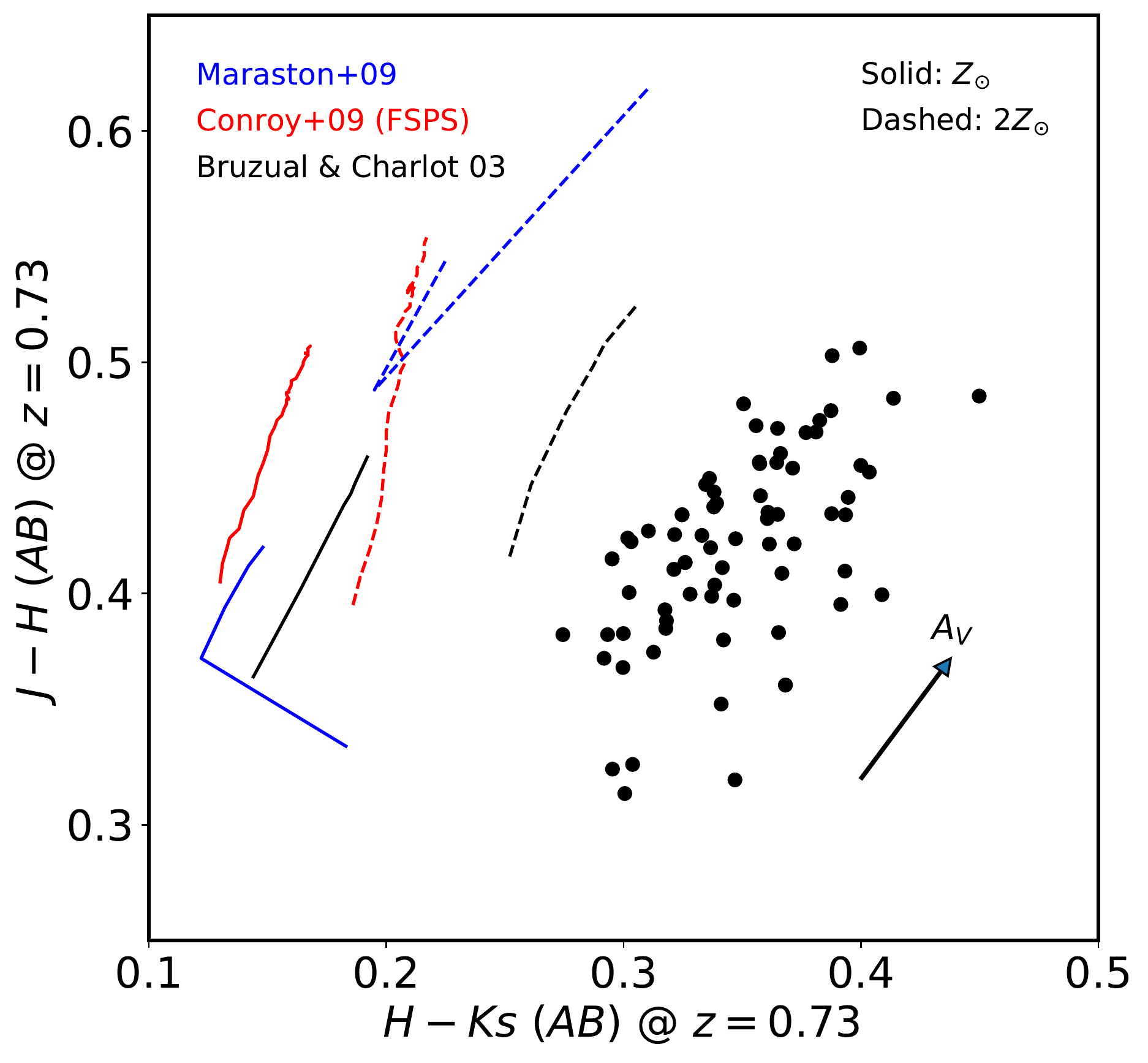}{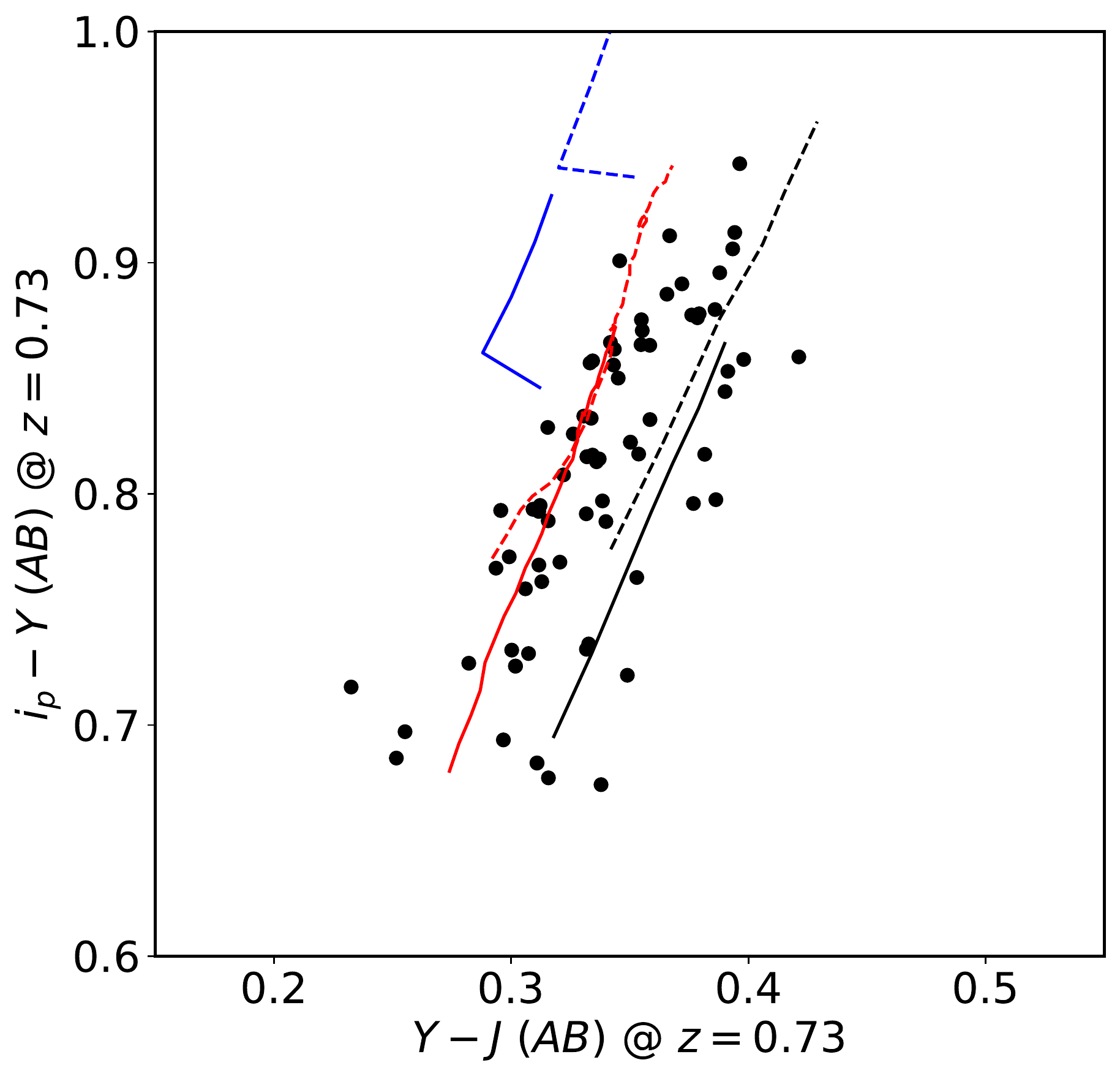} 
    \caption{$J-H$ vs.~$H-K_s$ (left-hand panel) and $i_p - Y$ vs.~$Y-J$ colors (all with revised zero points) of massive, passive galaxies at $z=0.72-0.74$ selected from LEGA-C, compared with synthesized photometry for three SPS models (Maraston, FSPS and BC03) and two metallicities (solar and $2\times$ solar) in the age range 2-6 Gyr. \label{fig:jhk_galaxies}}
\end{figure}

Now that we have accurate zero points that are independent of stellar population synthesis models we are in a position to verify those models for systematic errors. In Figure \ref{fig:jhk_galaxies} (left-hand panel) we show the $J-H$ and $H-K_s$ colors with revised zero points for a sample of 76 quiescent galaxies (UVJ selected) with stellar velocity dispersions $\sigma > 200~km~s^{-1}$ and in the redshift range $0.72<z<0.74$, chosen to contain a significant overdensity of massive galaxies.   The colors are in the observed frame at $z=0.73$ and the filters correspond to rest-frame wavelengths of $\sim$0.7, 0.95 and 1.3$\mu m$. In the right-hand panel we show the $i_p - Y$ and $Y-J$ colors for the same sample (rest-frame: $\sim$0.44, 0.6, and 0.7$\mu$m). The majority of these galaxies have been shown on the basis of their high-quality optical spectra from LEGA-C to have little or no star formation in the past few Gyr and ages typically in the range of 3-5 Gyr \citep{chauke18}: their colors should resemble those of single stellar populations (SSPs). We compare these colors with synthesized photometry for a variety of SPS models with ages in the range of 2 to 6 Gyr.  The FSPS models \citep{conroy09} and the BC03 models use different Padova evolutionary tracks: \citet{marigo07} and \citet{bertelli94}, respectively.   Neither solar nor super-solar models match the observed $H-K_s$ colors, and attenuation is unlikely to explain the offset both due to mismatched `direction' in color-color space and the direct evidence for low attenuation even at shorter wavelengths for this type of galaxy. Considering this set of photometric measurements in isolation, both the Maraston and FSPS models can match the data by adopting high metallicities, younger ages (0.5 - 1 Gyr) and a significant amount of attenuation, but these parameter ranges are firmly ruled out by independent constraints on age (from the LEGA-C spectra) and attenuation. In the rest-frame optical wavelength range (probed by the colors used in the right-hand panel of the figure) no such wholesale offsets are apparent and the model colors generally correspond well with the observed colors.  The result that no large changes in metallicity, age, and attenuation are required to match the optical spectral energy distribution (SED) suggests that the problems lie with the models in the near-IR.

\begin{figure}[h]
\epsscale{.55}
\plotone{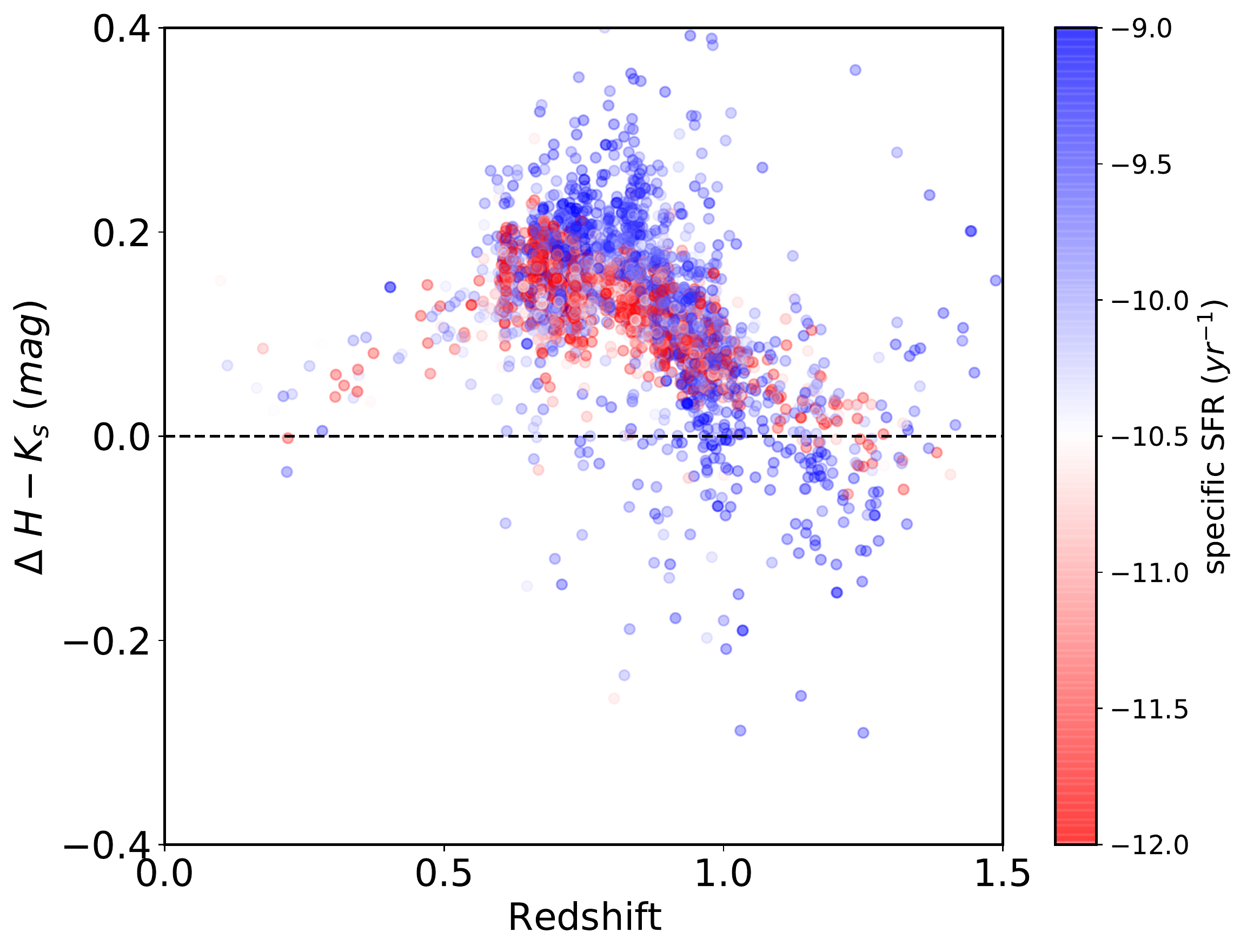} 
    \caption{Difference between observed $H-K_s$ color and inferred $H-K_s$ based on fits to the $BVrizYJ$ photometry. Color coding is with specific star formation rate: red colors indicate passive galaxies; blue colors indicate high star-formation activity. The strong redshift dependence indicates the offsets are not due to a simple calibration error in the photometry (which would lead to a redshift-independent offset). The small difference between young (star-forming) and old (quiescent) galaxies that the offsets are not strongly age dependent. \label{fig:residHK}}
\end{figure}

In order to analyse the near-infrared offsets further we use the fitting code Prospector \citep{leja19}. We provide the photometry in the wavelength range $B$ to $K_s$ for the filter set given in Table \ref{tab:zp} but $H$ and $K_s$ are not used in the fit; instead, residuals are calculated based on the models that are fit to the $B$ through $J$ photometry. \footnote{Note that the MIPS 24$\mu$m data are also used in this fit in order to constrain the star-formation rate.} In Figure \ref{fig:residHK} we show the residuals in $H-K$ across redshift. These differences between observed and inferred $H-K_s$ colors are to first order independent of galaxy type (star-formation activity) but strong change with redshift. The implication is that the color offsets seen for old galaxies (Figure \ref{fig:jhk_galaxies}) are present in all galaxies in the redshift range $z=0.6-1$: the models systematically underpredict the $H-K_s$ colors. The redshift dependence suggest that the photometric measurements are not the problem: these would be redshift independent.

We conclude that using photometry longward of rest-frame $\sim0.8\mu\rm{m}$ leads to systematic errors in fits to the spectral energy distribution of $\sim20\%$, which will propagate into small but systematic errors in the stellar mass, star-formation rate, attenuation, and other parameters that are inferred.   For the purpose of calibrating our spectra we choose the $BVrizYJ$ filter set. We note that if $H$ and $K_s$ are included in the fits, then the attenuation parameters are adjusted to accommodate for the systematic effect. We do not fully explore this here, but one illustrative result is that passive galaxies typically require $A_V\sim 0.5$, which is ruled out by other means (optical depth estimates; attenuation measurements from spectra, ...). However, stellar mass estimates differ by no more than $\sim$0.05 dex on average if $H$ and $K_s$ are included in SED fits.

\begin{deluxetable}{c|ccccccccc}
\tabletypesize{\scriptsize}
\tablecaption{Zero points shifts in magnitude for 9 filters. A positive number corresponds with decreasing the flux density / increasing the magnitude.  The shift in $K_s$ is zero by construction. \label{tab:zp}}
\tablewidth{0pt}
\tablehead{
\colhead{Filter} & 
\colhead{$B$} & \colhead{$V$} & \colhead{$r$} & \colhead{$i$} & \colhead{$z$} & \colhead{$Y$} & \colhead{$J$} & \colhead{$H$} & \colhead{$K_s$}}
\startdata
Zero point shift & 0.12 & 0.05 & 0.04 & 0.04 & 0.05 & 0.07 & 0.07 & 0.07 & 0.00 \\
\enddata
\end{deluxetable}

\restartappendixnumbering
\section{Uncertainties in Spectral Properties: Comparing Duplicate Observations}
\label{AppendixB}

The formal uncertainties, based on the noise spectra, may not fully capture the full random error budget due to correlations in the noise spectra and systematics in the spectra that propagate to random errors.  For specific measurements there may be specific contributions: template mismatch for $\sigma_*$, the quasi-random wavelength coverage of emission lines for $\sigma_g$, flux calibration uncertainties for spectral index measurements, and so forth.

By design, LEGA-C targeted hundreds of galaxies multiple times, providing an automatic assessment of the complete budget of spectrum-to-spectrum random uncertainties. We compare the variance in the differences between duplicate measurements to the expected variance inferred from the formal measurement uncertainties, and correct the latter if necessary. This approach was already used in DR2 -- here we show that due to improvements in the data reduction and analysis techniques such corrections are generally smaller than before. For the analysis we choose the sample of 205 galaxies for which we have duplicate spectra with $S/N>5$ and successful redshift measurements. (We omit galaxies in mask 101, which were taken with the slits perpendicular to the rest of the survey. This leads to physically different velocity dispersions due to projection effects along the slit.) The scatter between the duplicate measurements is compared with the random uncertainty (added in quadrature), producing a $\sqrt{\chi^2/N}$ value to parameterize the excess scatter compared to the scatter that can be expected on the basis of the formal measurement uncertainties.  This number serves as the correction factor on our measurement uncertainties, propagated into our published catalog.

\begin{figure}[!h]
\epsscale{1.1}
\plottwo{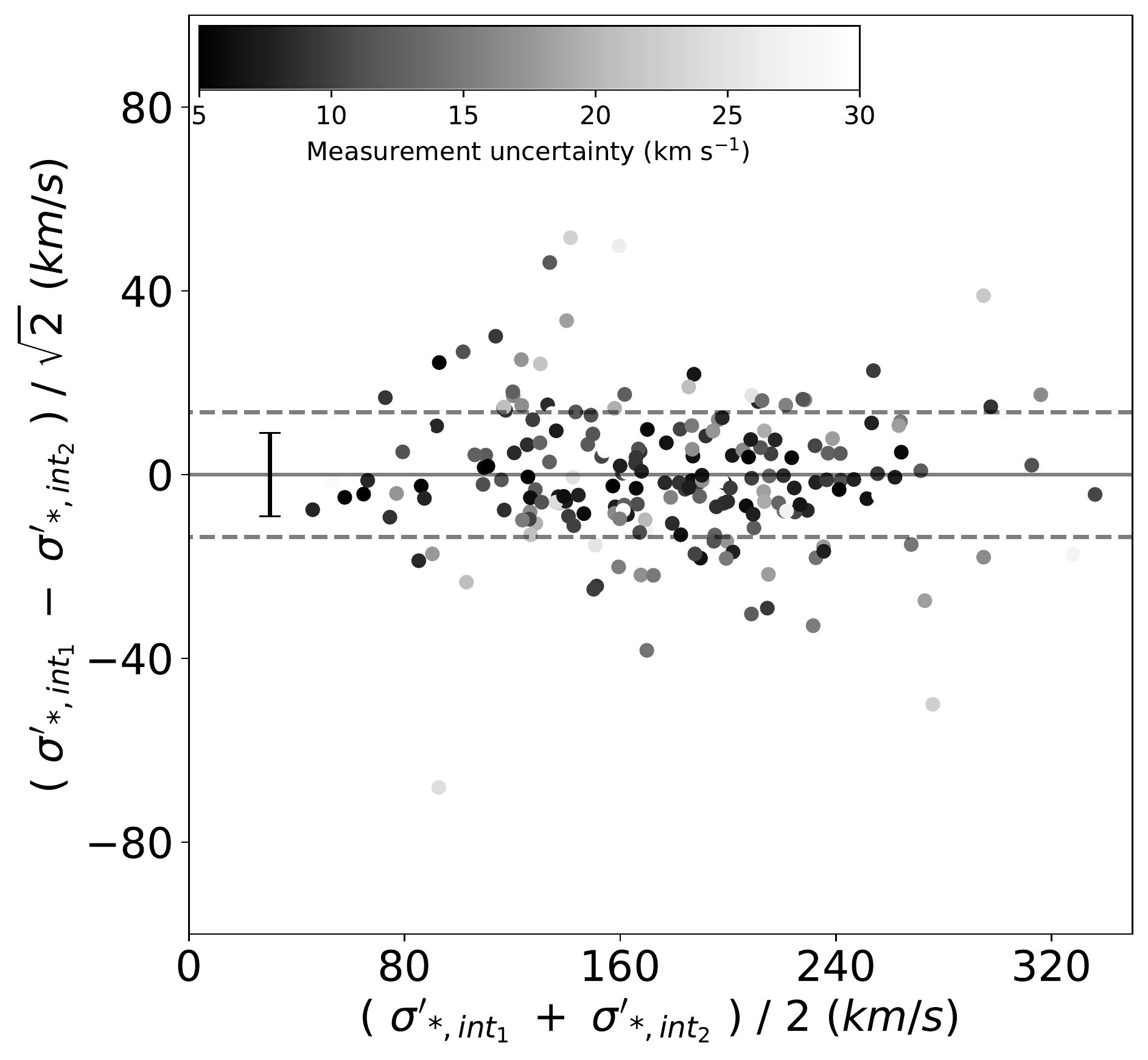}{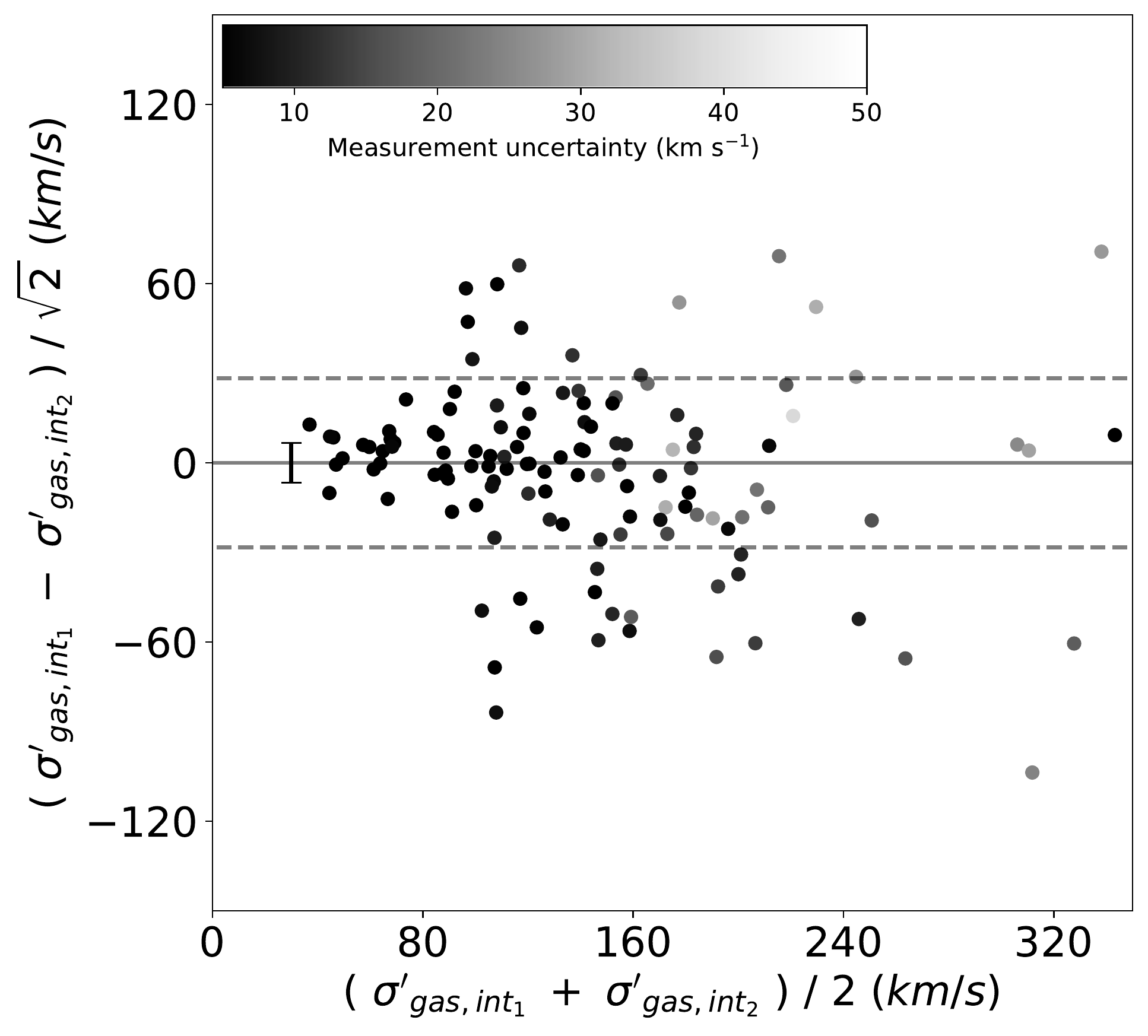} 
    \caption{Comparison of duplicate measurements of \sigs~(left) and \sigg~(right). The grey scaling reflects the measurement uncertainties, which serves to increase the focus on high-precision measurements isntead of uncertain outliers. The factor $\sqrt{2}$ on the y axis serves to convert the difference between the duplicate measurements into a number that reflects the uncertainty on the individual measurements. The dashed lines indicate the scatter; the error bar represents the typical formal uncertainty. The upward correction factor on the formal uncertainties serve to match those with the observed scatter among the duplicate obserations. \label{fig:sig_sig}}
\end{figure}

We first examine the duplicate measurements of \sigs~(Figure \ref{fig:sig_sig}, left-hand side), and conclude that the true uncertainties are 50\% higher than the formally derived uncertainties.  There are several possible causes: template mismatch, correlated noise, and wavelength coverage. For \sigg~(Figure \ref{fig:sig_sig}, right-hand side) we infer a similar correction of 1.5, bearing in mind the 15\% systematic uncertainty discussed in Figure \ref{section:emission} due to emission line-specific slit losses.   

\begin{figure}[!h]
\epsscale{1.17}
\plotone{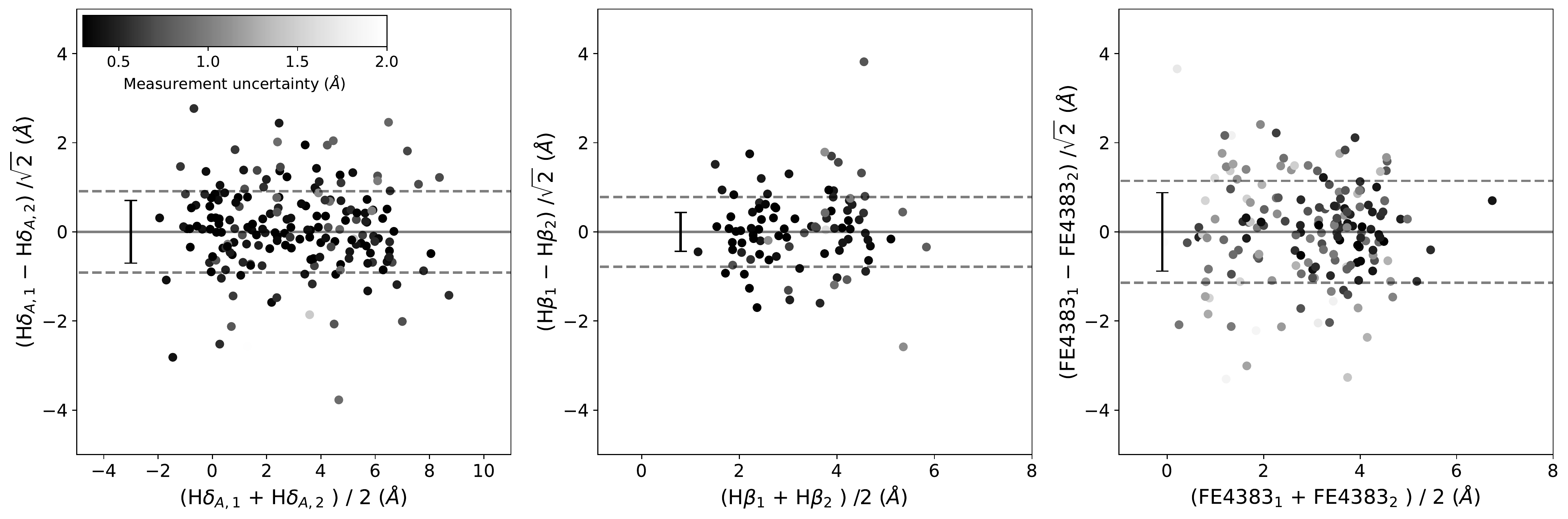}
    \caption{ Comparison of duplicate measurements of 3 absorption line indices (LICK\_H$\delta$\_A, LICK\_H$\beta$, LICK\_FE4383). See Fig.~\ref{fig:sig_sig} caption and text for details. \label{fig:idx_idx}}
\end{figure}

Figure \ref{fig:idx_idx} shows the results for duplicate index measurements for three spectral features. In all cases the formal uncertainties, indicated with the error bars, are somewhat smaller than the scatter among duplicates.  Analogous to the correction factors on the velocity dispersion uncertainties we measure correction factors for all absorption indices of a factor 1.3. Only LICK\_H$\beta$ (1.8) and D$_N$4000 (2.0) receive larger correction factors. The measurement of the former is complicated by the often-bright emission line that is subtracted before the absorption strength is measured. The precision of the measurement of the latter is limited by imperfections in the wavelength dependence of the flux calibration.

Formal uncertainties on the emission line fluxes and equivalent widths are more difficult to determine due to the aforementioned systematic problems with variable slit losses. The analysis of the LICK\_H$\beta$ absorption and emission uncertainties, driven in a large part by the decomposition of the spectrum into ionized gas emission and stellar light, led us to adopt a correction factor of 1.8.  This correction factor is applied to all emission line measurements in the catalog, again keeping in mind a 15\% systematic uncertainty that is not propagated. These results compare well with similar tests on emission lines in the Sloan Digital Sky Survey \citep{brinchmann08}. They find typical corrections to the uncertainty estimates are a factor $1.2-1.5$ for forbidden lines and a factor of $2.0$ for Balmer lines and OII\_3727.

\end{document}